\documentclass[english,latin9,longauth]{aa}
\usepackage[T1]{fontenc}
\usepackage{textcomp}
\usepackage{url}
\usepackage{amssymb}
\usepackage{graphicx}
\usepackage{esint}

\makeatletter

\newcommand{\noun}[1]{\textsc{#1}}

\usepackage{xcolor}

\makeatother

\usepackage{babel}
\begin{document}

\title{\emph{TESS} first look at evolved compact pulsators}
\subtitle{Discovery and asteroseismic probing of the \emph{g}-mode hot B subdwarf
pulsator EC 21494-7018}
\author{S. Charpinet\inst{1}\and P. Brassard\inst{2}\and G. Fontaine\inst{2}\and V.
Van Grootel\inst{3}\and W. Zong\inst{4}\and N. Giammichele\inst{1}\and U.
Heber\inst{5}\and Zs. Bogn\'ar\inst{6,7}\and S. Geier\inst{8}\and E.M.
Green\inst{9}\and J.J. Hermes\inst{10}\and D. Kilkenny\inst{11}\and R.H.
{\O}stensen\inst{12}\and I. Pelisoli\inst{8}\and R. Silvotti\inst{13}\and J.H.
Telting\inst{14}\and M. Vu\v{c}kovi\'c\inst{15}\and H.L. Worters\inst{16}\and A.S.
Baran\inst{17}\and K.J. Bell\inst{18,19}\and P.A. Bradley\inst{20}\and J.H.
Debes\inst{21}\and S.D. Kawaler\inst{22}\and P. Ko\l{}aczek-Szyma\'nski\inst{23}\and S.J.
Murphy\inst{24}\and A. Pigulski\inst{23}\and \`A. S\'odor\inst{6,7}\and M.
Uzundag\inst{15}\and R. Handberg\inst{25}\and H. Kjeldsen\inst{25}\and G.R.
Ricker\inst{26}\and R.K. Vanderspek\inst{26}}
\institute{Institut de Recherche en Astrophysique et Planétologie, CNRS, Université
de Toulouse, CNES, 14 avenue Edouard Belin, F-31400 Toulouse, France\\
\email{stephane.charpinet@irap.omp.eu}\and Département de Physique,
Université de Montréal, Québec H3C 3J7, Canada\and Space sciences,
Technologies and Astrophysics Research (STAR) Institute, Université
de Liège, 19C Allée du six-août, B-4000 Liège, Belgium\and Department
of Astronomy, Beijing Normal University, Beijing 100875, China\and Dr.
Karl Remeis-Observatory \& ECAP, Astronomical Institute, Friedrich-Alexander
University Erlangen-Nürnberg (FAU), Sternwartstr. 7, 96049, Bamberg,
Germany\and Konkoly Observatory, MTA Research Centre for Astronomy
and Earth Sciences, Konkoly Thege Miklós út 15-17, H-1121 Budapest\and MTA
CSFK Lend\"ulet Near-Field Cosmology Research Group\and Institute
for Physics and Astronomy, University of Potsdam, Karl-Liebknecht-Str.
24/25, D-14476 Potsdam, Germany\and Steward Observatory, University
of Arizona, 933 North Cherry Avenue, Tucson, AZ, 85721, USA\and Department
of Astronomy, Boston University, 725 Commonwealth Ave., Boston, MA
02215 - USA\and Department of Physics and Astronomy, University of
the Western Cape, Private Bag X17, Bellville 7535, South Africa\and Department
of Physics, Astronomy, and Materials Science, Missouri State University,
Springfield, MO 65897, USA\and INAF-Osservatorio Astrofisico di Torino,
strada dell'Osservatorio 20, I-10025 Pino Torinese, Italy\and Nordic
Optical Telescope, Rambla José Ana Fernández Pérez 7, 38711 Breña
Baja, Spain\and Instituto de Física y Astronomía, Universidad de
Valparaiso, Gran Bretaña 1111, Playa Ancha, Valparaíso 2360102, Chile\and South
African Astronomical Observatory, PO Box 9, Observatory, Cape Town
7935, South Africa\and Uniwersytet Pedagogiczny, Obserwatorium na
Suhorze, ul. Podchor\.{z}ych 2, 30-084 Kraków, Polska\and DIRAC
Institute, Department of Astronomy, University of Washington, Seattle,
WA 98195-1580, USA\and NSF Astronomy and Astrophysics Fellow and
DIRAC Fellow\and XCP-6, MS F699, Los Alamos National Laboratory,
Los Alamos, NM 87545 USA\and Space Telescope Science Institute, 3700
San Martin Drive, Baltimore, MD 21218, USA\and Department of Physics
and Astronomy, Iowa State University, Ames, IA 50011, USA\and Instytut
Astronomiczny, Uniwersytet Wroc\l awski, ul. Kopernika 11, 51-622
Wroc\l aw, Poland\and Sydney Institute for Astronomy (SIfA), School
of Physics, The University of Sydney, NSW 2006, Australia\and Stellar
Astrophysics Centre, Department of Physics and Astronomy, Aarhus University,
Ny Munkegade 120, DK-8000 Aarhus C, Denmark\and Department of Physics,
and Kavli Institute for Astrophysics and Space Research, Massachusetts
Institute of Technology, Cambridge, MA 02139, USA\\
}
\offprints{S. Charpinet}
\date{Received ...; Accepted...}
\abstract{The \emph{TESS} satellite was launched in 2018 to perform high-precision
photometry from space over almost the whole sky in a search for exoplanets
orbiting bright stars. This instrument has opened new opportunities to study variable
hot subdwarfs, white dwarfs, and related compact objects. Targets
of interest include white dwarf and hot subdwarf pulsators, both carrying
high potential for asteroseismology. }{We present the discovery and detailed asteroseismic analysis of a
new $g$-mode hot B subdwarf (sdB) pulsator, EC 21494-7018 (TIC 278659026),
monitored in \emph{TESS} first sector using 120-second cadence. }{The\emph{ TESS} light curve was analyzed with standard prewhitening
techniques, followed by forward modeling using our latest generation
of sdB models developed for asteroseismic investigations. By simultaneously best-matching all the observed frequencies with those computed from
models, we identified the pulsation modes detected and, more importantly,
we determined the global parameters and structural configuration of
the star.}{The light curve analysis reveals that EC 21494-7018 is a sdB pulsator
counting up to 20 frequencies associated with independent $g$-modes.
The seismic analysis singles out an optimal model solution in full
agreement with independent measurements provided by spectroscopy (atmospheric
parameters derived from model atmospheres) and astrometry (distance
evaluated from \emph{Gaia} DR2 trigonometric parallax). Several key
parameters of the star are derived. Its mass ($0.391\pm0.009$ $M_{\odot}$)
is significantly lower than the typical mass of sdB stars and suggests
that its progenitor has not undergone the He-core flash; therefore
this progenitor could originate from a massive ($\gtrsim2$ $M_{\odot})$ red giant,
which is an alternative channel for the formation of sdBs. Other
derived parameters include the H-rich envelope mass ($0.0037\pm0.0010$
$M_{\odot}$), radius ($0.1694\pm0.0081$ $R_{\odot}$), and luminosity
($8.2\pm1.1$ $L_{\odot}$). The optimal model fit has a double-layered
He+H composition profile, which we interpret as an incomplete but
ongoing process of gravitational settling of helium at the bottom
of a thick H-rich envelope. Moreover, the derived properties of the
core indicate that EC 21494-7018 has burnt $\sim43\%$ (in mass) of
its central helium and possesses a relatively large mixed core ($M_{{\rm core}}=0.198\pm0.010$
$M_{\odot}$), in line with trends already uncovered from other g-mode
sdB pulsators analyzed with asteroseismology. Finally, we obtain for
the first time an estimate of the amount of oxygen (in mass; $X({\rm O})_{{\rm core}}=0.16{}_{-0.05}^{+0.13}$)
produced at this stage of evolution by an helium-burning core. This
result, along with the core-size estimate, is an interesting constraint
that may help to narrow down the still uncertain $^{12}{\rm C}(\alpha,\gamma)^{16}{\rm O}$
nuclear reaction rate.}{}
\maketitle

\section{Introduction}

The NASA \emph{Transiting Exoplanet Survey Satellite} (\emph{TESS}),
successfully launched on 2018 April 18, is the latest instrument dedicated
to high-precision photometric monitoring of stars from space. Aside from
its main objective to identify new exoplanets transiting nearby stars
\citep{2014SPIE.9143E..20R}, \emph{TESS} is expected to contribute
significantly to the study of stellar variability, extending in particular
the use of asteroseismology for all types of pulsating stars. An important
asset of \emph{TESS}, compared to its predecessors \textendash{}
the satellites \emph{Kepler 2} (\emph{K2}; \citealt{2014PASP..126..398H}),
\emph{Kepler} \citep{2010Sci...327..977B,2010PASP..122..131G},
\emph{CoRoT} (\emph{Convection, Rotation et Transits planétaires}; \citealt{2006cosp...36.3749B}),
and \emph{MOST} (\emph{Microvariability and Oscillations of Stars}; \citealt{2003PASP..115.1023W})
\textendash{} is the much more extended sky coverage because during the
two-year duration of the main mission, the satellite will survey over
90\% of the sky, avoiding only a narrow band around the ecliptic already
explored, in part, by \emph{K2}.

The \emph{TESS }nominal two-year survey is sectorized, with each sector
consisting of a nearly continuous observation of the same $24^{\circ}\times\ensuremath{90^{\circ}}$
field for $\sim27$ days. Some overlap between sectors exists for
the highest northern and southern ecliptic latitudes, meaning that
some stars can be observed longer. In particular, stars located in
the continuous viewing zone (CVZ), close to the ecliptic caps, could
be monitored for as much as a year. The\emph{ TESS} data products include
full-frame images (FFI) taken every 30 minutes and containing the
entire field of view, as well as short-cadence observations sampled
every 120 seconds, providing a better time resolution for a selection
of approximately 16,000 stars per sector. An even faster 20 s cadence
mode is also considered for a small selection of rapidly varying objects
(including evolved compact pulsators), but this ``ultra short''
sampling rate will only be available for the extended mission that
will follow the original two-year survey.

Efforts have been made since 2015 to assemble lists of evolved compact
stars, mostly white dwarfs and hot subdwarfs, to be submitted for
the shortest cadence modes. This was coordinated through the \emph{TESS
Asteroseismic Science Consortium} (TASC)\footnote{ \url{https://tasoc.dk}}
Working Group 8 (WG8), which ultimately proposed an extensive variability
survey using the \emph{TESS} 120 s cadence mode for all known evolved
compact stars brighter than $\sim16$$^{{\rm th}}$ magnitude. A shorter
list of selected objects, mostly fast white dwarf and hot subdwarf
pulsators that critically depend on the planned 20 s-sampling, is
also kept updated for the upcoming extended mission. These target
lists were assembled from existing catalogs of hot subdwarf and white dwarf
stars, further enriched by discoveries of new objects of this kind
obtained by dedicated efforts conducted from ground-based facilities\footnote{Dedicated efforts in preparation for \emph{TESS} have been carried
out from Steward Observatory (U. of Arizona, USA), Nordic Optical
Telescope (Spain), South African Astronomical Observatories (South
Africa), and Piszk\'estet\H o (Konkoly Observatory, Hungary).}. To date, approximately 2,600 white dwarfs and 3,150 hot subdwarf
stars are scheduled to be observed by \emph{TESS} as part of the TASC
WG8 120 s cadence list, while the 20 s cadence list counts approximately
400 targets\footnote{Information can be found from TASC WG8 wiki pages accessible from
\url{https://tasoc.dk} after registration.}.

Monitoring pulsating hot B subdwarf (sdB) stars with \emph{TESS} is
one among several objectives pursued by TASC WG8. The occurrence of
nonradial pulsations in sdB stars provides an extraordinary way, through
asteroseismology, to probe their inner structure and dynamics. Hot
sdBs are associated with the so-called extreme horizontal branch
(EHB), forming a blue extension to the horizontal branch. These stars
correspond to low-mass (typically $\sim0.47$ $M_{\odot}$) objects
burning helium in their cores \citep[see ][for a recent review on the subject]{2016PASP..128h2001H},
and as such, they are representative of this intermediate phase of
stellar evolution. They differ from classical horizontal branch stars
mainly at the level of their residual H-rich envelope, which has been
strongly reduced during the previous stage of evolution, leaving only
a thin layer less massive than $\sim0.02$ $M_{\odot}$. As a consequence,
sdB stars remain hot and compact ($T_{{\rm eff}}\sim$ 22 000 \textendash{}
40 000 K, log g \ensuremath{\sim} 5.2 \textendash{} 6.2; \citealp{1994ApJ...432..351S})
throughout their He-burning lifetime ($\sim150$ Myr), and never ascend
the asymptotic giant branch before reaching the white dwarf cooling
tracks \citep[e.g.,][]{1993ApJ...419..596D}.

Two main classes of sdB pulsators have offered, so far, the opportunity
to use asteroseismology to investigate this intermediate evolutionary
stage. The \emph{V361 Hya} stars (also named sdBV$_{{\rm r}}$ or
\emph{EC14026} stars from the class prototype; \citealp{1997MNRAS.285..640K}
and see the nomenclature proposed by \citealp{2010IBVS.5927....1K})
were the first to be discovered and oscillate rapidly with periods
typically in the 80 \textendash{} 600 s range that correspond to low-order,
low-degree $p$-modes. These modes are driven by a classical $\kappa$-mechanism
produced by the accumulation of iron-group elements in the $Z$-bump
region \citep{1996ApJ...471L.103C}. This accumulation is triggered
by radiative levitation \citep{1997ApJ...483L.123C,2001PASP..113..775C}.
The second group is the\emph{ V1093 Her} stars (sdBV$_{{\rm s}}$ or
PG1716 stars; \citealp{2003ApJ...583L..31G}) that
pulsate far more slowly with periods typically in the 1 \textminus{}
4 hour range, corresponding to mid-order ($k\sim10-60$) gravity ($g$-)modes
driven by the same mechanism \citep{2003ApJ...597..518F,2006MNRAS.371..659J}.
A fraction of these stars belongs to both classes and are usually
referred to as hybrid pulsators (also known as the \emph{DW Lyn} or
sdBV$_{r{\rm s}}$ stars; \citealp{2006A&A...445L..31S}).

The advent of space-based, high-photometric-precision instruments
has played a fundamental role in unlocking the application of asteroseismology
to the long-period $g-$mode sdB pulsators. Prior to this space
age, detailed asteroseismology of sdB stars was limited to sdBV$_{{\rm r}}$
pulsators \citep[e.g.,][and references therein]{2008A&A...489..377C}.
Despite efforts carried out from the ground \citep{2006ApJ...643.1198R,2006ApJ...645.1464R,2009MNRAS.392.1092B},
it had proved extremely difficult to differentiate $g$-mode pulsation
frequencies from aliases introduced by the lack of continuous coverage,
particularly because of the long periods and low amplitudes ($\sim0.1\%$)
typically involved. This difficulty was overcome when \emph{CoRoT}
and \emph{Kepler} observations \citep{2010A&A...516L...6C,2010MNRAS.409.1470O,2011MNRAS.414.2860O}
first provided much clearer views of the $g$-mode spectrum in these
stars. Since then, many studies based mostly on \emph{Kepler} and
\emph{K2} data have enriched our general understanding of $g$-mode
pulsations in sdBs, and in some cases have provided new insights
into various properties of these stars \citep[e.g.,][]{2011MNRAS.414.2885R,2014A&A...569A..15O,2014A&A...570A.129T,2016A&A...594A..46Z,2017A&A...597A..95B,2017MNRAS.465.1057K,2017MNRAS.467..461K,2018ApJ...853...98Z,2019MNRAS.483.2282R},
such as their rotation rates \citep[see, e.g.,][and references therein]{2012MNRAS.422.1343P,2014MNRAS.440.3809R,2018OAst...27..112C}.

However, to date, detailed quantitative asteroseismic inferences
of the internal structure of sdB stars exist only for a dozen $p$-mode
pulsators \citep[see, e.g.,][and references therein]{2008A&A...483..875V,2008A&A...488..685V,2008A&A...489..377C,2009A&A...507..911R,2012A&A...539A..12F}
and three $g$-mode pulsators \citep{2010ApJ...718L..97V,2010A&A...524A..63V,2011A&A...530A...3C}.
The latter analyses have established the great potential of $g$-mode
asteroseismology that was envisioned for these stars. Gravity modes,
because they propagate far into the stellar interior, as opposed to
$p$-modes, which remain confined to the outermost layers \citep{2000ApJS..131..223C},
have the potential to reveal the structure of the deepest regions,
including the thermonuclear furnace and core boundary structure
\citep{2014IAUS..301..397C,2014ASPC..481..179C,2017MNRAS.465.1518G}.
\citet{2010ApJ...718L..97V,2010A&A...524A..63V} and \citet{2011A&A...530A...3C}
showed that important constraints on the inner core, such as its chemical
composition (related to the age of the star on the EHB) and its size,
are indeed accessible, suggesting in particular that the mixed core
may be larger than expected. This would imply that efficient extra
mixing processes (e.g., core convection overshoot, and semi-convection)
are effective. More analyses of this kind, with improved modeling
tools, are required to fully and objectively map the internal properties
of hot subdwarf stars. The \emph{Kepler} and \emph{K2} legacy are already
providing a remarkable set of seismic data that remains to be fully
exploited in that context, but \emph{TESS} adds another dimension
at this level and has the promise of considerably expanding the sample
of objects exploitable through asteroseismology.

As anticipated, the delivery of the \emph{TESS} data for hundreds
of selected TASC WG8 targets monitored in the first sectors has revealed
a wealth of photometric variations occurring in all types of evolved
compact stars, including many eclipsing or non-eclipsing binaries
and compact pulsators. These will be presented in forthcoming dedicated
publications. In this work, we focus on one of these objects, EC 21494-7018
(TIC 278659026), whose observation establishes that it is a bright,
long-period pulsating sdB star that was not previously known to pulsate.
The data reveal a particularly clean and rich pulsation spectrum,
making this star a perfect target to attempt a detailed asteroseismic
study. We present in Sect. 2, the analysis of the \emph{TESS} light
curve obtained for TIC 278659026, which constitutes the basis of the
detailed asteroseismic study that follows in Sect.~3. We summarize
our results and conclusions in Sect.~4.

\section{Observations}

\subsection{About TIC 278659026}

\begin{figure}
\begin{centering}
\includegraphics[scale=0.45]{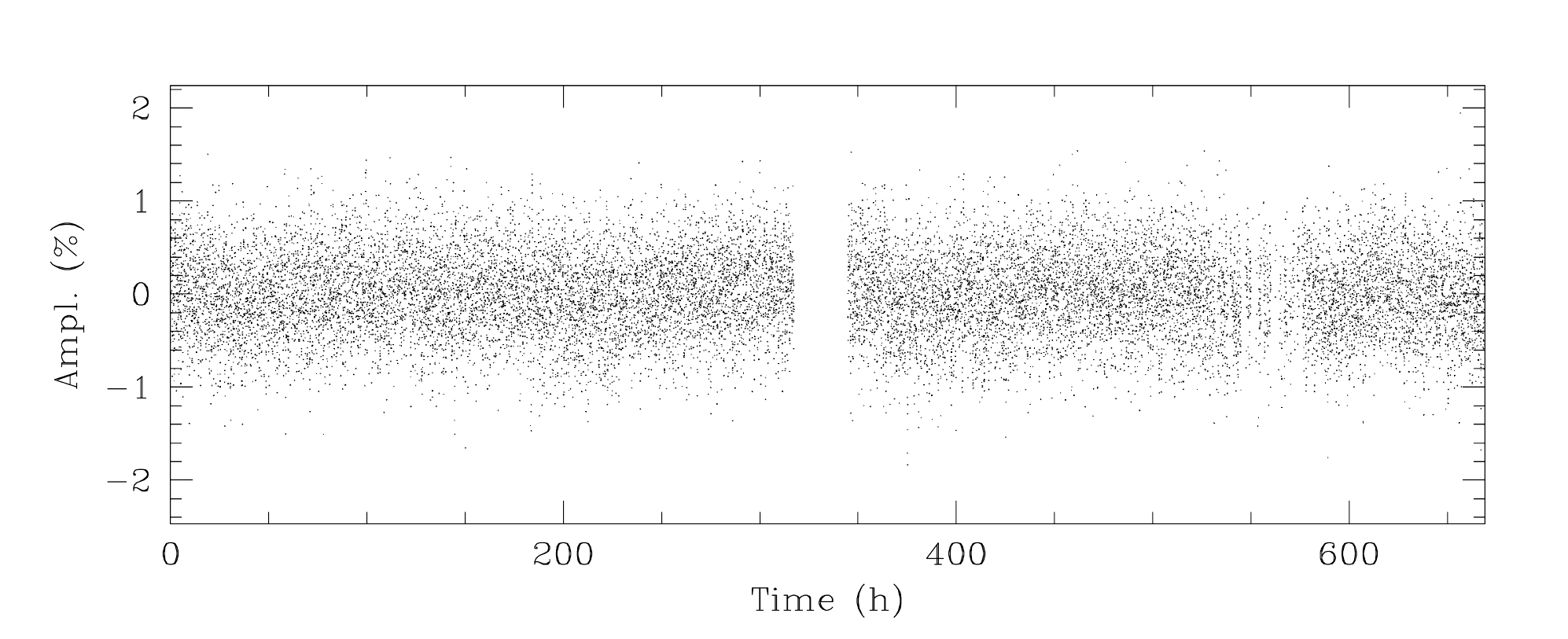}
\par\end{centering}
\begin{centering}
\includegraphics[scale=0.45]{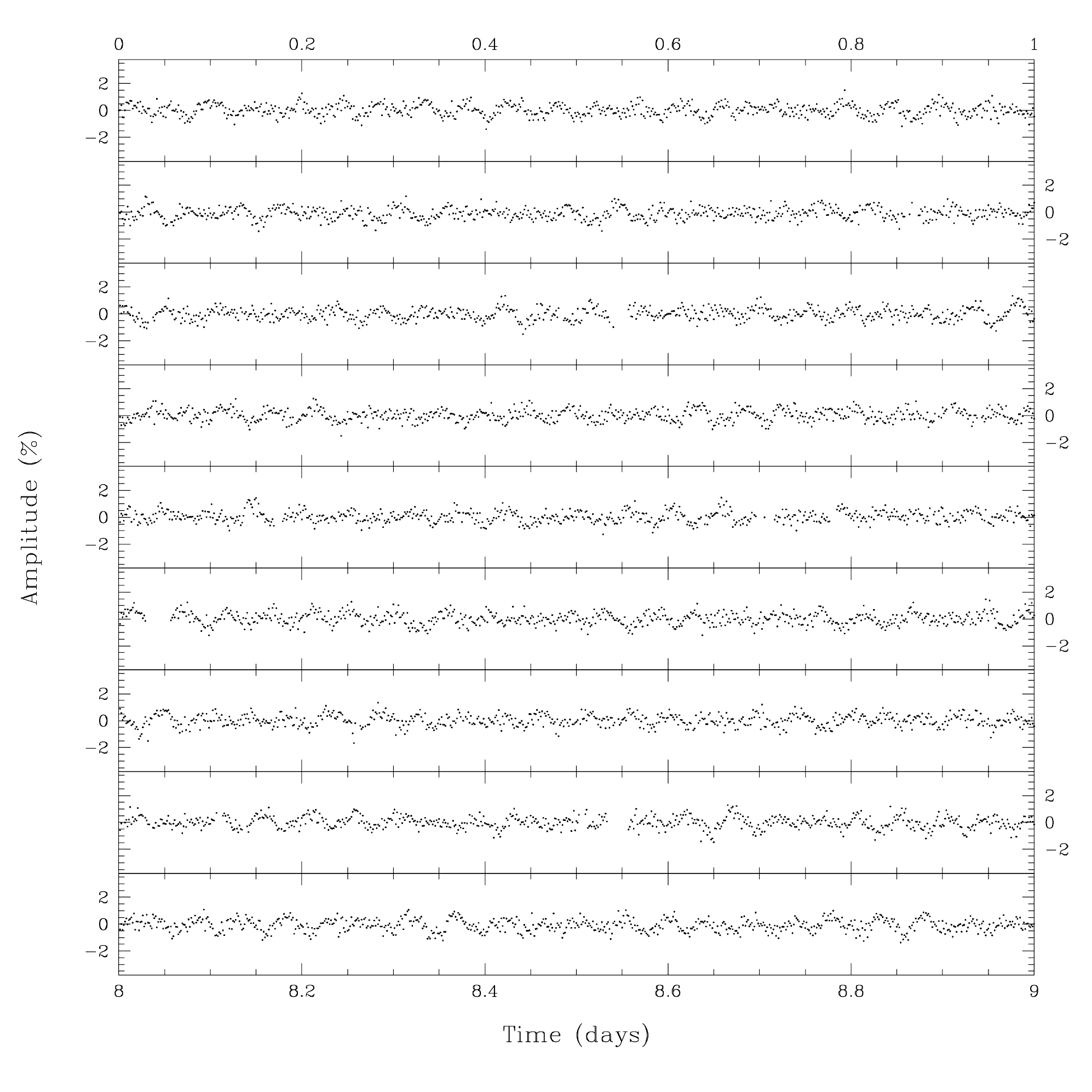}
\par\end{centering}
\begin{centering}
\includegraphics[scale=0.45]{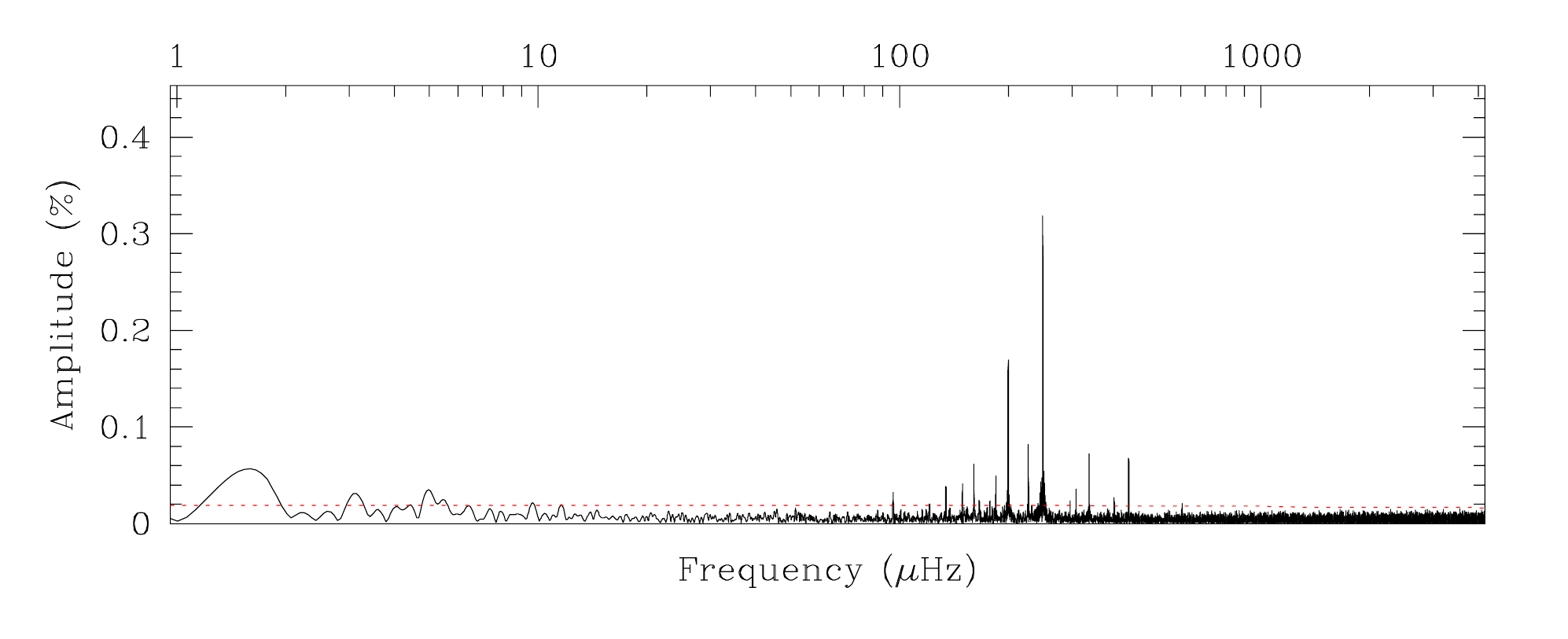}
\par\end{centering}
\caption{Illustration of \emph{TESS} photometry obtained for TIC 278659026.
The top panel shows the entire light curve (amplitude is in percent
of the mean brightness of the star) spanning 27.88 d sampled every
120 s. Gaps in this time series are due to the mid-sector interruption
during data download and missing points removed from the light curve
because of a non-optimal quality flag warning. The middle panel
shows an expanded view of the light curve covering the first 9 days,
where modulations due to pulsations are clearly visible.The bottom
panel shows the LSP of the light curve
up to the Nyquist frequency limit of the 2 min sampling rate ($\sim4467$
$\mu$Hz). The horizontal dotted line indicates 4 times the median
noise level. Significant activity well above this threshold and in
a frequency range corresponding to g-mode pulsations is clearly detected.\label{fig1}}
\end{figure}

TIC 278659026 (also known as EC 21494-7018; \citealp{2013MNRAS.431..240O})\footnote{Other names for this star are TYC 9327-1311-1, GSC 09327-01311, 2MASS
J21534125-7004314, GALEX J215341.2-700431, and \emph{Gaia} DR2 6395639996658760832.} is a bright, $V=11.57\pm0.09$ \citep{2000A&A...355L..27H} or $G=11.5928\pm0.0009$
\citep{2016A&A...595A...1G,2018A&A...616A...1G}, blue high-proper-motion
object first identified as a potential hot subdwarf by \citet{2011A&A...525A..29J}.
Its classification was later confirmed by \citet{2012MNRAS.427.2180N}
on the basis of a detailed spectroscopic analysis of its atmosphere
using non local thermodynamic equilibrium (NLTE) models that gives $T_{{\rm eff}}=23720\pm260$ K, $\log g=5.65\pm0.03$,
and $\log({\rm He/H})=-3.22{}_{-1.13}^{+0.13}$ for this star. The
rather high surface gravity and low effective temperature imply a
position for TIC 278659026 below the standard, $M=0.47$ $M_{\odot}$,
zero age EHB, indicating that it might be less
massive than typical sdB stars. \citet{2015MNRAS.450.3514K}
indeed suggested that it could be a rare extremely low-mass (ELM)
white dwarf progenitor, although they did not detect any significant
radial velocity variations indicating the presence of a companion.
Their average radial velocity measurement and dispersion is $43.4\pm4.2$
km s$^{-1}$, which is consistent with the independent measurement
of $39.4\pm7.5$ km s$^{-1}$ from \citet{2011MNRAS.415.1381C}, who
did not find any variability either. The photometry available for
this star (see Sect. 3.3.1) rules out any main sequence companion
earlier than type M5 or M6.

\citet{2012A&A...543A.149G} reported a projected rotational velocity
of $v_{{\rm rot}}\sin i=8.6\pm1.8$ km s$^{-1}$ for EC 21494-7018,
which is close to the average value measured in their sample of 105
sdB stars. These estimates are obtained by modeling the broadening
of metal lines in high-resolution optical spectra, as described by
\citet{2010A&A...519A..25G}. A critical look at the spectrum
of EC 21494-7018 obtained with the Fiber-fed Extended Range Optical
Spectrograph (FEROS) reveals, however, that its S/N is not excellent,
and only three metal lines could be used to estimate the projected
rotational velocity, due to the low temperature of the star. Therefore,
caution is advised concerning this particular measurement (see also
the discussion of Sect. 3.1).

\subsection{TESS photometry}

Despite having no previously known variability due to pulsations or
other causes, TIC 278659026 was positioned at a high priority rank
among the stars proposed by the TASC Working Group 8 because its
atmospheric parameters, $T_{{\rm eff}}$ and $\log g$, places this source
well within the sdB $g$-mode instability region (see,
e.g., Figure 4 of \citealp{2009AIPC.1170..585C}). It was observed
in \emph{TESS} Sector 1, from July 25 to August 22, 2018, and the
time series consists of 18102 individual photometric measurements\footnote{Only data points without any warning flag are considered and no additional
processing of the light curve is done.} obtained with the 120 s cadence mode, covering almost continuously
27.88 days (669.12 hours) of observation. We based our analysis on
the corrected time series extracted with the \emph{TESS} data processing
pipeline developed by the Science Processing Operations Center (SPOC)
at NASA Ames Research Center. These light curves are delivered publicly
along with pixel data at the \emph{Mikulski} Archive for Space Telescope
(MAST).

The top panel of Fig.~\ref{fig1} illustrates the TIC 278659026 light
curve in its entirety. The apparent amplitude scatter visible at this
scale shows up as a clearly multiperiodic signal in the close-up view
focusing on 9 d of observation presented in the middle panel of Fig.~\ref{fig1}.
A LSP of the time series\footnote{All L-S periodograms presented in this study are computed using oversampling
by a factor of 7. } (bottom panel of Fig.~\ref{fig1}) confirms the presence of
highly coherent low-amplitude modulations of the star brightness; timescales range from 27 minutes to 2.89 hours. Such variations
are typical of $g-$mode oscillations in hot subdwarf stars. Additional
weak signals are also possibly present at the very low frequency range
($<5$ $\mu$Hz), but could be of instrumental origin. In contrast,
no significant peaks are found in the LSP at higher frequencies, up
to the Nyquist limit. Hence, these \emph{TESS} data clearly establish
that TIC 278659026 is a new bright pulsating sdB star of the \emph{V1093
Her} class.

\subsection{Pulsation spectrum in g-mode }

\begin{figure}
\begin{centering}
\includegraphics[scale=0.45]{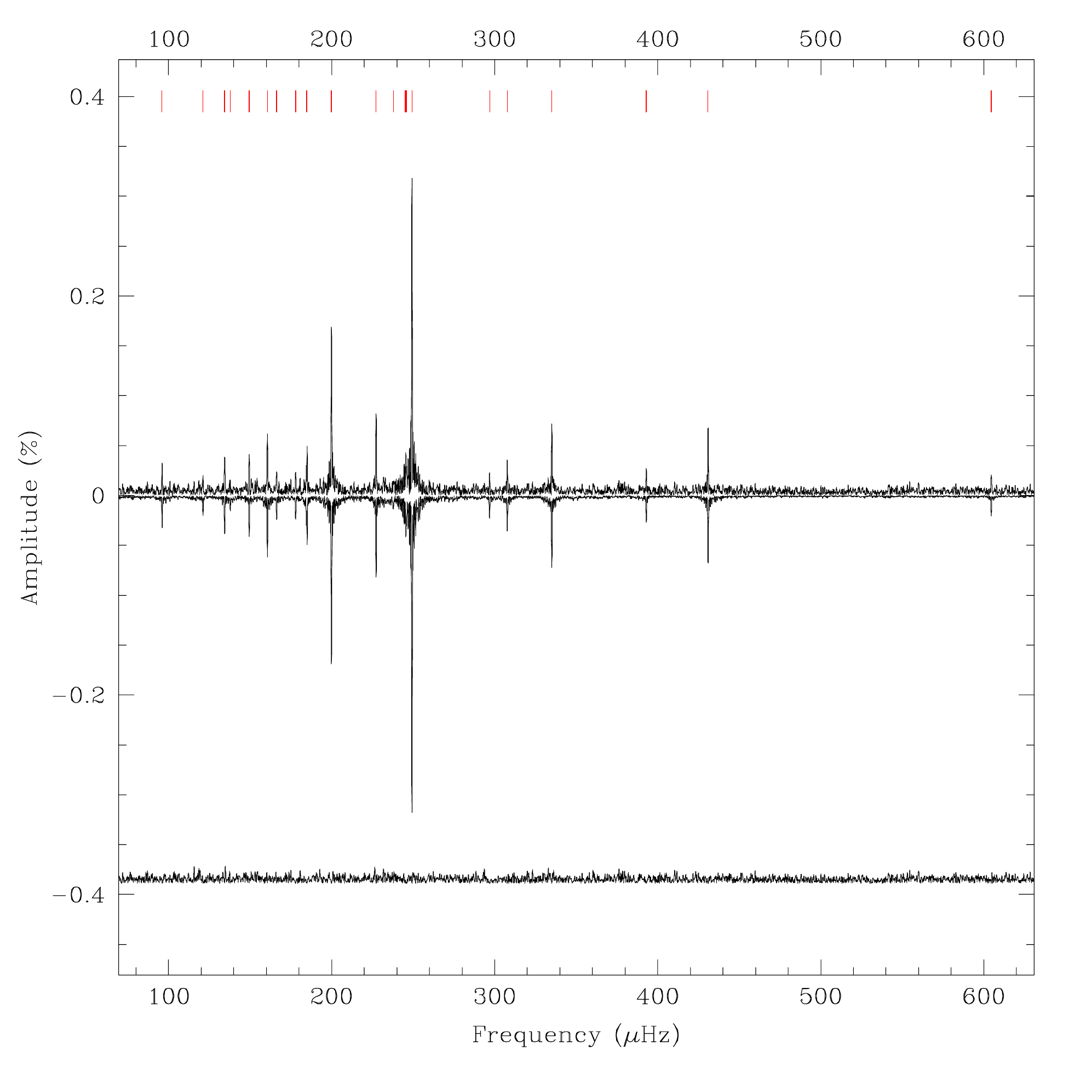}
\par\end{centering}
\caption{Frequencies detected in the g-mode pulsation range. The top curve
shows a close-up view of the LSP computed
from TIC 278659026 light curve. Each small vertical segment indicates
a frequency extracted during the prewhitening and nonlinear least-squares
fitting analysis. All frequencies and their properties are tabulated
in Table \ref{tab1}. The curve plotted upside-down is a reconstruction
of the LSP based on the data summarized in this table, and the curve
at the bottom (shifted vertically by an arbitrary amount for visibility)
is the residual containing only noise after removing all the signal
from the light curve. No peak above $4$ times the average noise level
is left in this residual.\label{fig2}}
\end{figure}

\begin{table*}
\caption{List of frequencies above 4 times the median noise level extracted
from TIC 278659026 light curve and their parameters.$^{a}$\label{tab1}}

\raggedright{}\vspace{-20bp}
\small
\begin{center}\begin{tabular}{lccccccccccl}
\hline\hline
\\
Id. & Frequency & $\sigma_f$ & Period & $\sigma_P$ & Amplitude & $\sigma_A$ & Phase & $\sigma_{\rm Ph}$ & S/N & Comments\tabularnewline
  & ($\mu$Hz) & ($\mu$Hz) & (d) & (d) & ($\%$) & ($\%$) &   &   & &\tabularnewline
\\
\hline
\\
{ $f_{A}$ } & { 1.563} & { 0.019} & { 7.407} & { 0.088} & { 0.0574} & { 0.0047} & { 0.300} & { 0.017} & { 12.3} & Instr.?$^{\ddag}$  \tabularnewline
{ $f_{B}$ } & { 3.173} & { 0.035} & { 3.648} & { 0.040} & { 0.0304} & { 0.0047} & { 0.031} & { 0.015} & { 6.5} & Instr.?; $2f_A $$^{\ddag}$\tabularnewline
{ $f_{C}$ } & { 4.966} & { 0.033} & { 2.331} & { 0.015} & { 0.0324} & { 0.0047} & { 0.819} & { 0.009} & { 6.9} & Instr.?$^{\ddag}$\tabularnewline
\\
\hline
\\
Id. & Frequency & $\sigma_f$ & Period & $\sigma_P$ & Amplitude & $\sigma_A$ & Phase & $\sigma_{\rm Ph}$ & S/N & Comments\tabularnewline
  & ($\mu$Hz) & ($\mu$Hz) & (s) & (s) & ($\%$) & ($\%$) &   &   & &\tabularnewline
\\
\hline
\\
{ $f_{11}$$^*$ } & { 96.073}  & { 0.030} & {10408.72} & { 3.29} & { 0.0324} & { 0.0043} & { 0.3310} & { 0.0004} & { 7.5} &  \tabularnewline
{ $f_{22}$$^*$ } & { 121.144} & { 0.051} & { 8254.62} & { 3.46} & { 0.0194} & { 0.0043} & { 0.6661} & { 0.0006} & { 4.5} &  99.5\% real signal$^{\dag}$ \tabularnewline
{ $f_{10}$$^*$ } & { 134.387} & { 0.026} & { 7441.22} & { 1.46} & { 0.0375} & { 0.0043} & { 0.8808} & { 0.0003} & { 8.7} &  \tabularnewline
{ $f_{23}$$^*$ } & { 137.912} & { 0.052} & { 7251.01} & { 2.72} & { 0.0192} & { 0.0043} & { 0.6056} & { 0.0005} & { 4.4} &  99\% real signal$^{\dag}$ \tabularnewline
{ $f_{8}$$^*$ }  & { 149.432} & { 0.024} & { 6692.01} & { 1.07} & { 0.0416} & { 0.0043} & { 0.6680} & { 0.0002} & { 9.6} &  \tabularnewline
{ $f_{6}$$^*$ }  & { 160.622} & { 0.015} & { 6225.81} & { 0.59} & { 0.0655} & { 0.0043} & { 0.5316} & { 0.0001} & { 15.1} &  \tabularnewline
{ $f_{12}$$^*$ } & { 166.256} & { 0.037} & { 6014.82} & { 1.35} & { 0.0266} & { 0.0043} & { 0.0062} & { 0.0003} & { 6.1} &  \tabularnewline
{ $f_{16}$$^*$ } & { 177.869} & { 0.044} & { 5622.13} & { 1.40} & { 0.0223} & { 0.0043} & { 0.1369} & { 0.0003} & { 5.2} &  \tabularnewline
{ $f_{18}$ }     & { 184.479} & { 0.046} & { 5420.68} & { 1.36} & { 0.0214} & { 0.0043} & { 0.5000} & { 0.0003} & { 5.0} &  linked to $f_7$\tabularnewline
{ $f_{7}$$^*$ }  & { 185.014} & { 0.022} & { 5404.99} & { 0.64} & { 0.0452} & { 0.0043} & { 0.2034} & { 0.0002} & { 10.4} &  \tabularnewline
{ $f_{2}$$^*$ }  & { 199.913} & { 0.006} & { 5002.18} & { 0.14} & { 0.1706} & { 0.0043} & { 0.4374} & { 0.0001} & { 39.5} &  \tabularnewline
{ $f_{3}$$^*$ }  & { 227.285} & { 0.012} & { 4399.77} & { 0.24} & { 0.0810} & { 0.0043} & { 0.4979} & { 0.0001} & { 18.8} &  \tabularnewline
{ $f_{21}$$^*$ } & { 237.829} & { 0.049} & { 4204.70} & { 0.87} & { 0.0200} & { 0.0043} & { 0.7711} & { 0.0003} & { 4.6} &  99.7\% real signal$^{\dag}$\tabularnewline
{ $f_{20}$ }     & { 245.102} & { 0.047} & { 4079.93} & { 0.78} & { 0.0210} & { 0.0043} & { 0.8531} & { 0.0003} & { 4.9} &  linked to $f_{14}$\tabularnewline
{ $f_{14}$$^*$ } & { 245.609} & { 0.041} & { 4071.52} & { 0.68} & { 0.0240} & { 0.0043} & { 0.5993} & { 0.0002} & { 5.6} &  \tabularnewline
{ $f_{17}$ }     & { 246.027} & { 0.045} & { 4064.60} & { 0.75} & { 0.0219} & { 0.0043} & { 0.8582} & { 0.0003} & { 5.1} &  linked to $f_{14}$\tabularnewline
{ $f_{1}$$^*$ }  & { 249.269} & { 0.003} & { 4011.74} & { 0.05} & { 0.3184} & { 0.0043} & { 0.7808} & { 0.0001} & { 73.7} &  \tabularnewline
{ $f_{15}$$^*$ } & { 296.822} & { 0.044} & { 3369.02} & { 0.50} & { 0.0224} & { 0.0043} & { 0.5569} & { 0.0002} & { 5.2} &  \tabularnewline
{ $f_{9}$$^*$ }  & { 307.720} & { 0.026} & { 3249.70} & { 0.28} & { 0.0377} & { 0.0043} & { 0.6263} & { 0.0001} & { 8.7} &  \tabularnewline
{ $f_{4}$$^*$ }  & { 335.033} & { 0.013} & { 2984.78} & { 0.12} & { 0.0735} & { 0.0043} & { 0.9914} & { 0.0001} & { 17.1} &  \tabularnewline
{ $f_{13}$$^*$ } & { 393.032} & { 0.039} & { 2544.32} & { 0.25} & { 0.0251} & { 0.0043} & { 0.6513} & { 0.0001} & { 5.9} &  \tabularnewline
{ $f_{5}$$^*$ }  & { 430.875} & { 0.014} & { 2320.86} & { 0.08} & { 0.0686} & { 0.0043} & { 0.5398} & { 0.0001} & { 16.1} &  \tabularnewline
{ $f_{19}$$^*$ } & { 604.580} & { 0.046} & { 1654.04} & { 0.13} & { 0.0212} & { 0.0043} & { 0.7714} & { 0.0001} & { 5.0} &  \tabularnewline
\\
\hline

\end{tabular}\end{center}
{\footnotesize
$^{a}$Each modulation is modeled as $f_{i}=A_{i}\cos[2\pi(\nu_{i}t+\phi_{i})]$, where $A_{i}$,
$\nu_{i}$, and $\phi_{i}$ is the given amplitude, frequency, and phase, respectively.
The phase is relative to a starting time $T_{S}=2458325.297922$ BJD.\\
$^*$Frequencies interpreted as independent pulsation modes and used for the detailed asteroseismic analysis.\\
$^{\ddag}$These signals are difficult to interpret. Intrumental artifacts of
yet unclear origin may be present in this frequency range and contamination from a bright nearby object, TIC 278659026 (TYC 9327-275-1), cannot be excluded.\\
$^{\dag}$Evaluated from a specific Monte-Carlo test as described in \citet{2016A&A...585A..22Z}.
}
\normalsize

\end{table*}

\begin{figure*}
\begin{centering}
\includegraphics[scale=0.8]{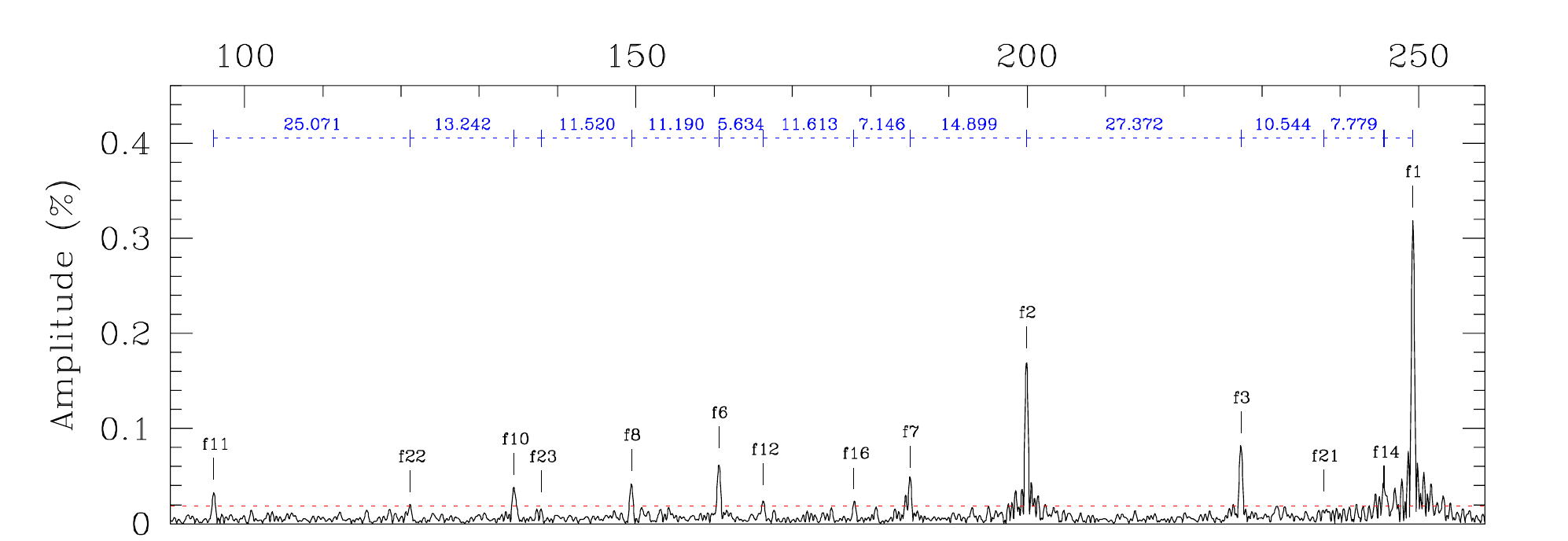}\\
\includegraphics[scale=0.8]{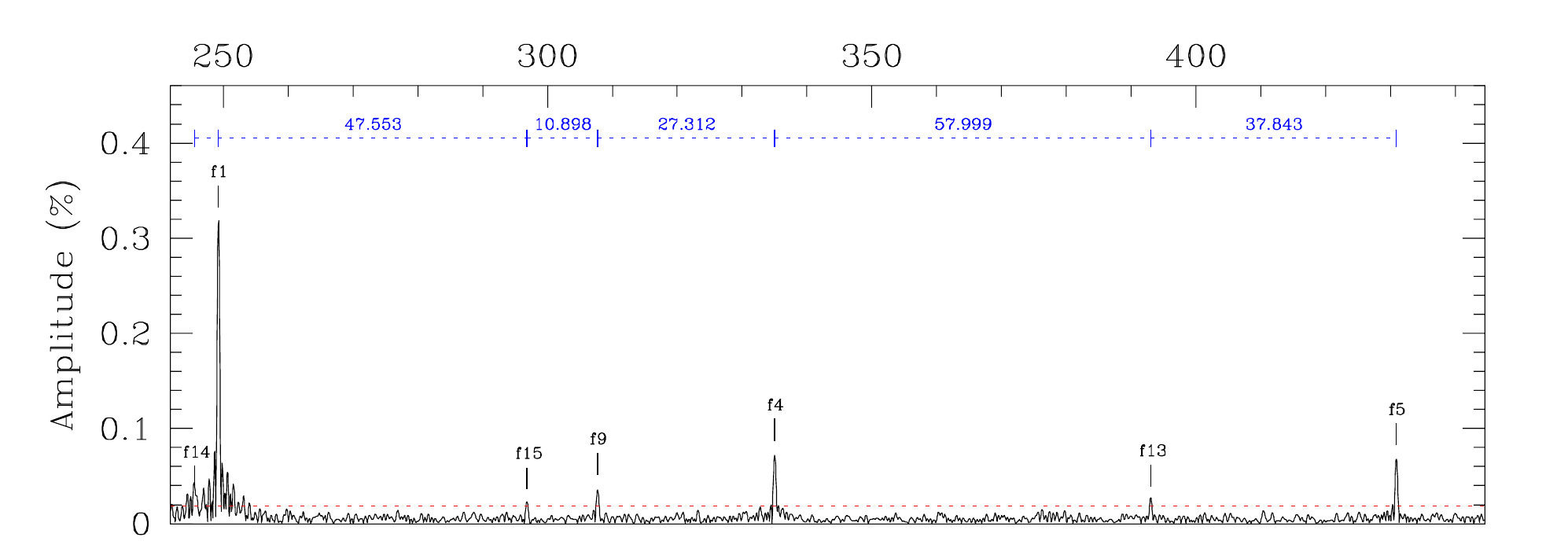}\\
\includegraphics[scale=0.8]{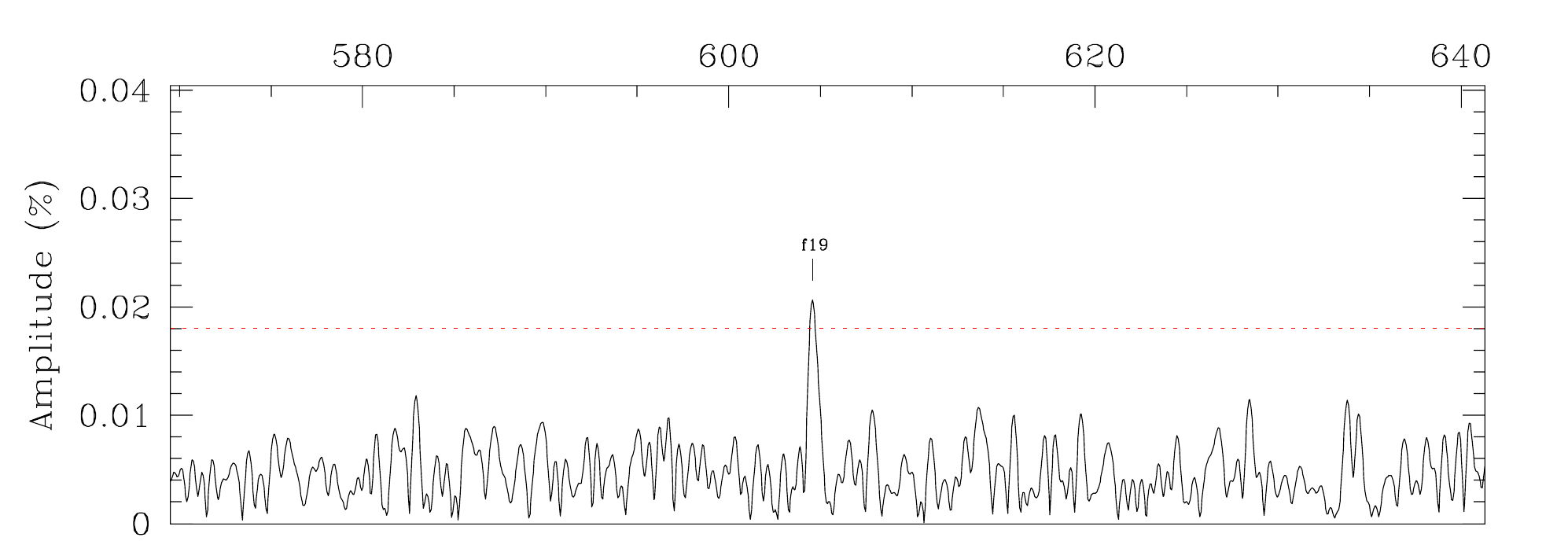}\\
\includegraphics[scale=0.8]{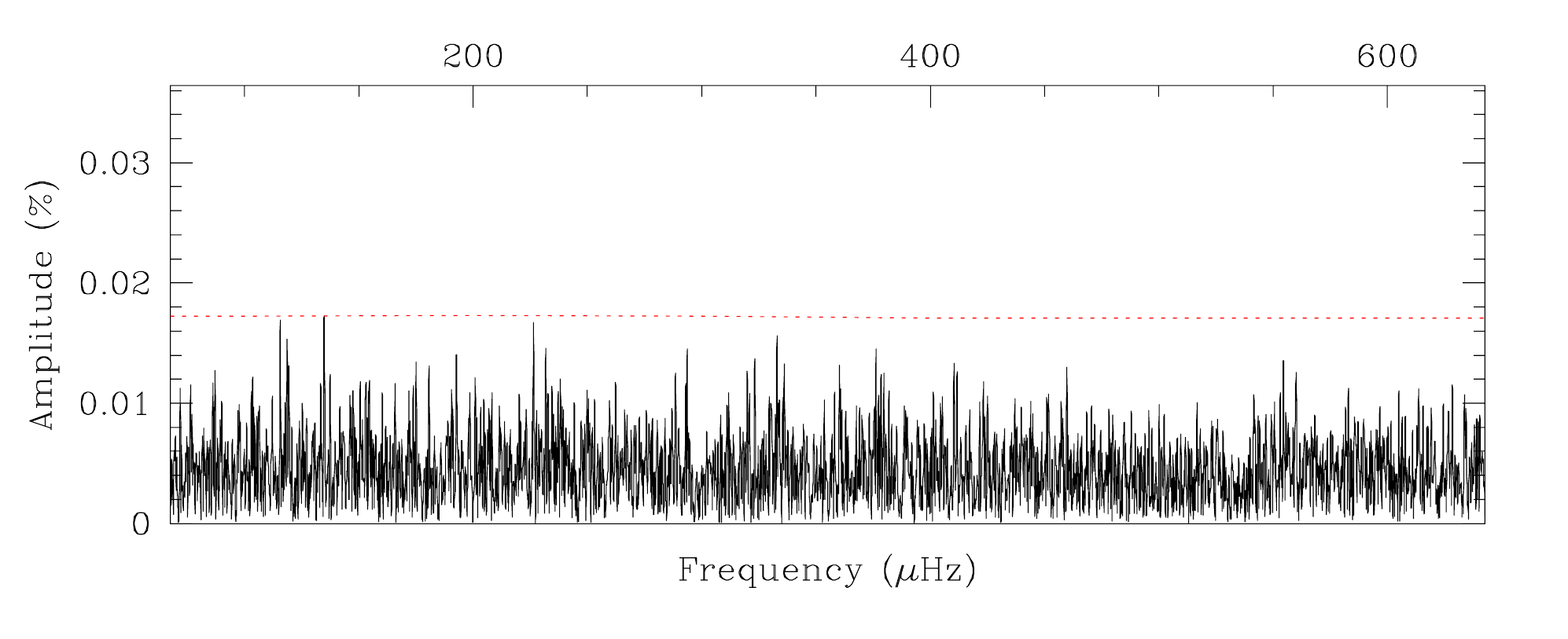}
\par\end{centering}
\caption{Three top panels show close-up views of the LSP
of TIC 278659026 in the $g$-mode frequency range. The main frequencies
listed in Table \ref{tab1} are indicated, except the three peaks
($f_{18}$, $f_{17}$, and $f_{20}$; see text). In the top two panels,
the separation in frequency between two consecutive peaks is indicated
(the horizontal dotted-line segments). The horizontal dotted line
in each panel indicates 4 times the median noise level used as an
initial significance criterion. The bottom panel shows the
residual after prewhitening all the frequencies. \label{fig3}}
\end{figure*}

We performed a standard prewhitening and nonlinear least-squares fitting
analysis to extract the frequencies present in the brightness modulations
of TIC 278659026 \citep{1976Ap&SS..42..257D}. For that purpose, we
used the dedicated software \noun{felix} (\citealp{2010A&A...516L...6C},
see also \citealp{2016A&A...594A..46Z}), which greatly facilitates
the application of this procedure. Most of the coherent variability
is found to reside in the 95 \textendash{} 650 $\mu$Hz range, where
we could easily extract up to 23 frequencies above a chosen detection
threshold at 4 times the median noise level\footnote{This threshold follows common practice in the field, but the significance
of the lowest S/N peaks is re-evaluated afterward.} (see Table \ref{tab1}). The result of this extraction is illustrated
in Fig.~\ref{fig2}, showing that the signal in Fourier space reconstructed
from the 23 identified frequencies (plotted upside-down) reproduces
very well the periodogram of the \emph{TESS} light curve and no significant
residual is found after subtraction of this signal. Close-up views
of the most relevant parts of the spectrum are also provided in Fig.~\ref{fig3}.
In effect, the lowest amplitude peak extracted has a $S/N$ ratio
of 4.4, well above the $4\sigma$ threshold. In order to estimate
the reliability of the lowest amplitude peaks of our selection given
in Table \ref{tab1}, we performed a Monte Carlo test as described
in Section 2.2 of \citet{2016A&A...585A..22Z}; our test is specifically tuned
to the \emph{TESS} light curve of TIC 278659026, i.e., using the exact
same time sampling and parameters to compute the LSP. This allowed
us to evaluate probabilities that a peak at given $S/N$-values or
above is real and not due to a random fluctuation of the noise. The
test indicates that our lowest S/N frequency retained, $f_{23}$ with
$S/N=4.4$, has 99\% chance to be real, while this probability goes
above 99.99\% (less than 1 chance out of 10,000 to be a false positive)
for $S/N\gtrsim5.1$. Since the frequency spectrum is very clean,
in particular around the few frequencies below $S/N\sim5$, we consider
in the following analysis that even $f_{23}$ is a securely established
frequency of TIC 278659026 pulsation spectrum.

It is important to note that, while most frequencies in the LSP could
be prewhitened without leaving any residual behind, thus indicating
strong stability over the 27.88 days of observation, significant remaining
peaks could be found on two occasions. The first case occurred for
$f_{7}$ at $185.014$ $\mu$Hz, which has $f_{18}$ at $184.479$
$\mu$Hz as a close-by companion, considering that the formal frequency
resolution in Fourier space for these data is $0.415$ $\mu$Hz ($1/T$,
where $T$ is the time baseline of the run). The other case is $f_{14}$
at $245.609$ $\mu$Hz which has $f_{20}$ at $245.102$ $\mu$Hz
and $f_{17}$ at $246.027$ $\mu$Hz as close neighbors. From a pulsation
perspective, these frequencies could very well be real independent
modes of different degree $\ell$ that happen to have very close frequencies,
barely resolved with the current data set. However, they may also
be the signature of intrinsic amplitude and/or frequency modulations
of the dominant peak, as discussed by \citet{2016A&A...585A..22Z,2016A&A...594A..46Z,2018ApJ...853...98Z}.
Therefore, as a precaution, we retain only the main (highest
amplitude) peak of each complex, namely $f_{7}$ and $f_{14}$, for
the detailed asteroseismic study of TIC 278659026.

Finally, we emphasize from Table \ref{tab1} a few numbers characterizing
the quality of these first \emph{TESS} data obtained for a pulsating
sdB star. With only one sector (27.88 days), targeting
a bright ($V=11.57$), hot, and blue object (while \emph{TESS} CCDs
are red sensitive), the amplitude of the noise in Fourier space is
around 0.0043\% (43 ppm) in the $g$-mode frequency range, which is
a remarkable achievement, especially considering the modest aperture
of the instrument. We can expect this noise to be reduced further
by up to a factor of $\sim3.5$ for similar targets located in or
near the CVZ, which could be observed for up to
one year. We also note that the uncertainty on the measured frequencies
is typically around one-tenth of the formal frequency resolution.

\section{Asteroseismic analysis}

\subsection{Interpretation of the observed spectrum}

\begin{figure*}
\begin{centering}
\includegraphics[scale=0.9]{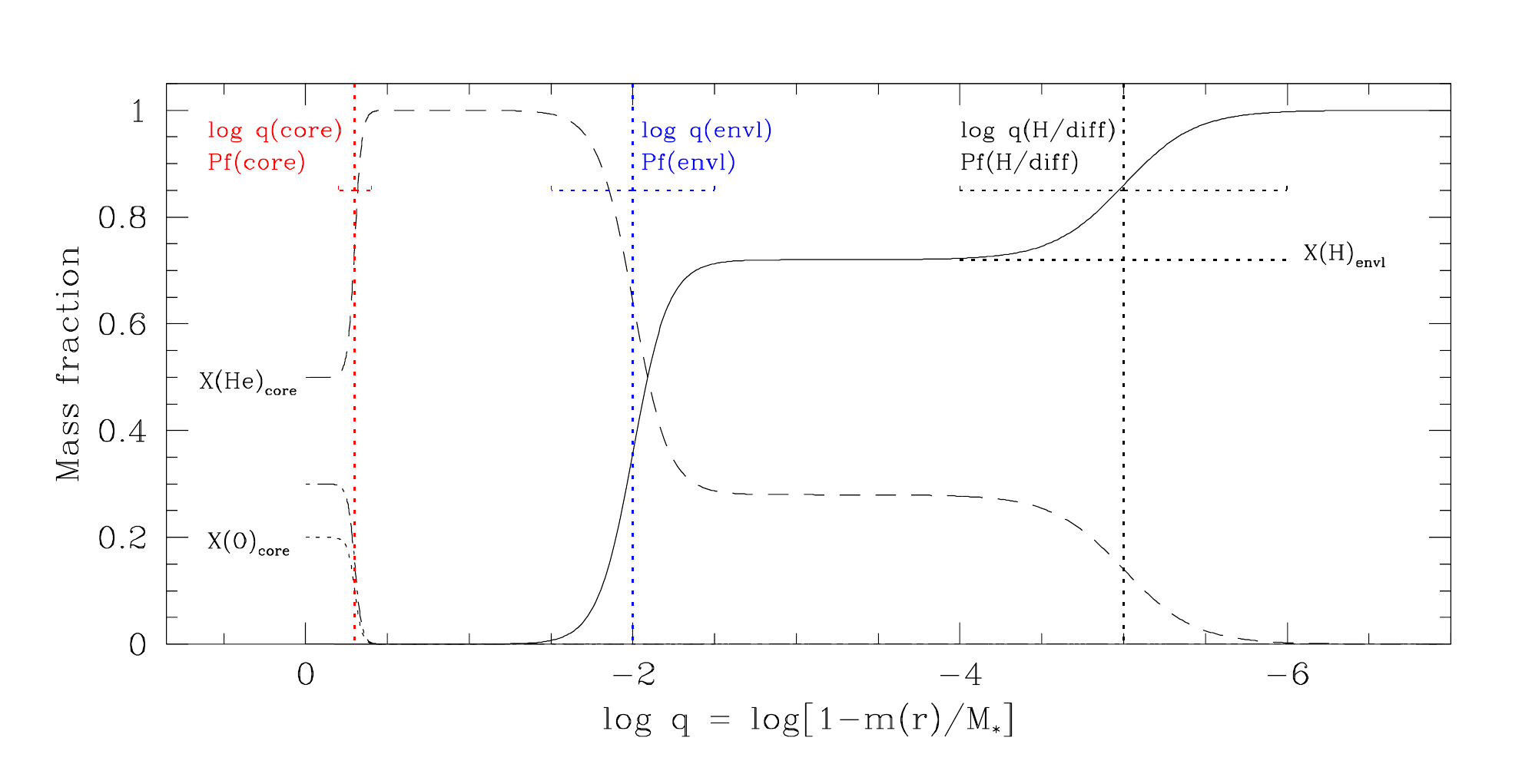}
\par\end{centering}
\caption{Illustration of the model parameters specifying the chemical stratification
(see text for a full description of these parameters). The hydrogen,
helium, carbon, and oxygen mass fractions are shown as a function
of fractional mass depth, $\log q$, with plain, dashed, dot-dashed,
and dotted curves, respectively.\label{fig4}}
\end{figure*}

Figure \ref{fig3} shows a pulsation spectrum whose frequency distribution
is very typical of $g$-mode pulsators observed from space \citep[e.g.,][]{2010A&A...516L...6C,2010MNRAS.409.1470O}.
The frequency spacings between consecutive peaks do not show quasi-symmetric
structures that could clearly correspond to the signature of rotational
splitting. Yet, we could argue that the projected rotational velocity
($v\sin i$) of $\sim8$ km s$^{-1}$ proposed for TIC 278659026,
if true (see Sect. 2.1), would correspond to a rotation period that is at most slightly less than
one day ($P\sin i$) for a star of radius 0.15
$R_{\odot}$. Assuming a median statistical inclination of
60°, this period would be $\sim0.82$ day, and the corresponding splittings
$\sim7$ $\mu$Hz for dipolar ($\ell=1$) $g$-modes and $\sim11$
$\mu$Hz for quadrupolar ($\ell=2$) $g$-modes. Values close to these
splittings can be seen between some of the observed peaks (see top
panels of Fig. \ref{fig3}), but without the regularity that usually
characterizes splittings at such slow rotation rates. Moreover, the
typical average period spacings between consecutive $g$-modes of
same degree in sdB stars generate frequency spacings in the range
at which these pulsations are observed that are also of the same order
without invoking rotation. To help distinguish between these two very
different interpretations of the spectrum, it is important to recall
that an unexplained discrepancy remains between the typical rotation
rates inferred for sdB stars from spectroscopy \citep{2010A&A...519A..25G,2012A&A...543A.149G}
and the rotation periods measured from well-identified rotational
splittings observed in pulsating sdB stars, in particular in those
with very long time series delivered by the \emph{Kepler} satellite
(see, e.g., \citealp{2017MNRAS.465.1057K} and other references provided
by \citealp{2018OAst...27..112C}). These splittings indicate in a
very robust way that nearly all sdB pulsators, in particular the
single pulsators, are extremely slow rotators with periods typically well
above 10 days, at odds with the average $\sim1$ day rotation timescale
suggested by \citet{2012A&A...543A.149G}. Asteroseismology is clearly
much more sensitive than line broadening analyses to very slow rotation rates.
The question then arises whether the $\sim8$ km s$^{-1}$
projected velocity found to be typical of sdB stars, which is not
far above the limit of this method with the available spectra, is
an actual measurement of the surface rotation of the star, an upper
limit of it, or reflects an effect other than rotation causing additional
broadening of the metal lines that has not been taken into account,
such as line broadening by pulsation. This issue is well beyond the
scope of the present paper, but clearly needs further investigation,
possibly with the acquisition and analysis of very high-resolution
spectra for a selection of pulsating and non-pulsating sdB stars.

For the present seismic study of TIC 278659026, we rely on previous
observations of pulsating sdB stars, indicating extremely slow rotation
rates. We interpret 20 observed frequencies (those marked with a ``{*}''
sign in Table \ref{tab1}) as independent sectoral ($m=0$) $g$-modes
(first assumption) of degree $\ell=1$ or $\ell=2$ (second assumption).
The first assumption follows from the above discussion and is further
justified by the fact that the time baseline of the \emph{TESS} run
would probably be too short to resolve eventual $m\ne0$ components
in this framework. The observed frequencies can therefore be compared
with the model frequencies computed assuming a nonrotating star, without
any impact on the solution. The second assumption limits our search
to dipole ($\ell=1$) and quadrupole ($\ell=2$) modes only, and is
justified by the fact that cancellation effects over the stellar surface
very strongly reduce the apparent amplitude of $\ell>2$ g-modes in
sdB stars, except for very specific inclination angles of the pulsation
axis relative to the line of sight. This has been clearly illustrated
in Figure 11 of \citet{2011A&A...530A...3C}, which shows that $\ell=3$
and 4 modes have apparent amplitudes already reduced by a factor of
at least 10 owing to this effect. Such modes are therefore much less
likely to be seen in the current data.

\subsection{Method and models}

The asteroseismic analysis of TIC 278659026 is based on the forward
modeling approach developed over the years for hot subdwarf and white dwarf
asteroseismology and described by, for example,
\citet{2005A&A...437..575C,2008A&A...489..377C},
\citet{2013A&A...553A..97V},
\citet{2014ASPC..481..105C,2015ASPC..493..151C}, and
\citet{2016ApJS..223...10G}.
The method is a multidimensional optimization of a set of parameters,
$\{p_{1},...,p_{n}\}$, defining the stellar structure of the star
with the goal to minimize a ``merit function'' that measures the
ability of that model to reproduce simultaneously all the observed
pulsation frequencies. In the present analysis, the quality of the
fit is quantified using a $\chi^{2}$-type merit function defined
as
\[
S^{2}(p_{1},...,p_{n})=\sum\limits _{i}^{N_{{\rm obs}}}\left(\frac{\nu_{{\rm obs,}i}-\nu_{{\rm th,}i}}{\nu_{{\rm obs,}i}}\right)^{2}\qquad,
\]
where $N_{{\rm obs}}$ is the number of observed frequencies and $(\nu_{{\rm obs,}i},\nu_{{\rm th,}i})$
is a pair of associated observed and computed frequencies. We recall that
this quantity, $S^{2}$, is minimized both as a function of the frequency
associations, referred to as the first combinatorial minimization
required because the observed modes are not identified a priori,
and as a function of the model parameters, $\{p_{1},...,p_{n}\}$
(the second minimization). This optimization is carried out by the
code \noun{lucy} \citep[see][]{2008A&A...489..377C}, a Real-Coded
Genetic Algorithm (RCGA)\footnote{A class of Genetic Algorithms (GAs), RCGAs encode
solutions into chromosomes whose genes are real numbers, typically between 0
and 1, instead of bits of value 0 or 1 used in binary-coded GAs. Apart
from this distinction, the solutions are evolved similarly by applying
selection, mating, and mutation operators specifically designed for
real coded chromosomes. RCGAs overcome several issues typically encountered
with classical binary-coded GAs, such as the Hamming cliff problem,
the difficulty to achieve arbitrary precision, and problems related
to uneven schema significance in the encoding. }\textbf{ }that can pursue multimodal optimization (i.e., searching
simultaneously for multiple solutions) in any given multidimensional
parameter space. This optimizer is coupled with a series of codes
dedicated to the computation of the stellar structure itself (given
a set of parameters), its pulsation properties, and to the
matching of the computed frequencies with the observed frequencies.

Our current version of stellar models used for quantitative asteroseismology
of sdB stars is an update of the Montréal third-generation (3G)
models designed for an accurate computation of $g$-mode frequencies
\citep{2008ASPC..392..261B,2009JPhCS.172a2016B,2010ApJ...718L..97V,2011A&A...530A...3C}.
These models are complete hydrostatic stellar structures in thermal
equilibrium computed from a set of parameters that define the main
structural properties of the star. The implemented parameterization
is inspired from full evolutionary calculations, which provide the
guidelines to reproduce, for example, the general shape of the composition
profiles in these stratified stars, but with additional flexibility.
Such parameterized static structure models offer greater versatility,
faster computations, and allow for much greater exploration of structural
configurations in the spirit of constraining these from the pulsation
modes themselves. The derived optimized structures are therefore seismic
models that are mostly independent of the uncertainties still present
in the physics that drives stellar evolution and shapes the chemical
structure of the star over time, such as mixing and diffusion processes, mass
losses, and nuclear reaction rates.\footnote{Of course, full independence from stellar evolution theory is not
achieved (and may never be), because parameterized static models still
assume some general shapes for composition profiles that are motivated
by stellar evolution calculations (the double-layered H+He envelope
expected from the action of gravitational settling, for instance).
Moreover, static models are still subject, as are evolutionary models,
to uncertainties associated with the constitutive microphysics, such
as equation-of-state and opacities. }

\begin{table}[!tbph]
\caption{Parameter space explored during optimization.\label{tab:ranges}}

\vspace{-20bp}
\normalsize
\begin{center}\begin{tabular}{lcc}
\hline
\\
           & First exploration   &  Second exploration       \tabularnewline
 Parameter & Range covered       &  Range covered            \tabularnewline
\\
\hline\hline
\\
 $M_*/M_{\odot}$            & $[0.30 , 0.60]$  & $[0.30 , 0.45]$     \tabularnewline
 $\log q ({\rm env})$       & $[-4.0 , -1.5]$  & $[-3.0 , -1.5]$     \tabularnewline
 $\log q ({\rm H/diff})$    & $[-5.0 , -2.0]$  & $[-5.0 , -2.0]$     \tabularnewline
 $X({\rm H})_{\rm envl}$    & $[0.70 , 0.75]$  & $[0.70 , 0.75]$     \tabularnewline
 Pf(envl)                   & $[1 , 10]$       & $[1 , 10]$          \tabularnewline
 Pf(H/diff)                 & $[1 , 10]$       & $[1 , 10]$          \tabularnewline
 $\log q ({\rm core})$      & $[-0.5 , -0.1]$  & $[-0.4 , -0.15]$    \tabularnewline
 Pf(core)                   & $[1 , 150]$      & $[50 , 150]$        \tabularnewline
 $X({\rm He})_{\rm core}$   & $[0.0 , 1.0]$    & $[0.0 , 1.0]$       \tabularnewline
 $X({\rm O})_{\rm core}$    & $[0.0 , 1.0]$    & $[0.0 , 1.0]$       \tabularnewline
 \\
 \hline
\end{tabular}\end{center}
\normalsize

\end{table}

Compared to the former 3G models used by, for example,
\citet{2010ApJ...718L..97V,2010A&A...524A..63V},
\citet{2011A&A...530A...3C}, and
\citet{2013A&A...553A..97V},
the most important improvement is that our current stellar structures
for sdB stars now implement a more realistic hydrogen-rich envelope
with double-layered hydrogen and helium composition profiles. This addition
was motivated by the fact that diffusion of helium through gravitational
settling does not have time, during the typical duration of the core
helium burning phase, to completely sink below the envelope when this
envelope is relatively thick (see, e.g., \citealp{2009A&A...508..869H,2010A&A...511A..87H}).
This is most likely to be relevant for $g$-mode sdB pulsators that
are found among the cooler sdB stars with thicker envelopes, while
the hotter $p$-mode sdB pulsators all have thinner envelopes that
quickly become fully hydrogen-pure. Consequently, in the present analysis,
the assumption of a pure hydrogen envelope used so far is better replaced
by an envelope that may still contain some helium, with a transition
from pure hydrogen (at the top) to a mixture of H+He (at its base).
The technical implementation of this double-layered H+He composition
profile is identical in its principle to the implementation of a double-layered
helium envelope in DB white dwarf models, as described recently in
the Methods section of \citet{2018Natur.554...73G}.

\begin{figure*}[tp]
\begin{centering}
\includegraphics[scale=0.44]{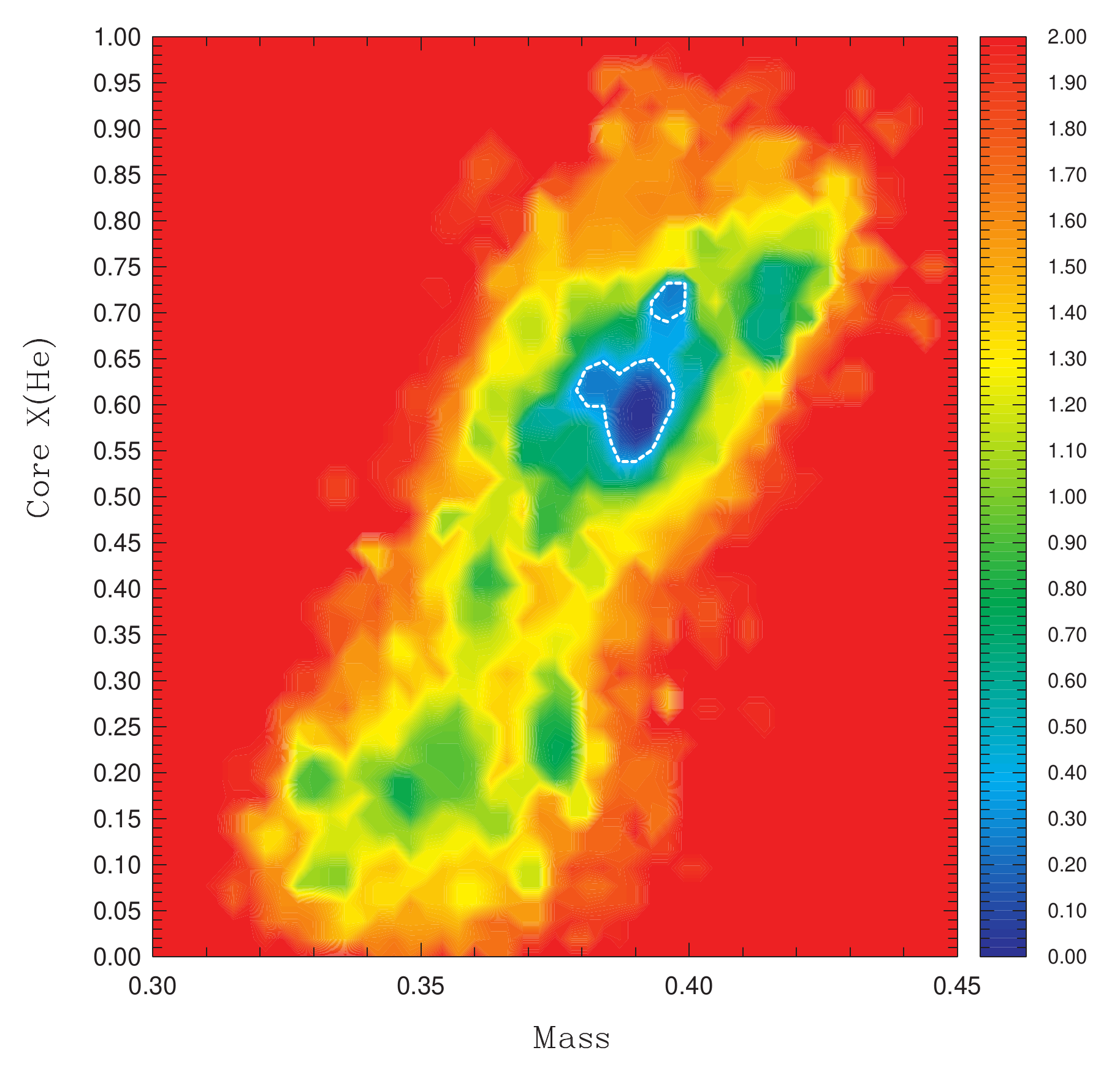}~\includegraphics[scale=0.44]{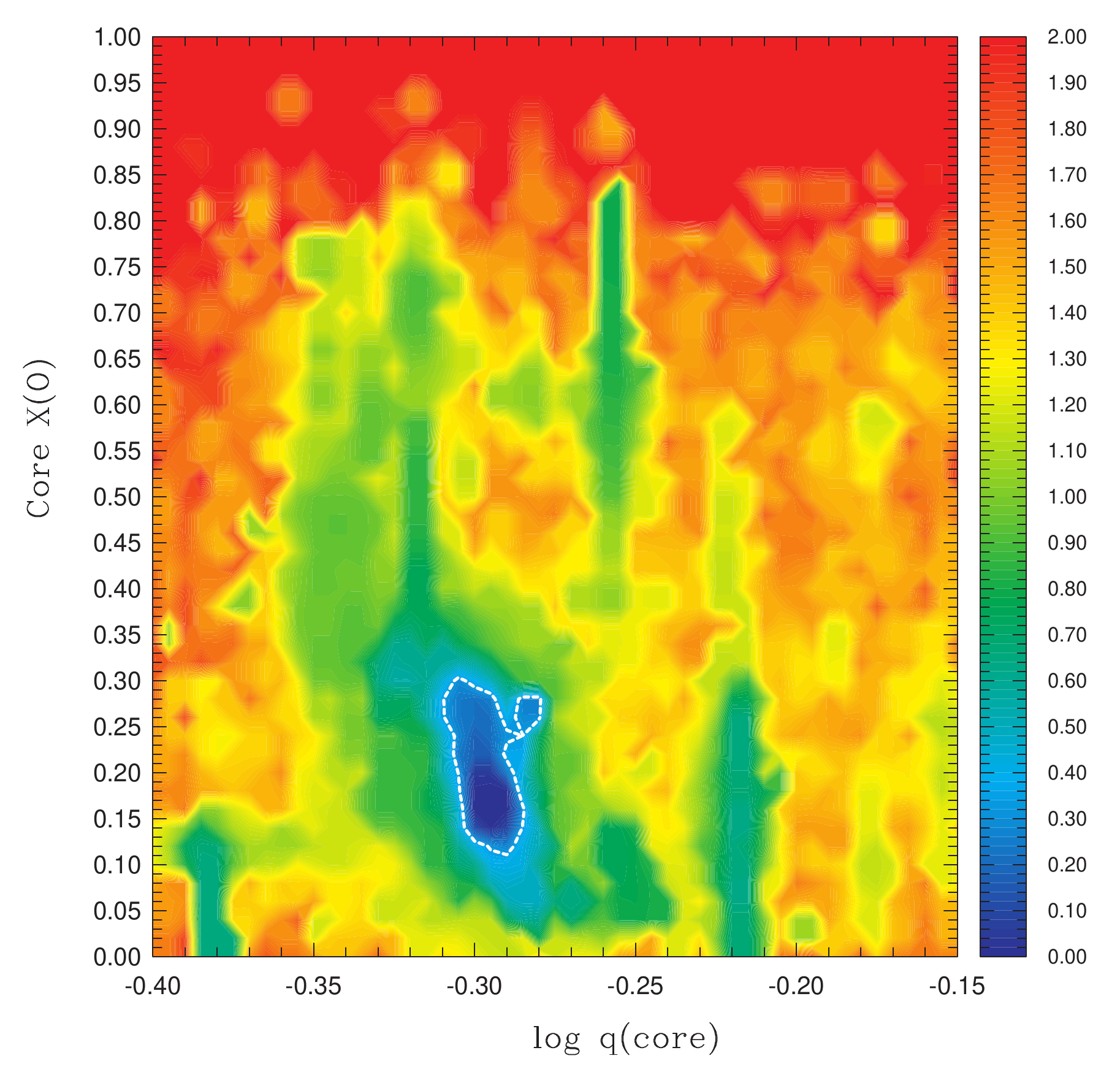}\\
\includegraphics[scale=0.44]{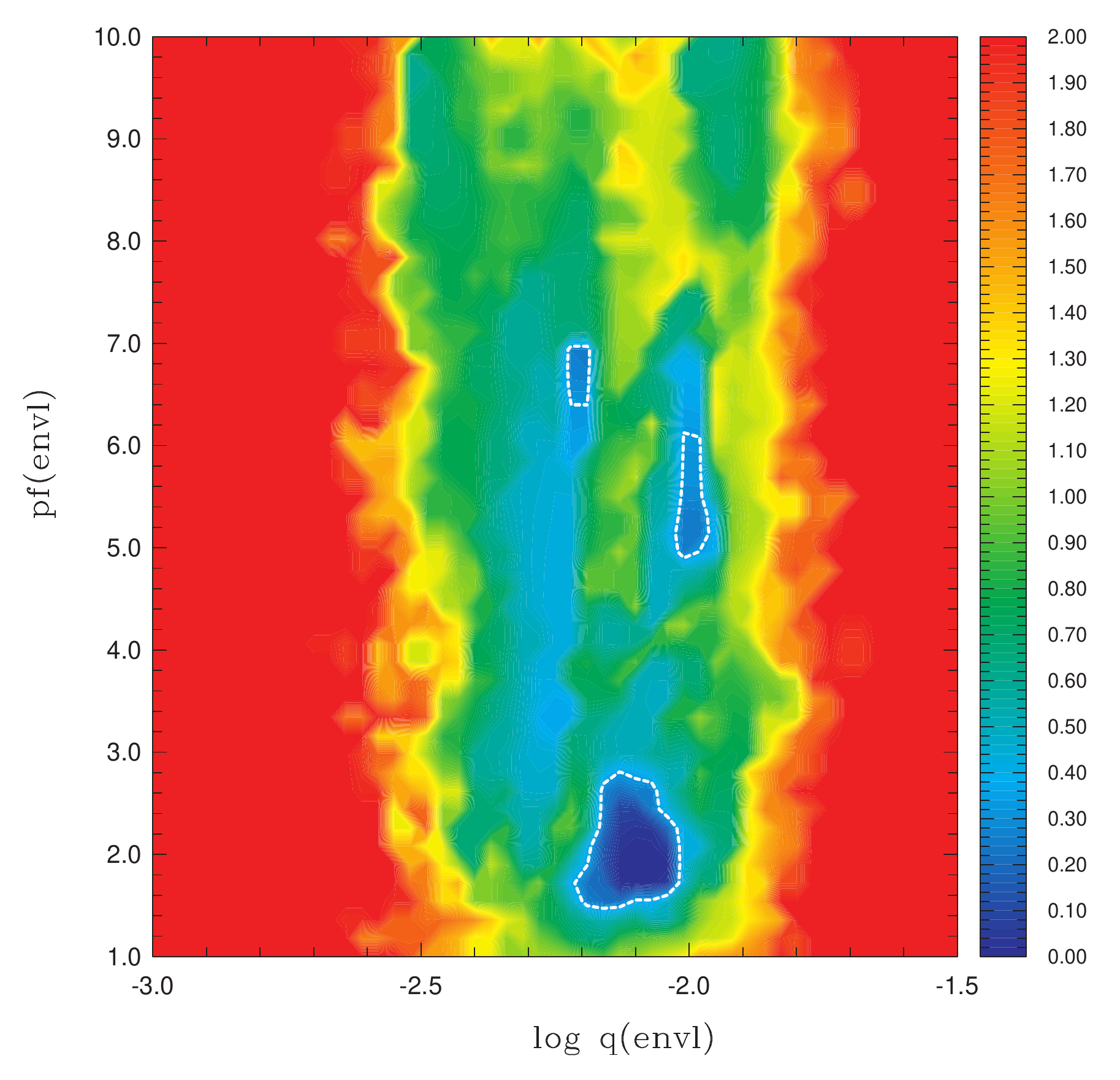}~\includegraphics[scale=0.44]{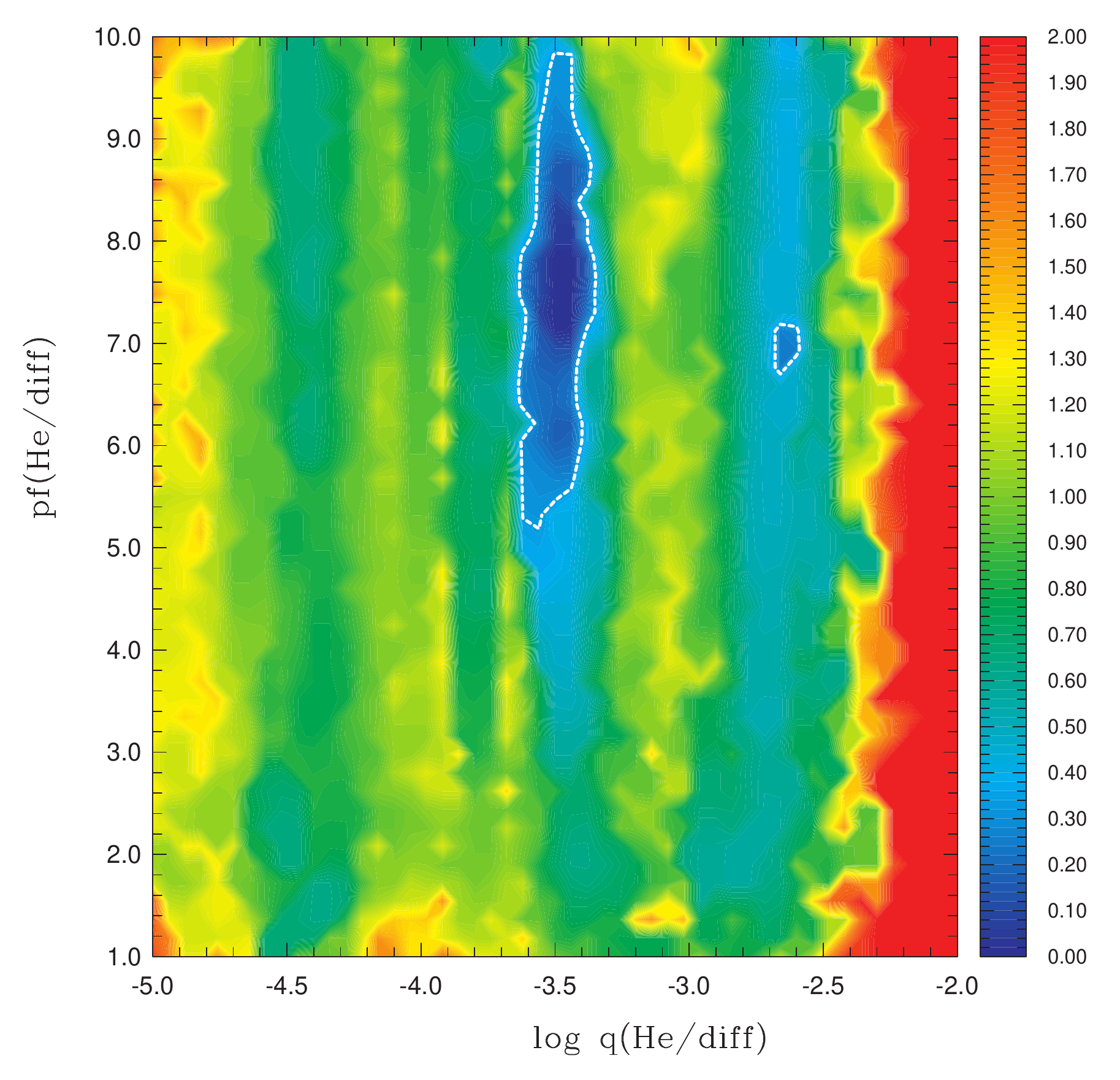}\\
\includegraphics[scale=0.44]{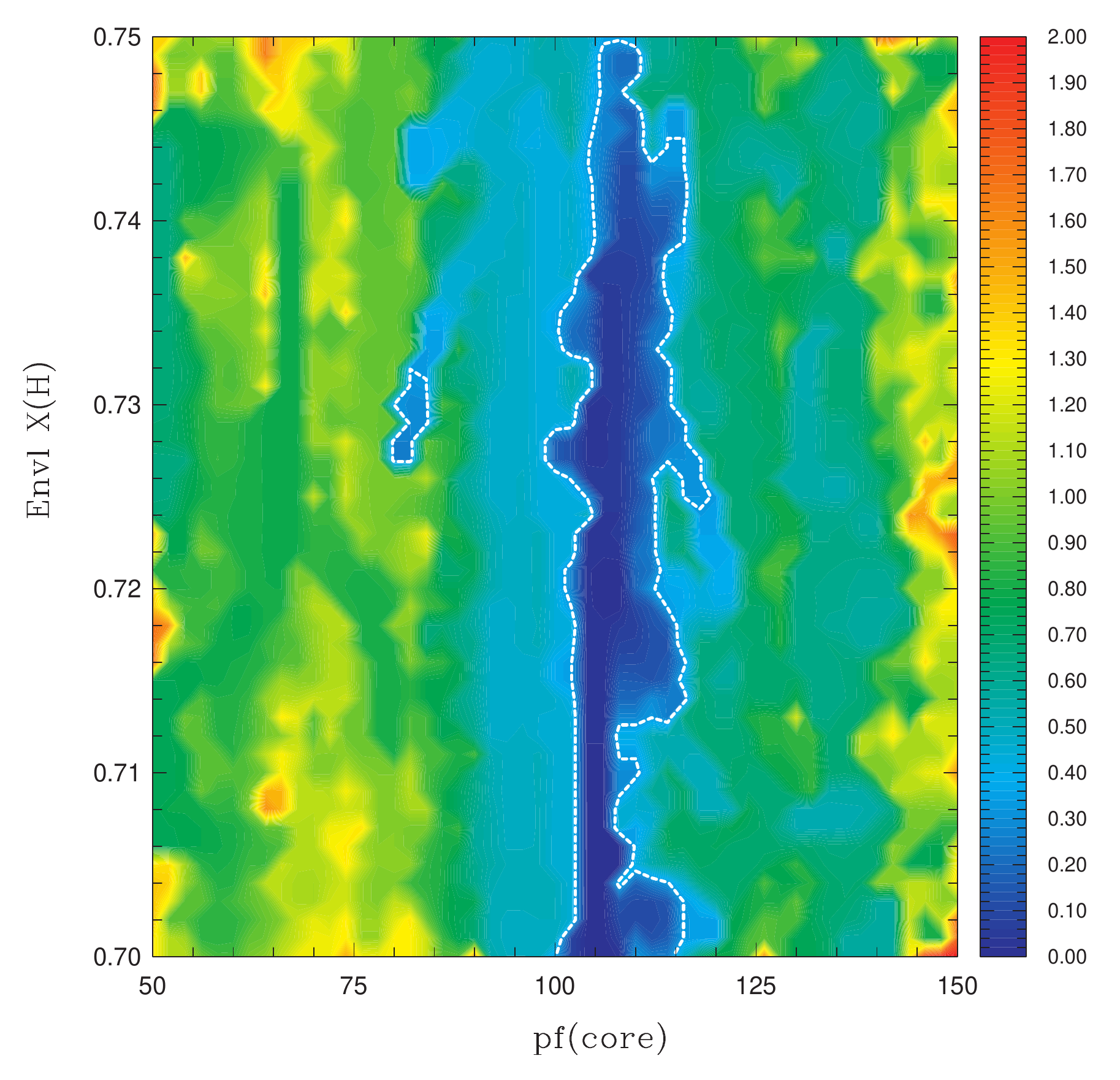}
\par\end{centering}
\caption{$\log S^{2}$ projection maps for pairs of primary model parameters
showing location and shape of best-fitting regions in parameter space
(see text for more details). The value $S^{2}$ is normalized to one
at minimum and the color scale is logarithmic. Dark blue indicates
the best-fit regions and the dotted contour line is an estimate of
the $1\sigma$ confidence level, obtained in a similar way to that
described by \citet{2001ApJ...563.1013B}. \label{fig5}}
\end{figure*}

\begin{figure}[!tbph]
\begin{centering}
\includegraphics[scale=0.44]{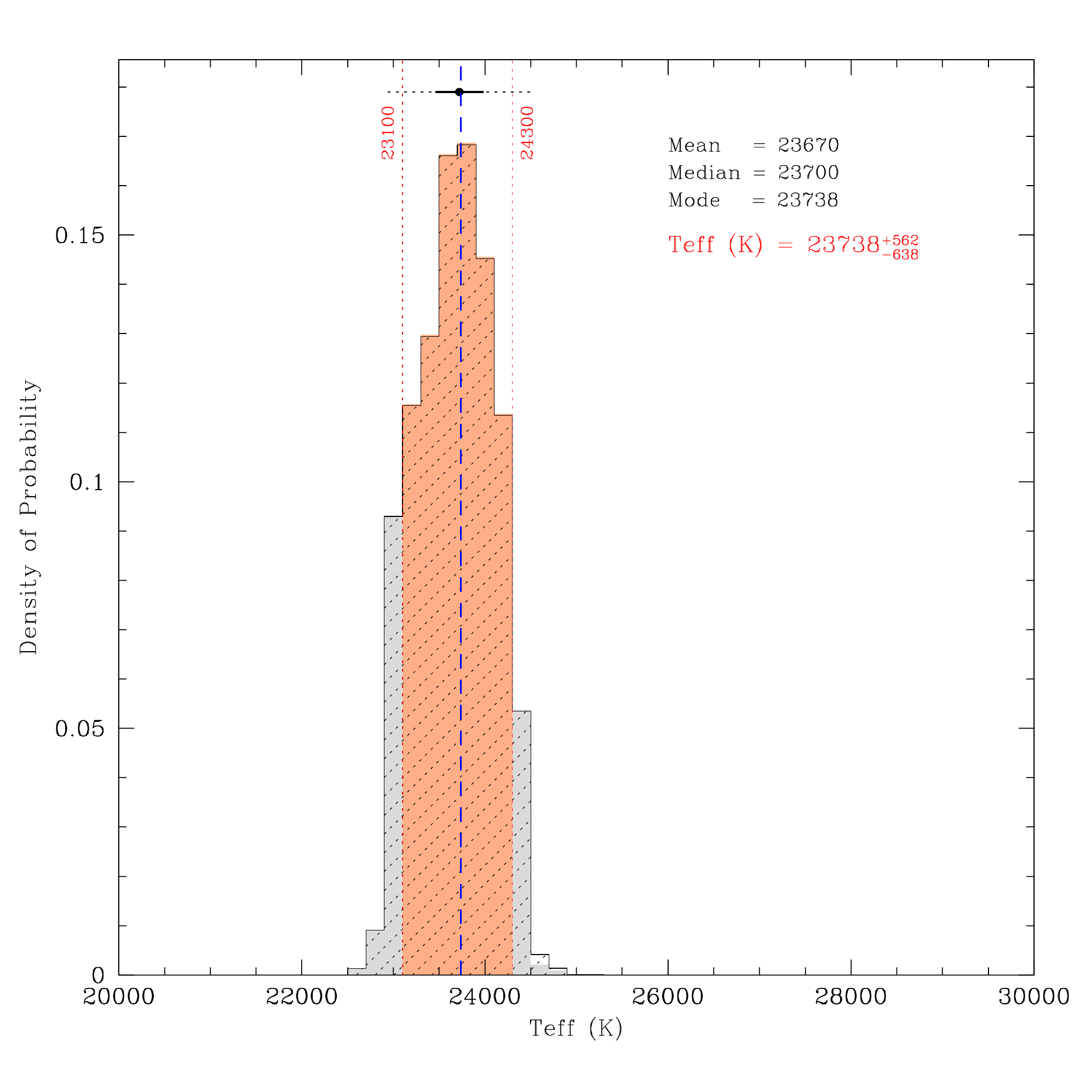}
\par\end{centering}
\begin{centering}
\includegraphics[scale=0.44]{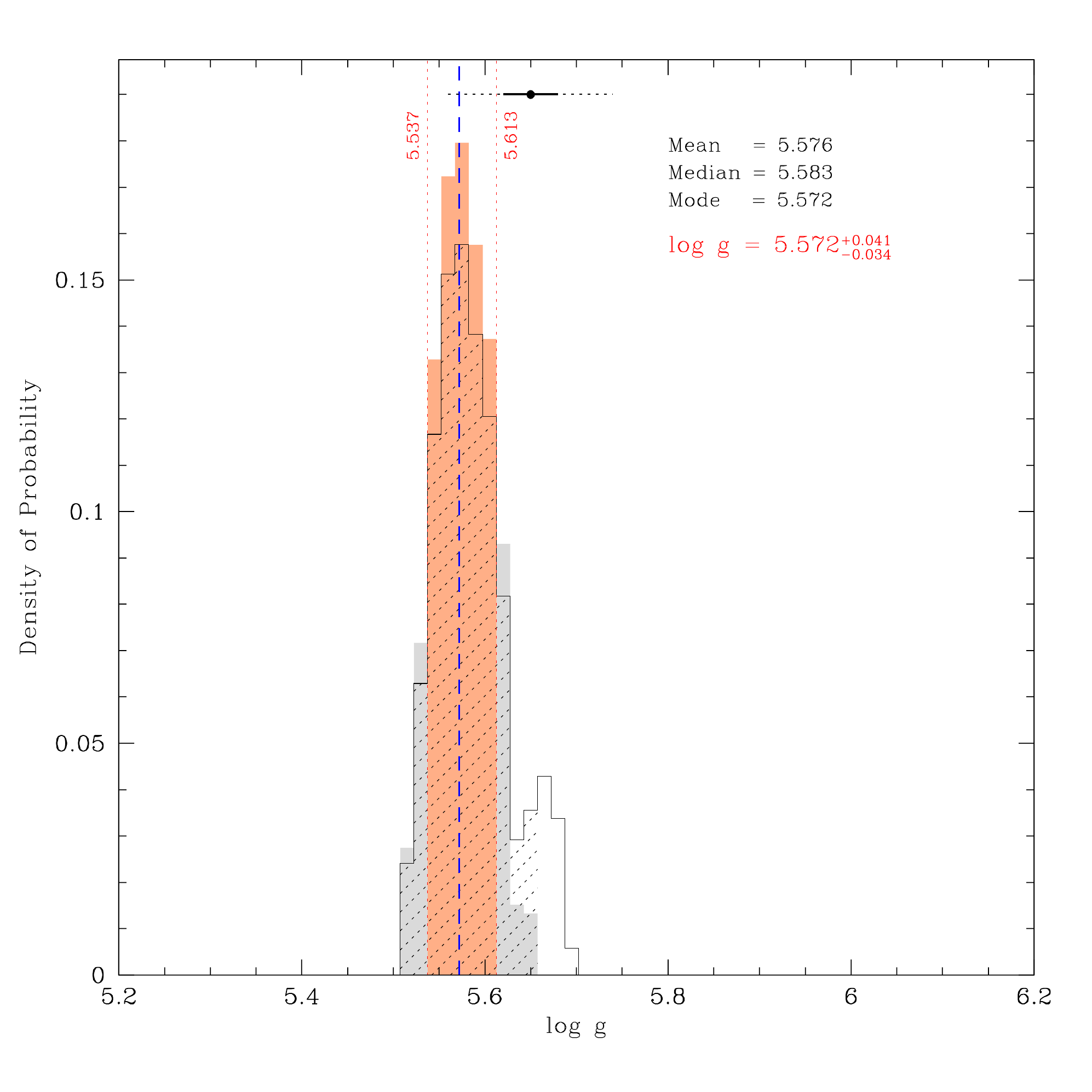}
\par\end{centering}
\caption{Probability distributions for $T_{{\rm eff}}$ (\emph{top panel})
and $\log g$ (\emph{bottom panel}) around the optimal seismic solution
for TIC 278659026. Two distributions are superimposed: one resulting
from the full sample including all secondary optima (black histograms)
and the other computed from a restricted sample focusing on the global
optimum (red and gray shaded histogram). This restricted sample is
built from excluding parts of the parameter space as indicated in
the \emph{bottom panel} for example and in some panels of Fig.~\ref{fig7},
\ref{fig:hist-envl}, \ref{fig:hist-xh}, and \ref{fig:hist-core},
by bins that do not show the dotted hatched areas. The red-shaded areas
delimited by the red vertical dotted lines contain 68\% of the restricted
probability distribution and provide our estimate of the $1\sigma$ error
range for the quantity considered. For internal physical consistency
of the derived parameters, the adopted value is that from the optimal
model (blue vertical dashed line), which closely corresponds to the
mode (maximum) of the distribution. The black dots, and their associated
horizontal lines, indicate the measurement obtained independently
from spectroscopy, and the $1\sigma$ (solid) and $3\sigma$ (dotted)
ranges around that value.\label{fig6}}
\end{figure}

The input parameters needed to fully define the stellar structure
with these improved models are the following. First the total mass,
$M_{*}$, of the star has to be given. All additional parameters specify
the chemical stratification inside the star, as illustrated in Fig.~\ref{fig4}.
We start with five parameters defining the double-layered He and H envelope
structure. The first parameter is the fractional mass of the outer hydrogen-rich
envelope, $\log q({\rm env})=\log(M_{{\rm env}}/M_{*})$, corresponding
in effect to the location of the deepest transition of the double-layered
structure, from the pure He mantle to the mixed He+H region at the
bottom of the envelope. The second parameter is the fractional mass of the
pure hydrogen layer at the top of the envelope, $\log q({\rm H/diff})=\log(M_{{\rm H/diff}}/M_{*})$,
which fixes the location of the transition between the He+H mixed
layer to the pure hydrogen region. The shape (or extent) of both transitions
are controlled by two additional parameters, $Pf({\rm envl})$ and
$Pf({\rm H/diff})$ (see the Methods section of \citealp{2018Natur.554...73G}
for details). The remaining parameter, $X({\rm H})_{{\rm envl}}$,
closes the specification of the envelope structure by fixing the mass
fraction of hydrogen in the mixed He+H region of the envelope. We
also note for completeness that the stellar envelope incorporates
a nonuniform iron distribution computed assuming equilibrium between
radiative levitation and gravitational settling (see \citealp{1997ApJ...483L.123C,2001PASP..113..775C}).
In addition to this set of parameters, the core structure has to be
defined by specifying the fractional mass of the convectively mixed
core, $\log q({\rm core})=\log(1-M_{{\rm core}}/M_{*})$, which sets
the position of the transition between the CO-enriched core (owing to
helium burning) with the surrounding helium mantle. How steep (or
wide) is this transition is specified by the profile factor, $Pf({\rm core})$,
and the composition of the core is determined by the two parameters
$X({\rm He})_{{\rm core}}$ and $X({\rm O})_{{\rm core}}$ providing
the mass fraction of helium and oxygen (the complement being carbon,
with $X({\rm C})_{{\rm core}}=1-X({\rm He})_{{\rm core}}-X({\rm O})_{{\rm core}}$).
A full EHB stellar structure is therefore entirely specified with
the ten primary parameters mentioned above. Other secondary quantities, such as the effective temperature, surface gravity, or
stellar radius, simply derive from the converged stellar model in
hydrostatic and thermal equilibrium.

Finally, we recall that the seismic properties of a given stellar
model are computed using the Montréal pulsation code \citep{1992ApJS...80..725B,2008Ap&SS.316..107B}.
Calculations in the adiabatic approximation are sufficient for the
present purposes and much more efficient numerically than the full
nonadiabatic treatment.

\subsection{Search for an optimal seismic model}

\begin{figure}
\begin{centering}
\includegraphics[scale=0.44]{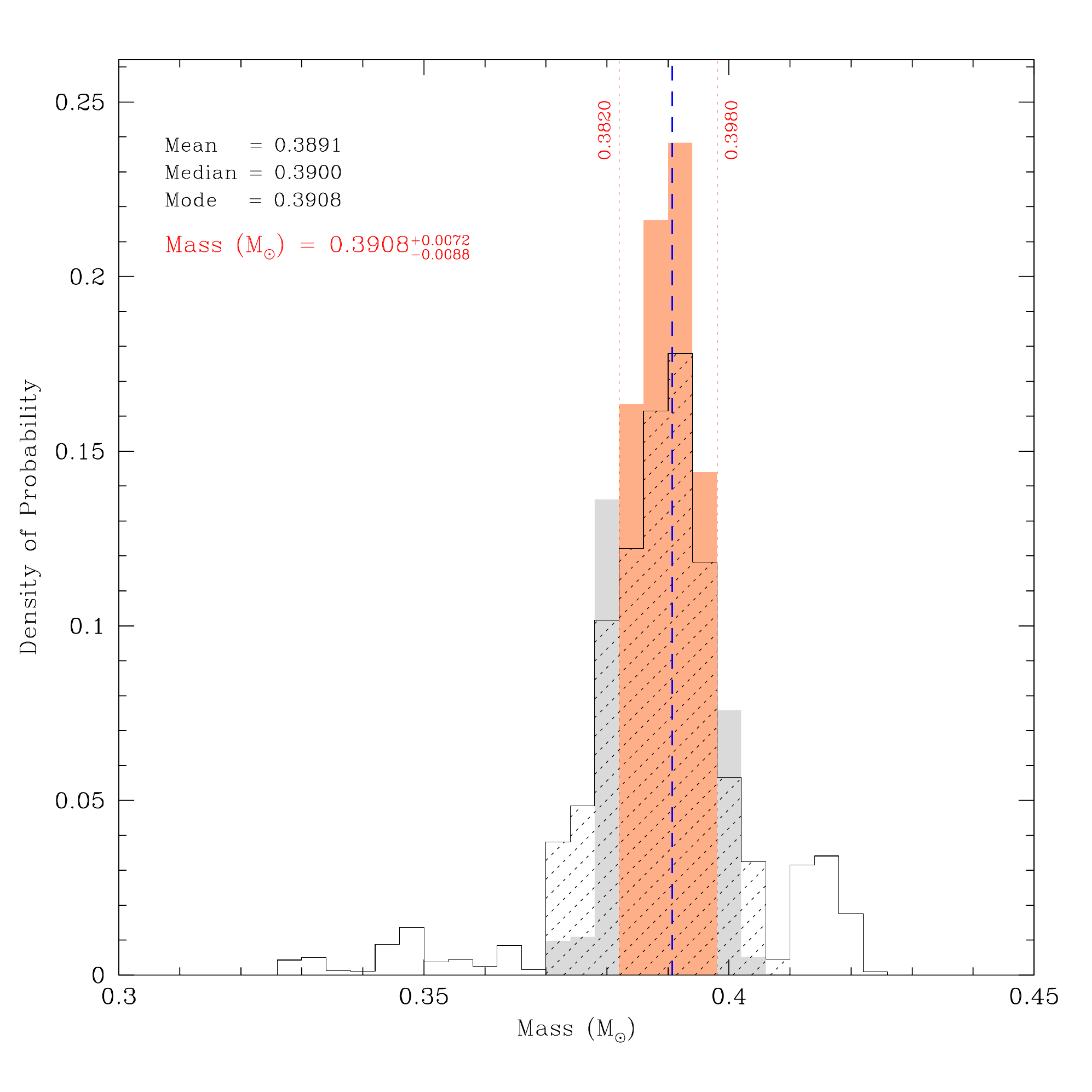}
\par\end{centering}
\begin{centering}
\includegraphics[scale=0.44]{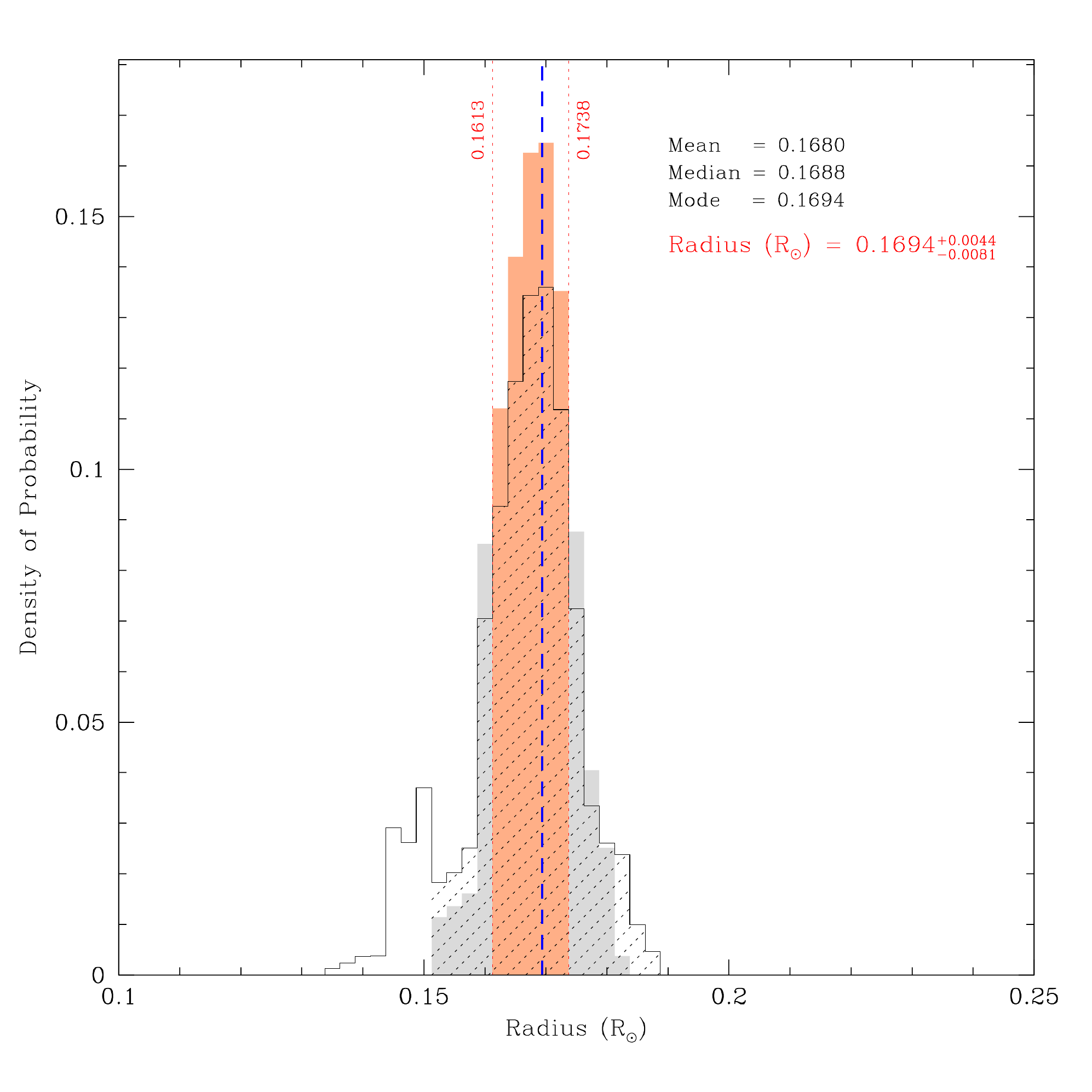}
\par\end{centering}
\caption{Same as Figure \ref{fig6} but for the mass (\emph{top panel}) and
radius (\emph{bottom panel}) of TIC 278659026.\label{fig7} }
\end{figure}

\begin{figure*}
\begin{centering}
\includegraphics[scale=0.45]{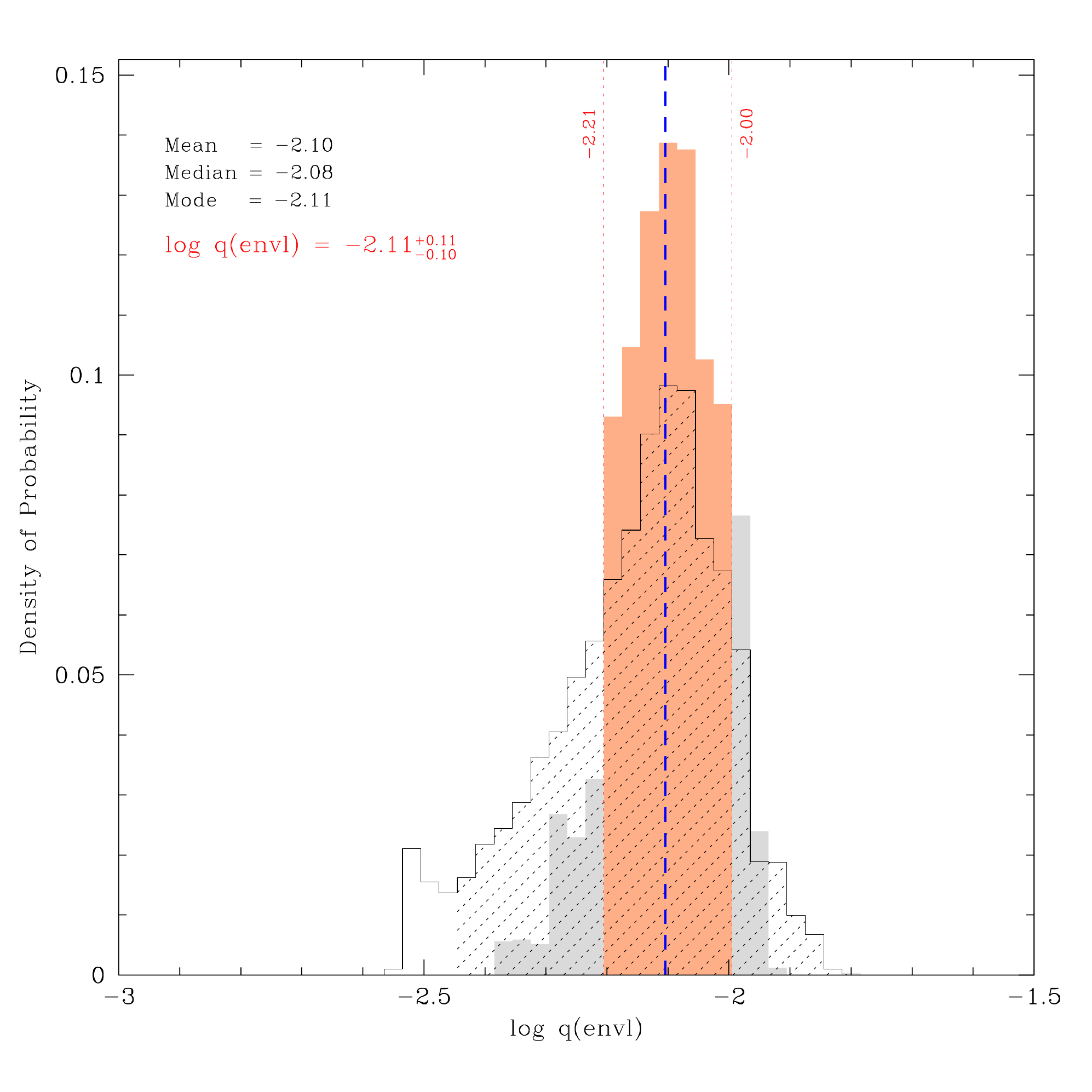}\includegraphics[scale=0.45]{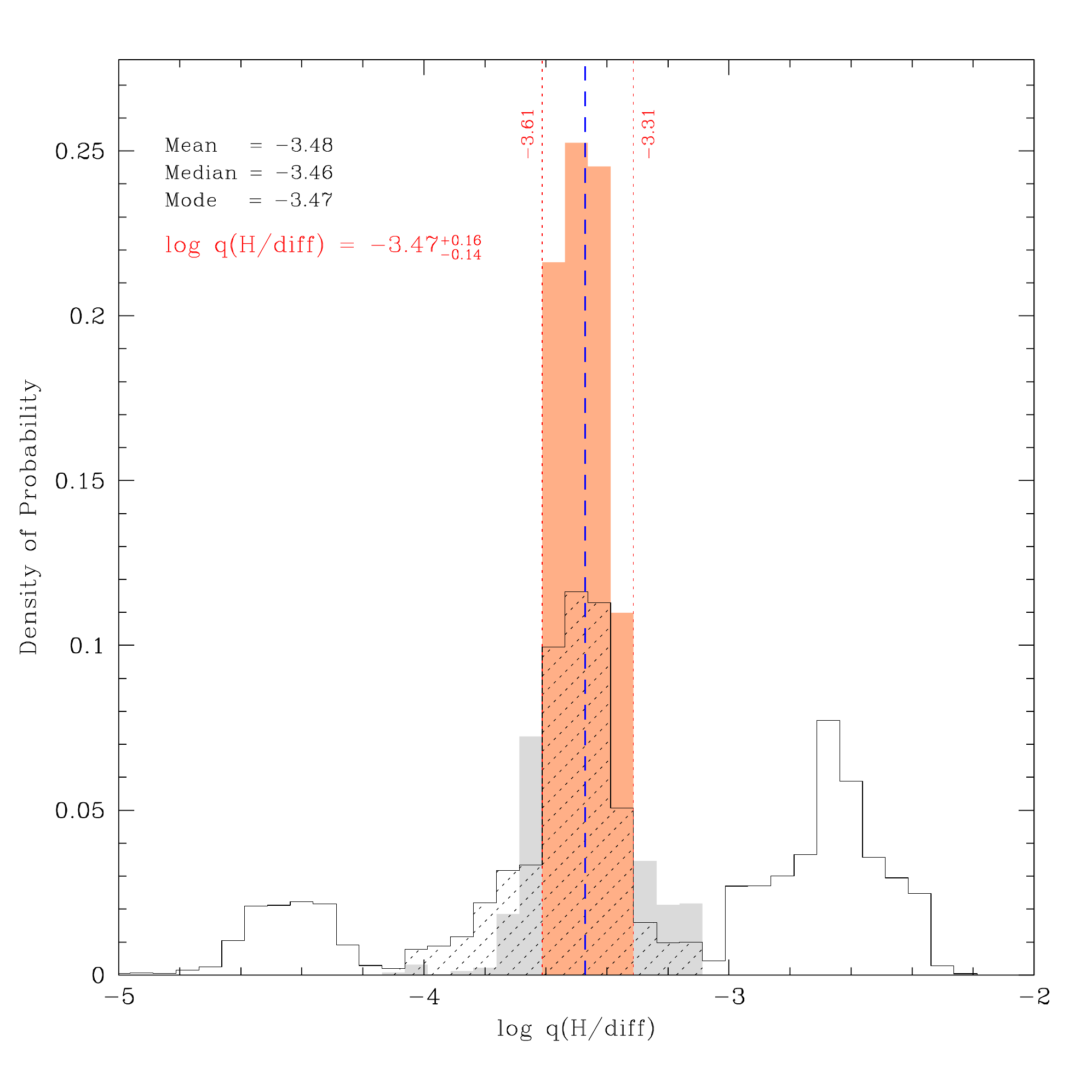}\\
\includegraphics[scale=0.45]{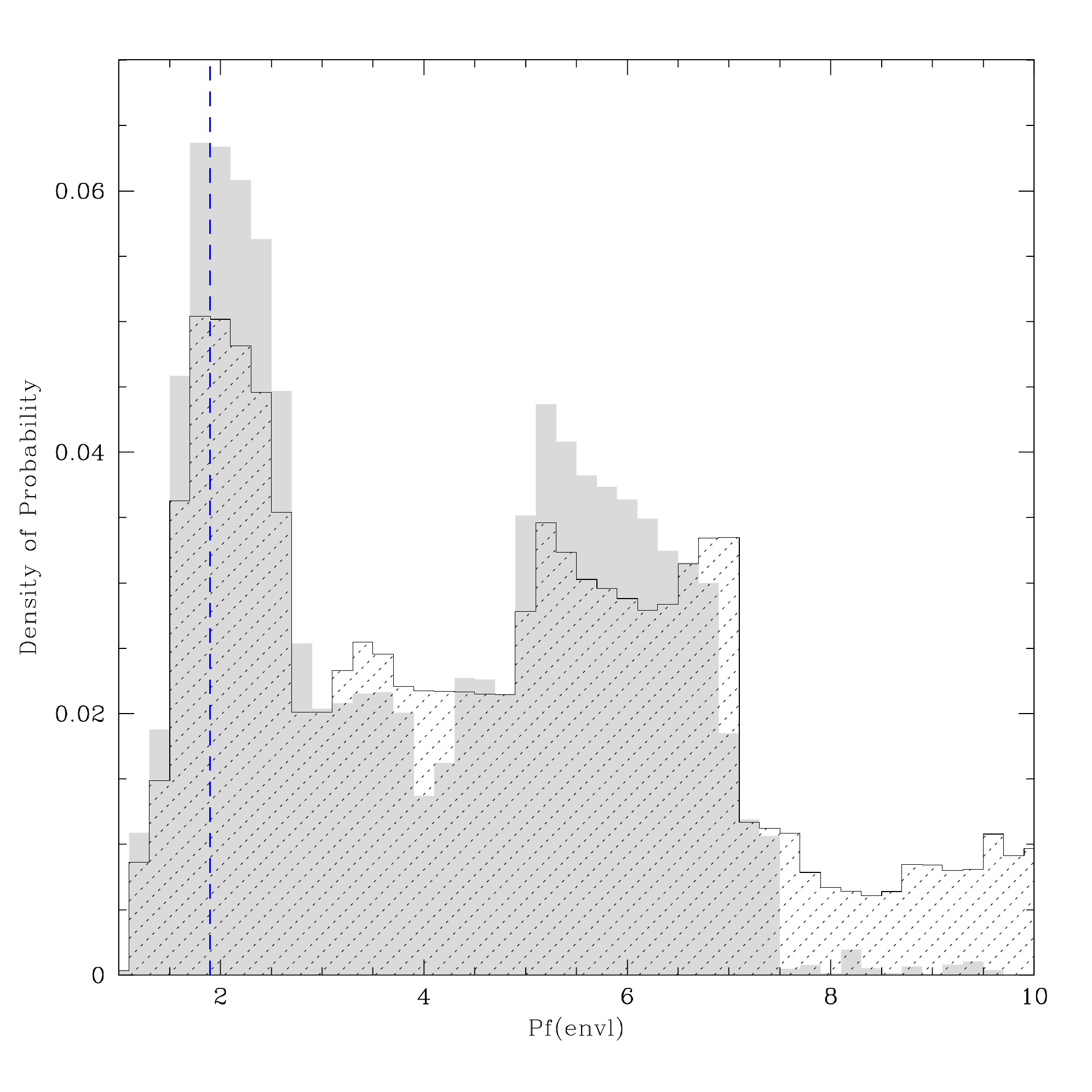}\includegraphics[scale=0.45]{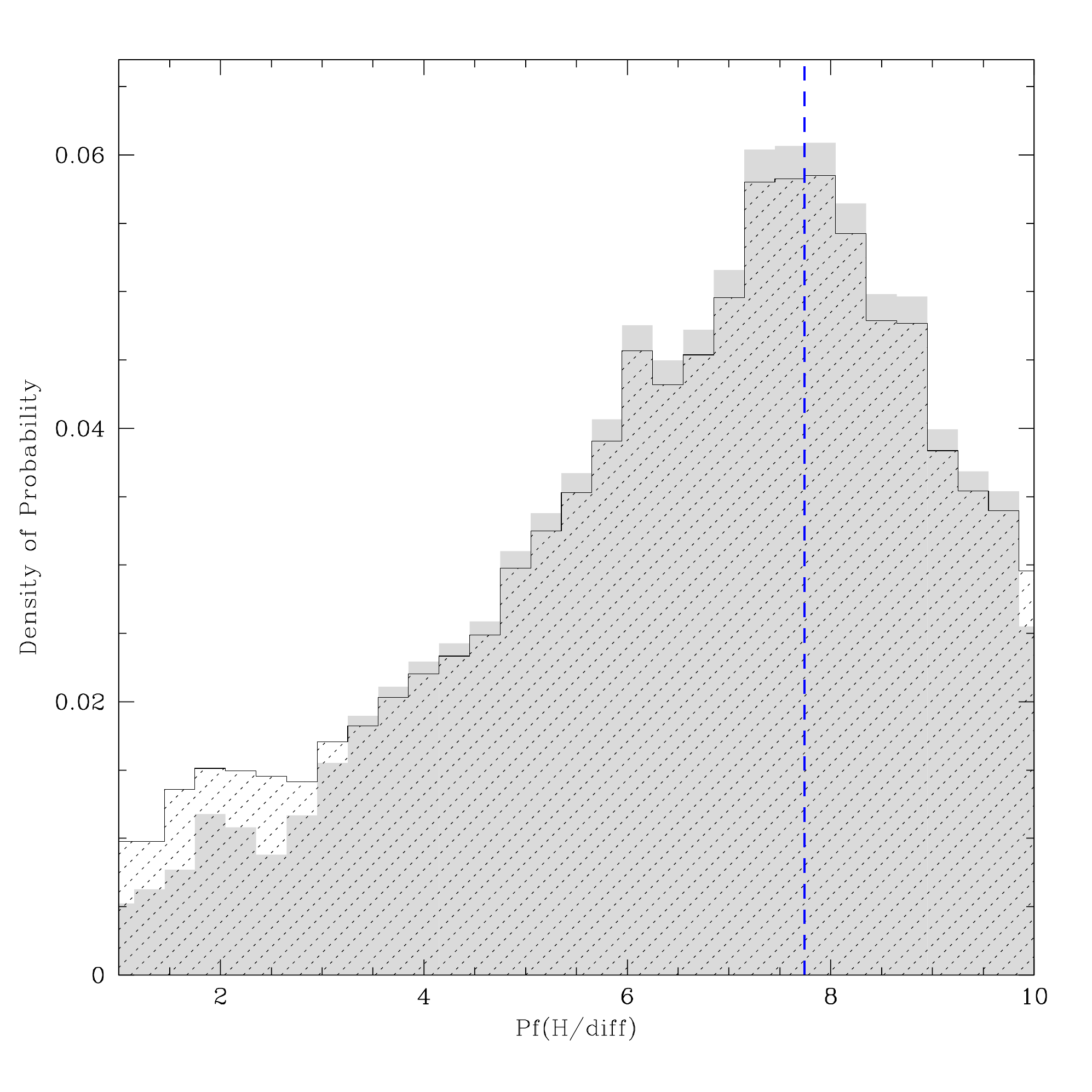}\\
\par\end{centering}
\caption{Same as Figure \ref{fig6} but for the remaining envelope structural
parameters defining the seismic model of TIC 278659026. No errors
are estimated for Pf(envl) and Pf(H/diff) considering the spread of
their distribution.\label{fig:hist-envl}}
\end{figure*}

The search for an optimal solution that reproduces the seismic properties
of TIC 278659026 was conducted in the multidimensional space defined
by the ten primary model parameters discussed in the previous subsection.
A first exploratory optimization with the code \noun{lucy} was performed
to cover the largest possible domain in parameter space, relevant
for virtually all configurations that could potentially correspond
to a typical sdB star. This wide domain is defined in the
second column of Table \ref{tab:ranges} and the search was done without
considering the constraint available from spectroscopy at this stage.
Remarkably, this first calculation pointed toward a basin of solutions
located very close to the spectroscopic values of $\log g$ and $T_{{\rm eff}}$,
suggesting a sdB star of relatively low mass. This helped to reduce
the size of the search space for the ultimate convergence toward the
best-fit solution.

\begin{figure}
\begin{centering}
\includegraphics[scale=0.45]{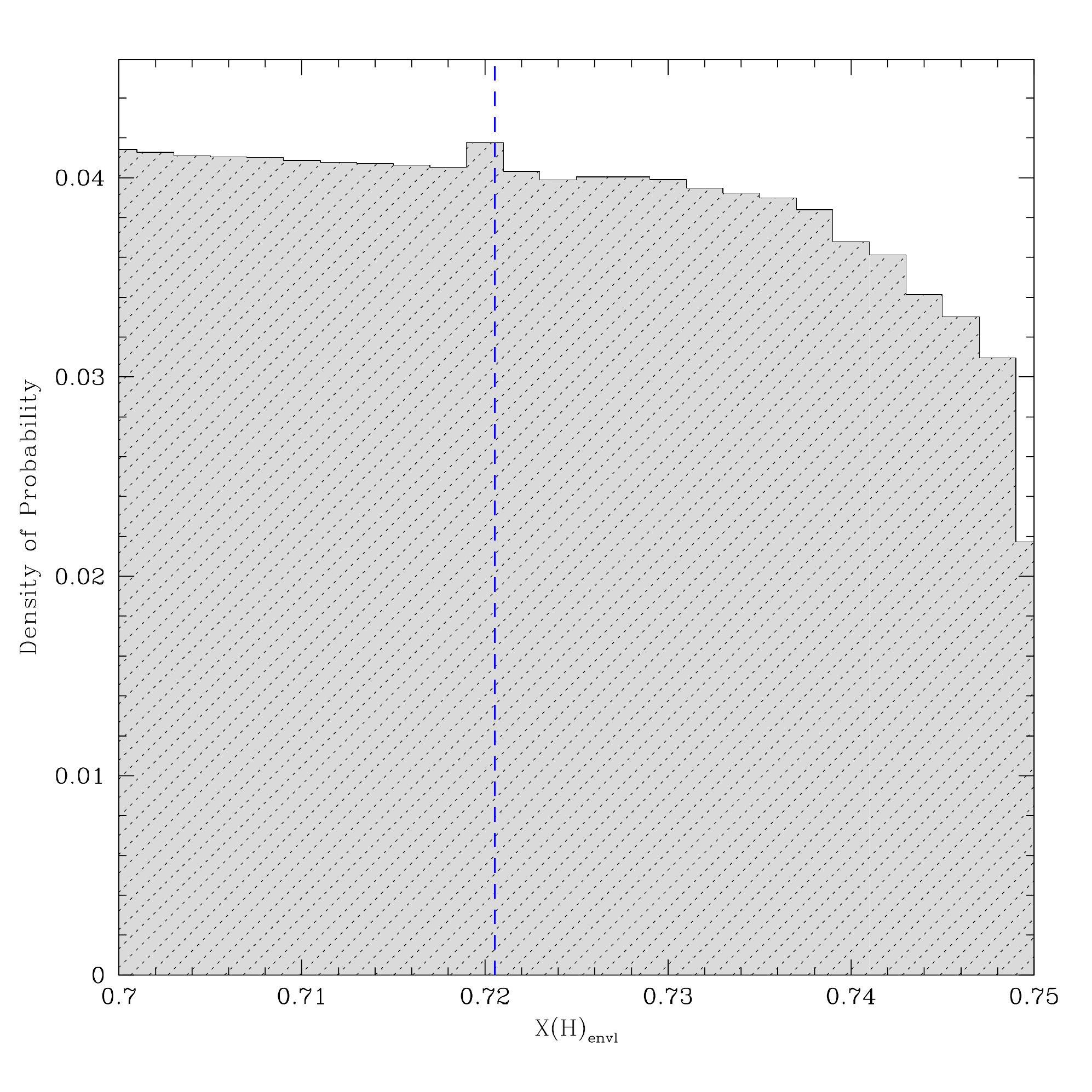}
\par\end{centering}
\caption{Same as Figure \ref{fig:hist-envl} but for the mass fraction of hydrogen,
$X({\rm H})_{{\rm envl}}$, at the bottom of the hydrogen-rich envelope.
\label{fig:hist-xh}}
\end{figure}

\begin{figure*}
\begin{centering}
\includegraphics[scale=0.45]{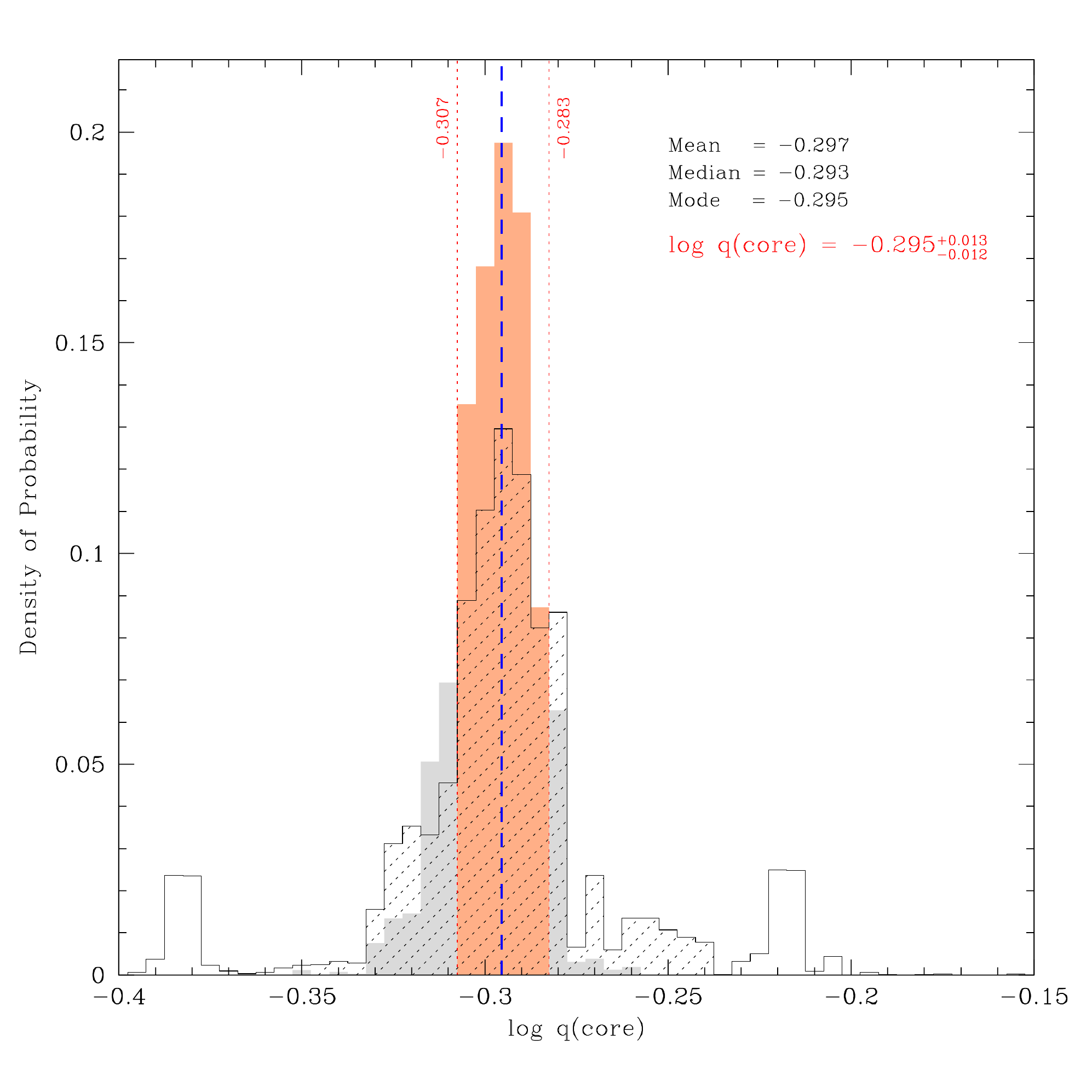}\includegraphics[scale=0.45]{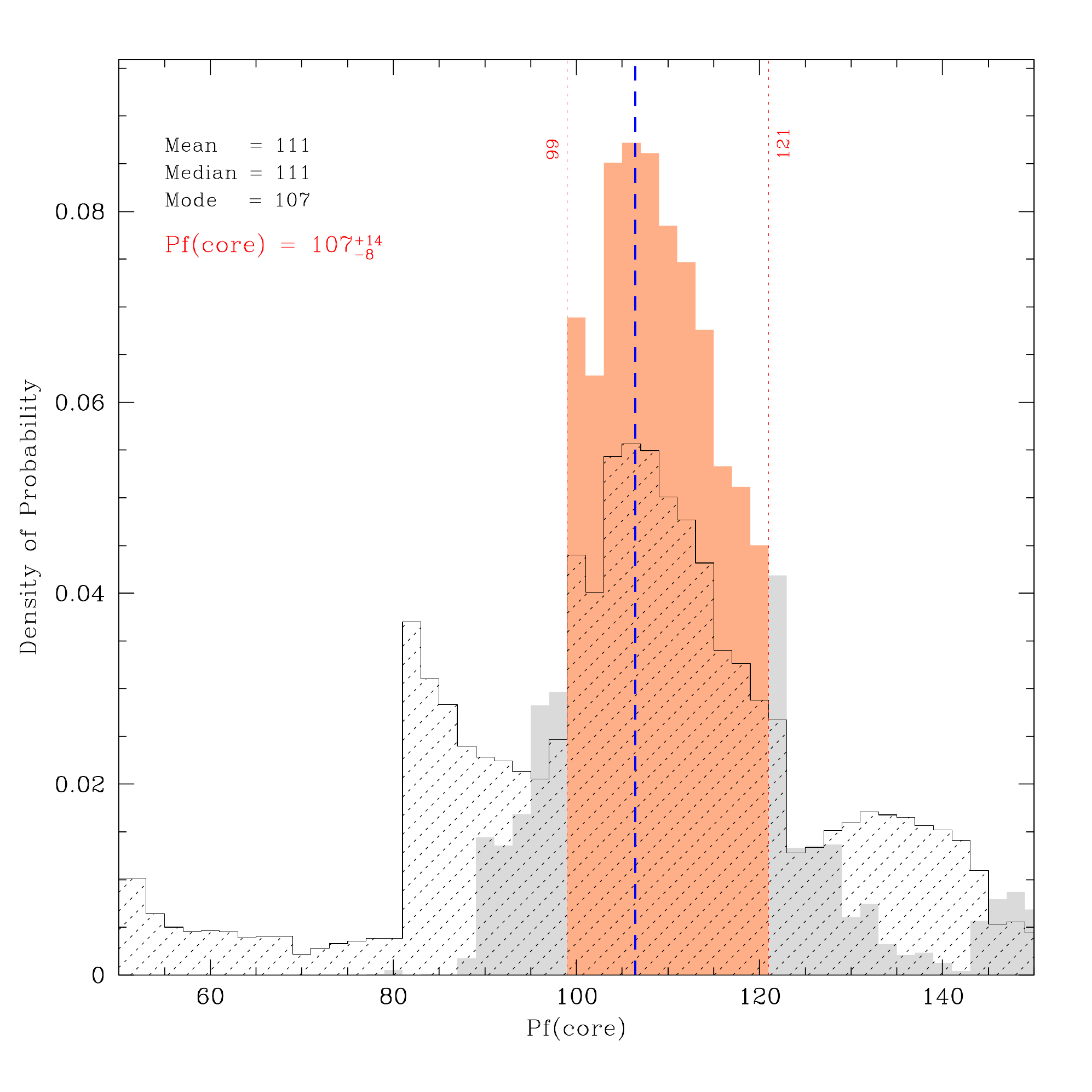}\\
\includegraphics[scale=0.45]{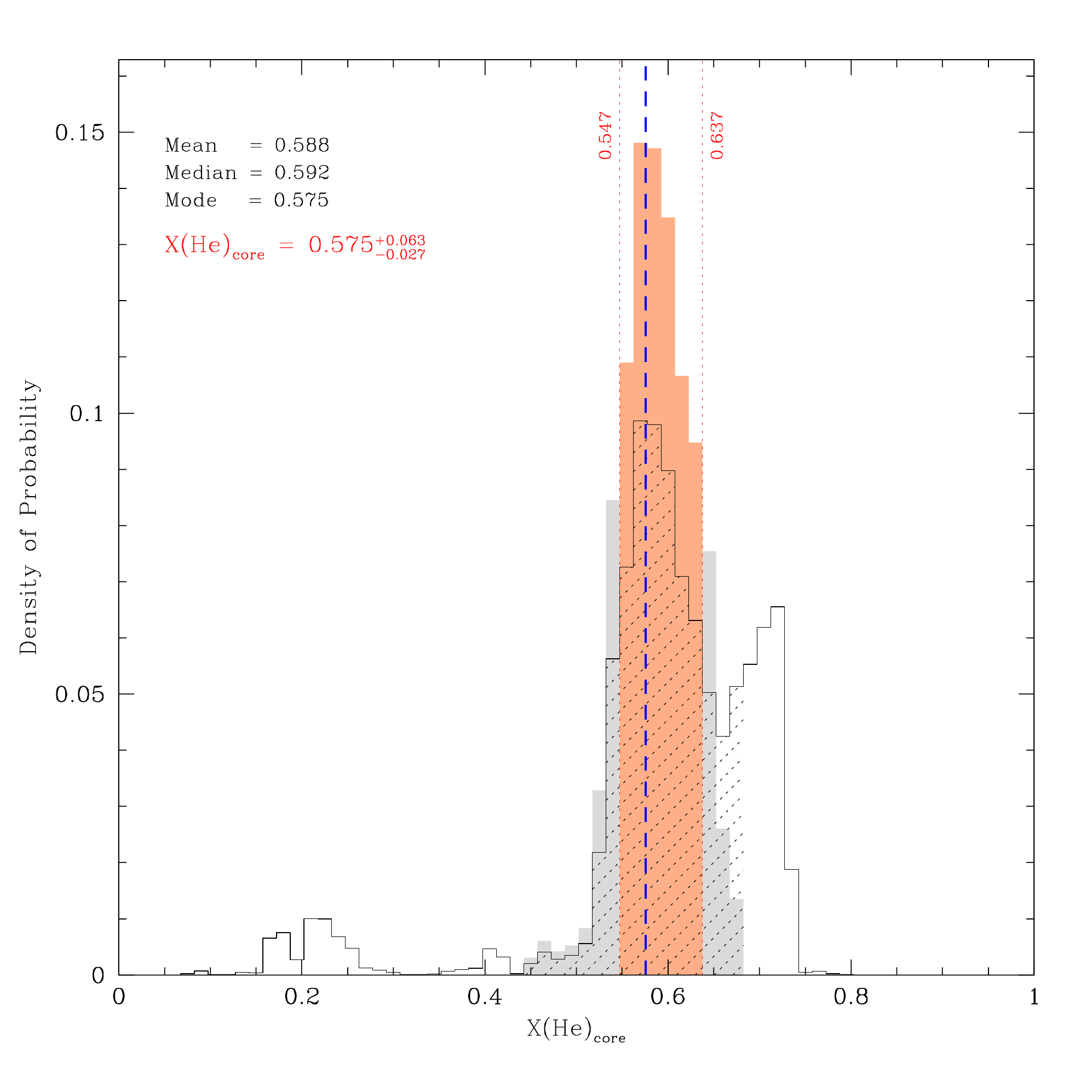}\includegraphics[scale=0.45]{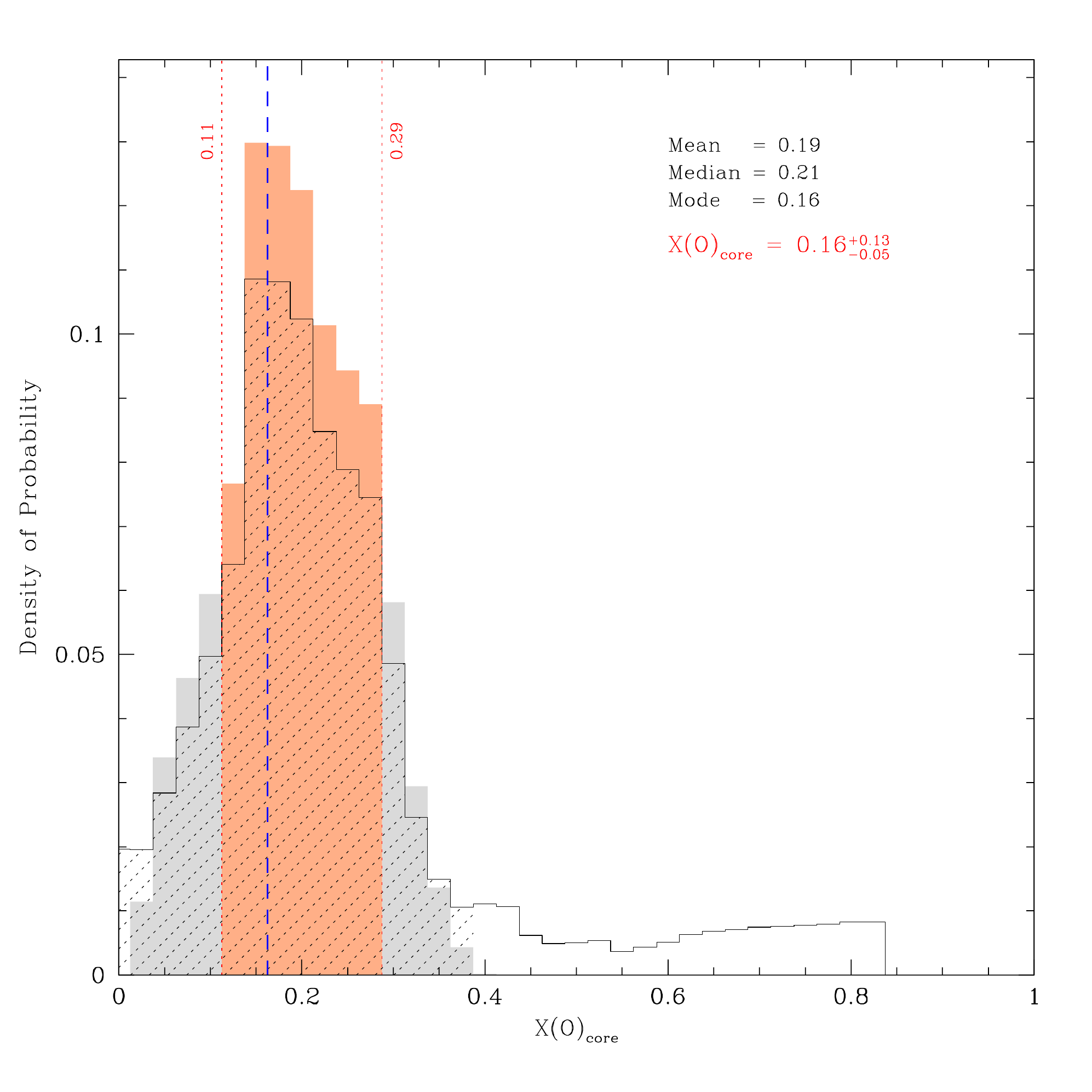}
\par\end{centering}
\caption{Same as Figure \ref{fig6} but for the core structural parmameters
defining the seismic model of TIC 278659026. \label{fig:hist-core}}
\end{figure*}

The second and ultimate round of optimization was conducted within
the ranges given in the third column of Table \ref{tab:ranges}. This
time the spectroscopic values of $\log g$ and $T_{{\rm eff}}$ were
introduced in the optimization process as additional external constraints
degrading progressively the merit function when a computed model has
$\log g$ and/or $T_{{\rm eff}}$ that differ from spectroscopy by
more than a tolerance range set to be $3\sigma$ (three times the spectroscopic
error). The optimizer \noun{lucy} was configured to aggressively converge
each potential model solution found in parameter space by combining
the GA search with a simplex optimization round. This step ensures
that the global and eventual local minima of the merit function are
effectively reached. This second round of calculation required the
computation and comparison with observed frequencies of 1,555,425
seismic models\footnote{This calculation was performed on the high-performance cluster OLYMPE
at the CALMIP computing center using 480 CPU cores in parallel for
approximately three days.}. We point out that the parameter $X({\rm H})_{{\rm envl}}$ was kept
within the narrow $0.70-0.75$ range from the assumption that the
progenitor of TIC 278659026 had a homogeneous H+He envelope close
to solar composition when it entered the EHB phase. Yet, the possibility
that the star may have a significantly different H/He ratio in the
envelope as a consequence of passed episodes of mixing related to binary evolution
is not totally excluded. The impact of widening this range is discussed
in Sect. 3.3.3 and Appendix A.

\begin{table}[!tbph]
\caption{Derived properties of TIC 278659026.\label{tab2}}

\vspace{-20bp}
\small
\begin{center}\begin{tabular}{lll}
\hline
\\
 Quantity & Value derived   & Other        \tabularnewline
          & from seismology & Measurement  \tabularnewline
\\
\hline\hline
\\
 \multicolumn{3}{l}{Primary quantities$^{a}$} \\
 \\
 \hline
 \\
 $M_*/M_{\odot}$            & $0.391 \pm 0.009$         & $0.390 \pm 0.091^{\ddag}$     \tabularnewline

 $\log q ({\rm env})$       & $-2.11 \pm 0.11$          & ...                           \tabularnewline
 $\log q ({\rm H/diff})$    & $-3.47 \pm 0.16$          & ...                           \tabularnewline
 $X({\rm H})_{\rm envl}$    & $\sim 0.72$               & ...                           \tabularnewline
 Pf(envl)                   & $\sim 1.87$               & ...                           \tabularnewline
 Pf(H/diff)                 & $\sim 7.7$                & ...                           \tabularnewline

 $\log q ({\rm core})$      & $-0.295 \pm 0.013$        & ...                           \tabularnewline
 Pf(core)                   & $107 \pm 14$              & ...                           \tabularnewline
 $X({\rm He})_{\rm core}$   & $0.575_{-0.027}^{+0.063}$ & ...                           \tabularnewline
 $X({\rm O})_{\rm core}$    & $0.16_{-0.05}^{+0.13}$    & ...                           \tabularnewline

\\
 \hline
 \\
 \multicolumn{3}{l}{Secondary quantities$^{b}$}\\
 \\
 \hline
 \\
 $T_{\rm eff}$ (K)          & $23740 \pm 640$           & $23720 \pm 260^{c}$        \tabularnewline
 $\log g$                   & $5.572 \pm 0.041$         & $5.650 \pm 0.030^{c}$      \tabularnewline
 $R/R_{\odot}$              & $0.1694 \pm 0.0081$       & ...                           \tabularnewline
 $M_{\rm H}/M_\odot$        & $0.0037\pm 0.0010^{\dag}$ & ...                           \tabularnewline
 $M_{\rm core}/M_\odot$     & $0.198\pm 0.010$          & ...                           \tabularnewline
 $X({\rm C})_{\rm core}$    & $0.27_{-0.14}^{+0.06}$    & ...                           \tabularnewline
& \tabularnewline
 $L/L_{\odot}$ ($T_{\rm eff}$, $R$)            & $8.2 \pm 1.7$          & $8.2 \pm 1.1^{c}$     \tabularnewline

 $M_V$ ($g$, $T_{\rm eff}$, $M_*$)             & $4.92 \pm 0.15^{d}$    & ...                   \tabularnewline
 $M_{B_p}$ ($g$, $T_{\rm eff}$, $M_*$)         & $4.70 \pm 0.16^{d}$    & ...                   \tabularnewline
 ($B$-$V$)$_0$                                 & $-0.225 \pm 0.007^{d}$ & ...                   \tabularnewline
 ($B_p$-$R_p$)$_0$                             & $-0.404 \pm 0.011^{d}$ & ...                   \tabularnewline

& \tabularnewline
 $V$                                    & ...                   & $11.57 \pm 0.09^{e}$  \tabularnewline
 ($B$-$V$)                              & ...                   & $-0.22 \pm 0.10^{e}$  \tabularnewline
 $B_p$                                  & ...                   & $11.433 \pm 0.008^{f}$  \tabularnewline
 ($B_p$-$R_p$)                          & ...                   & $-0.398 \pm 0.009^{f}$  \tabularnewline
& \tabularnewline
 $E$($B$-$V$)                           & $0.005 \pm 0.107$     & ...                      \tabularnewline
 $E$($B_p$-$R_p$)                       & $0.005 \pm 0.020$     & ...                      \tabularnewline
& \tabularnewline
 $d_{\rm parallax}$ (pc)                & ...                   & $203.7 \pm 2.1^{g}$      \tabularnewline
 $d_V$ (pc)                             & $212.2 \pm 57.0^{h}$  & ...                      \tabularnewline
 $d_{B_p}$ (pc)                         & $220.9 \pm 20.4^{i}$  & ...                      \tabularnewline
 \\
 \hline
\end{tabular}\end{center}
{\footnotesize $^{a}$Optimized model parameters (see text)\\
$^{\ddag}$From fitting the SED and using {\sl Gaia} DR2 parallax (see text)\\
$^{b}$Quantities derived from the computed models\\
$^{c}$From spectroscopy (N\'emeth et al. 2012)\\
$^{\dag}$$\log \frac{M(\rm H)}{M_*} = \log q({\rm envl})+C = -2.0251$;
$C$ comes from the model\\
$^{d}$From a model atmosphere with $\log(\frac{{\rm He}}{{\rm H}})=-3.0$\\
$^{e}$From H{\o}g et al. 2000\\
$^{f}$From {\sl Gaia} DR2 photometry\\
$^{g}$From {\sl Gaia} DR2 parallax, $\varpi = 4.910 \pm 0.051$ mas\\
$^{h}$$d=213.8\pm23.4$ without correction for extinction\\
$^{i}$$d=221.9\pm16.6$ without correction for extinction
}
\normalsize

\end{table}

The first result emerging from the optimization in the 10D
parameter space is the identification of a solution for TIC 278659026
that clearly dominates over other secondary options. This is illustrated
in Fig.~\ref{fig5}, showing the 2D projection maps of $\log S^{2}$
($S^{2}$ being normalized to one at minimum in this context) for
various pairs of the ten model parameters. The approximate reconstruction
of the topology of $S^{2}$ derives from the sampling of the merit
function by the optimizer during the search. While $\log S^{2}$ has
a rather complex structure with several local dips, as expected, a
dominant global minimum is found. Moreover, all parameters appear
well constrained (i.e., precisely localized around the optimum), except
for the hydrogen content in the mixed He+H region at the base of
the envelope, $X({\rm H})_{{\rm envl}}$, whose minimum is an elongated
valley (see also Appendix A), and, to a lesser extent, the shape factors
$Pf({\rm envl})$ and $Pf({\rm H/diff})$ that also tend to spread
more over their explored range than other parameters.

The identification of a well-defined global minimum dominating all
the other optima allows for the selection of an optimal seismic model
of TIC 278659026 and the determination of its parameters through the
series of plots provided in Figs.~\ref{fig6}, \ref{fig7}, \ref{fig:hist-envl},
\ref{fig:hist-xh}, and \ref{fig:hist-core}. These histograms show
the probability distributions for the most relevant parameters through
marginalization. As described in, for example, \citet{2013A&A...553A..97V},
these distributions are evaluated from the likelihood function ($\propto e^{-\frac{1}{2}S^{2}})$
and are useful to estimate the internal error associated with each
parameter value. In this work, slightly differing from previous similar studies,
the adopted parameter values are those given by the optimal model
solution, which closely corresponds to the mode (maximum) of their
associated distribution. This ensures internal physical consistency
between all values, since they exactly represent the optimal seismic
model. Moreover, since some histograms can differ significantly from
the normal distribution, for instance when secondary peaks are visible,
we chose to estimate these errors by constructing a second distribution
computed from a restricted sample in the vicinity of the dominant
solution, thus filtering out secondary optima (see the caption of
Fig.~\ref{fig6} for more details). All the values derived from the
optimal seismic model and their estimated errors are provided in Table
\ref{tab2}, along with other independent measurements of the same
quantity when available, for comparison purposes.

\subsubsection{Global parameters}

\begin{figure*}
\begin{centering}
\includegraphics[scale=0.65]{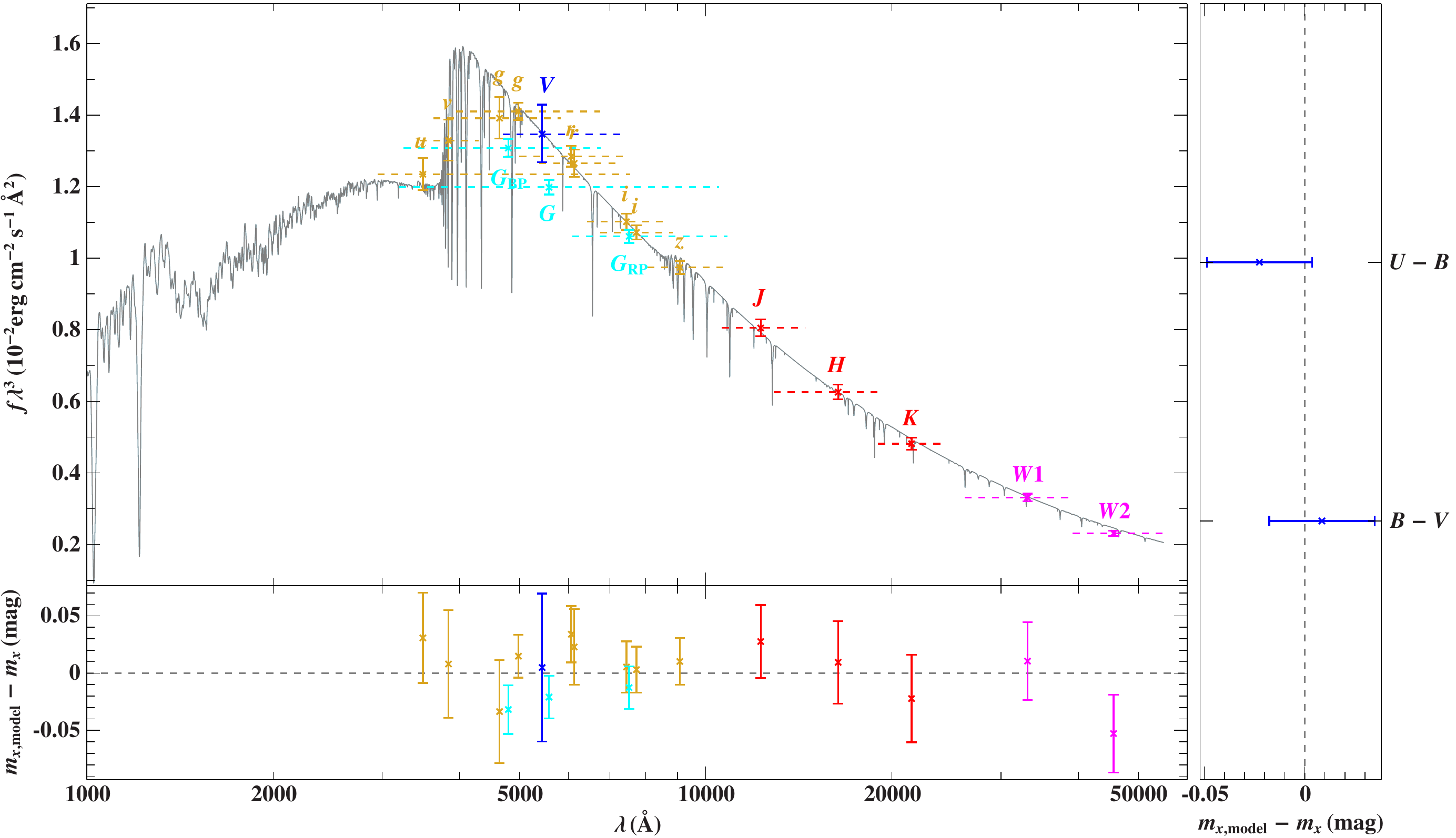}
\par\end{centering}
\caption{Comparison of synthetic and observed photometry: The \emph{top panel}
shows the SED. Colored data points represent
the filter-averaged fluxes, which were converted from observed magnitudes;
the respective filter width is indicated by dashed horizontal lines.
The gray solid line represents a synthetic spectrum computed
from a model atmosphere with the best-fit effective temperature. The
panels at the bottom and on the right-hand side show the differences
between synthetic and observed magnitudes and colors. The following color
codes are used to identify the photometric systems: Johnson (blue),
APASS $g$, $r$, $i$ (yellow), SkyMapper (yellow), \emph{Gaia} (cyan),
2MASS (red), and WISE (magenta). \label{fig:sed}}
\end{figure*}

The seismic model uncovered for TIC 278659026 has $T_{{\rm eff}}$
and $\log g$ values in very close agreement \textendash{} both within
or very close to the $1\sigma$ error bars \textendash{} with the
values measured independently from spectroscopy (Fig.~\ref{fig6}).
This demonstrates a remarkable consistency of the solution that was
not guaranteed a priori. We also find that TIC 278659026 has
a tightly constrained radius of $0.1694\pm0.0081$ $R_{\odot}$ and
a mass of $0.391\pm0.009$ $M_{\odot}$ (Fig.~\ref{fig7}). The latter
is significantly less than the canonical mass of $\sim0.47$ $M_{\odot}$
expected for typical sdB stars, which indeed explains why
TIC 278659026 has a rather high surface gravity (and small radius)
given its relatively low effective temperature, which places it somewhat
below the bulk of hot subdwarfs in a $\log g-T_{{\rm eff}}$ diagram.
It is the first $g$-mode pulsating sdB star (and only second pulsator,
if we include the \emph{V361 Hya} stars) with such a low mass determined
precisely from asteroseismology. Therefore, this object bears a particular
interest in the context of determining the empirical mass distribution
of field sdB stars \citep[see,][]{2012A&A...539A..12F}.

A second important test of the reliability of the seismic solution
is provided by the comparison, in Table \ref{tab2}, between the distance
of the star measured from the available \emph{Gaia} DR2 trigonometric
parallax and the ``seismic distance'' evaluated from the optimal
model properties. The latter is obtained by computing a representative
model atmosphere of TIC 278659026 based on $T_{{\rm eff}}$ and $\log g$
derived from the seismic model \citep[see,][]{2019ApJ...880...79F}.
The synthetic spectrum computed from this model atmosphere is then
used to derive absolute magnitudes and theoretical color indices in
the photometric band-passes of interest. Combined with photometric
measurements, these values give access to estimates of the interstellar
reddening and ultimately to the distance modulus (corrected by the
extinction, which is found to be almost negligible in the present
case). We find that the seismic distance obtained for TIC 278659026
is in very close agreement \textendash{} compatible with $1\sigma$
errors \textendash{} with the distance measured by \emph{Gaia}. This
indicates that the global properties of the star obtained from our
seismic analysis, in particular its effective temperature, mass, and
radius, are indeed accurate to the quoted precision.

\begin{table}[!tbph]
\caption{List of the observed magnitudes used to fit the SED (see Fig.~\ref{fig:sed}).\label{tab:phot}}

\vspace{-20bp}
\normalsize
\begin{center}\begin{tabular}{lll}
\hline
\\
 System & Bandpass & Magnitude        \tabularnewline
\\
\hline\hline
\\
{\sl Gaia}    &      G       &       $11.5928\pm0.0009$  \\
{\sl Gaia}    &    GRP       &       $11.8308\pm0.0018$  \\
{\sl Gaia}    &    GBP       &       $11.4327\pm0.0083$  \\
      WISE    &     W1       &       $12.413\pm0.023$    \\
      WISE    &     W2       &       $12.512\pm0.023$    \\
     2MASS    &      J       &       $12.152\pm0.021$    \\
     2MASS    &      H       &       $12.264\pm0.025$    \\
     2MASS    &      K       &       $12.375\pm0.027$    \\
      SDSS    &    $g$       &       $11.387\pm0.033$    \\
      SDSS    &    $r$       &       $11.792\pm0.022$    \\
      SDSS    &    $i$       &       $12.144\pm0.010$     \\
   Johnson    &    B-V       &       $-0.23$              \\
   Johnson    &    U-B       &       $-0.89$              \\
   Johnson    &      V       &       $11.62$              \\
 SkyMapper    &    $u$       &       $11.283\pm0.028$     \\
 SkyMapper    &    $v$       &       $11.309\pm0.035$     \\
 SkyMapper    &    $g$       &       $11.444\pm0.002$     \\
 SkyMapper    &    $r$       &       $11.767\pm0.013$     \\
 SkyMapper    &    $i$       &       $12.208\pm0.006$     \\
 SkyMapper    &    $z$       &       $12.521\pm0.007$     \\

  \\
 \hline
\end{tabular}\end{center}

\normalsize

\end{table}

\begin{figure}
\begin{centering}
\includegraphics[scale=0.48]{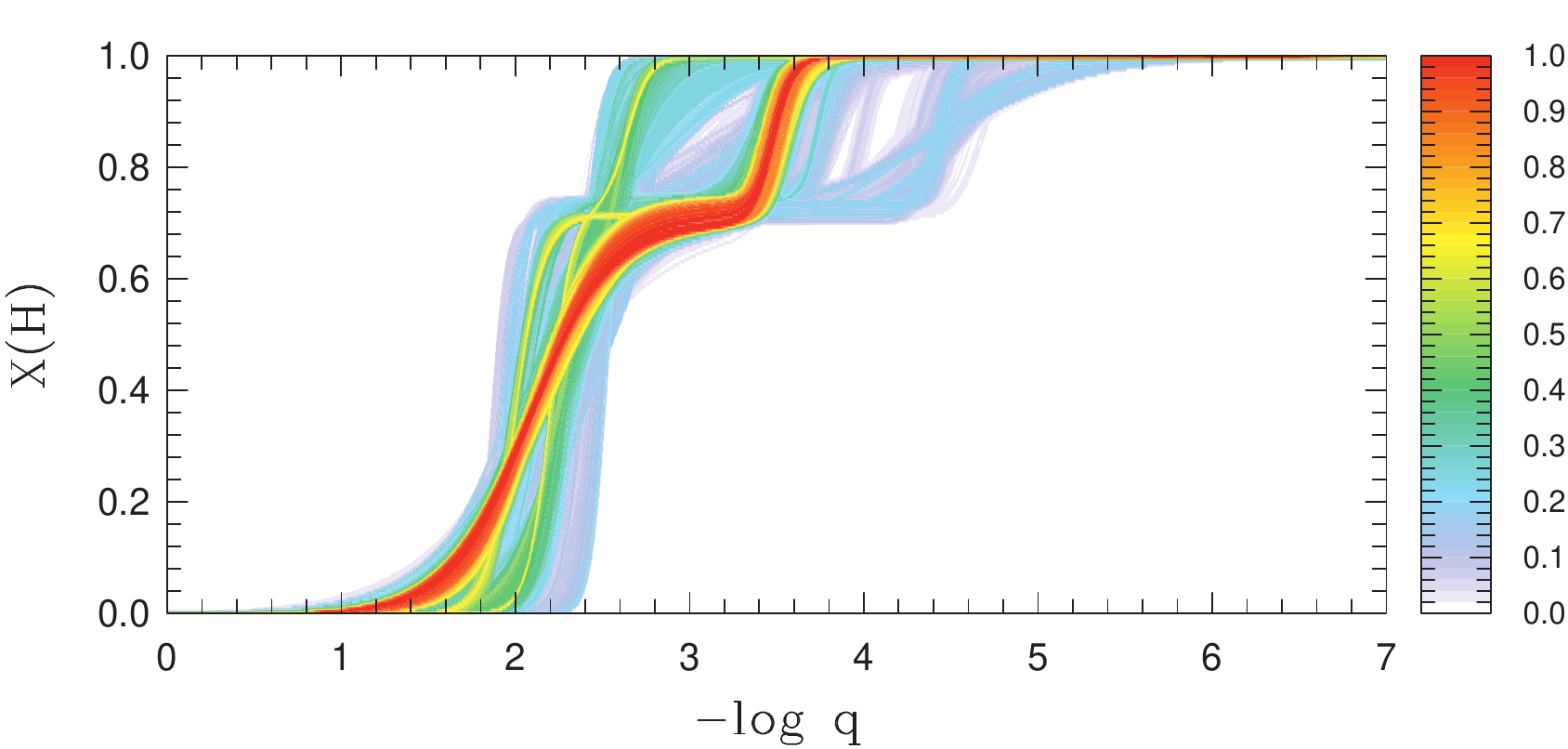}
\par\end{centering}
\begin{centering}
\includegraphics[scale=0.48]{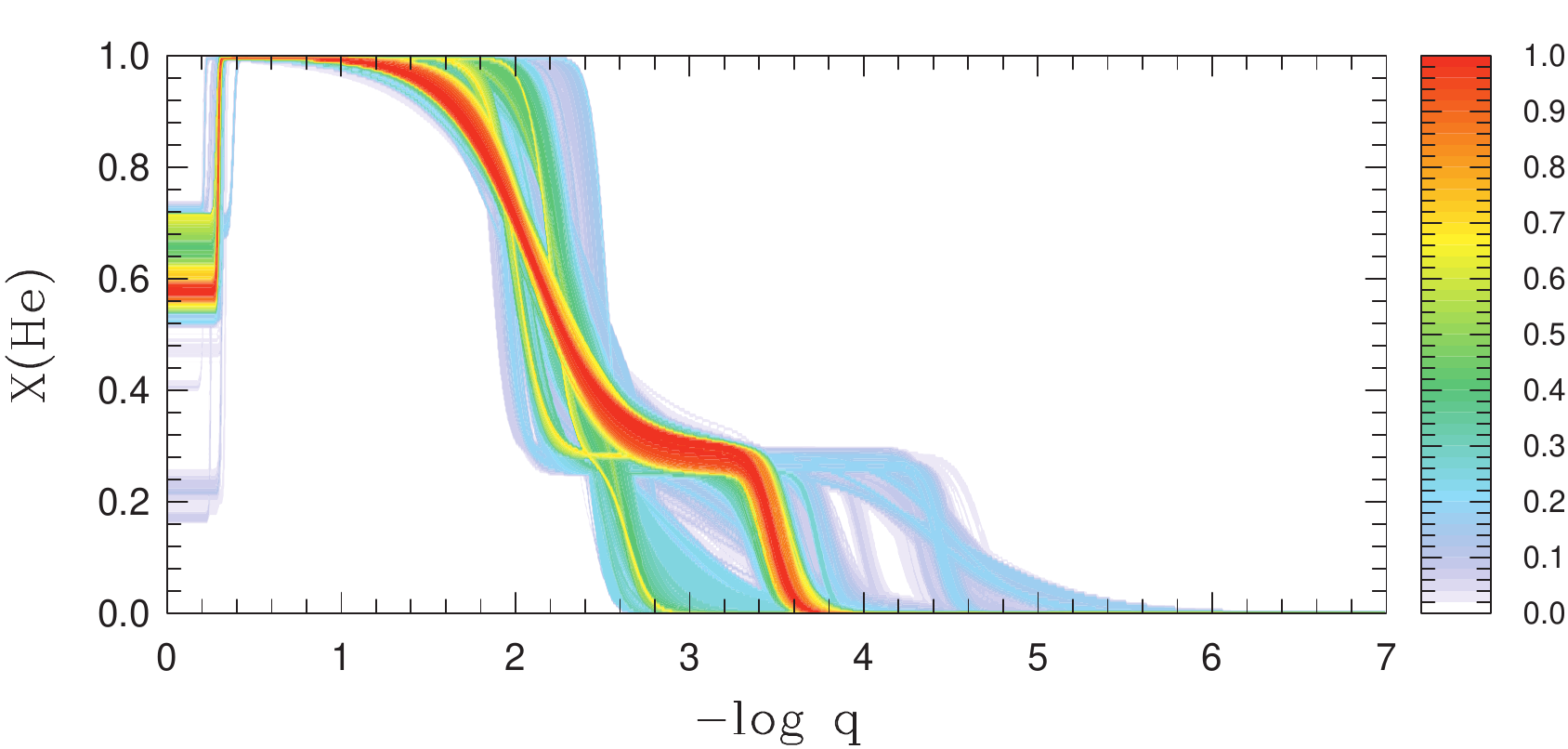}
\par\end{centering}
\begin{centering}
\includegraphics[scale=0.48]{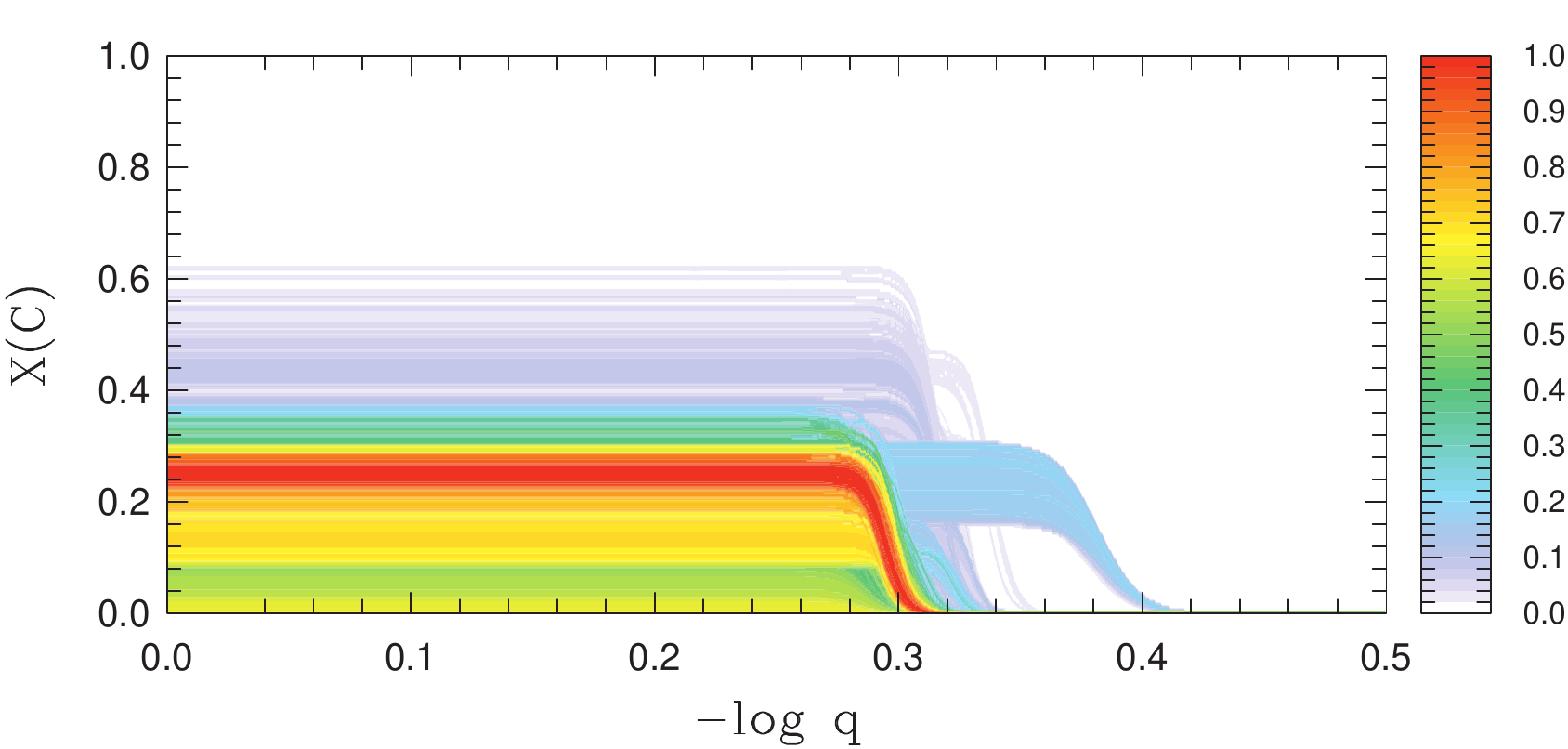}
\par\end{centering}
\begin{centering}
\includegraphics[scale=0.48]{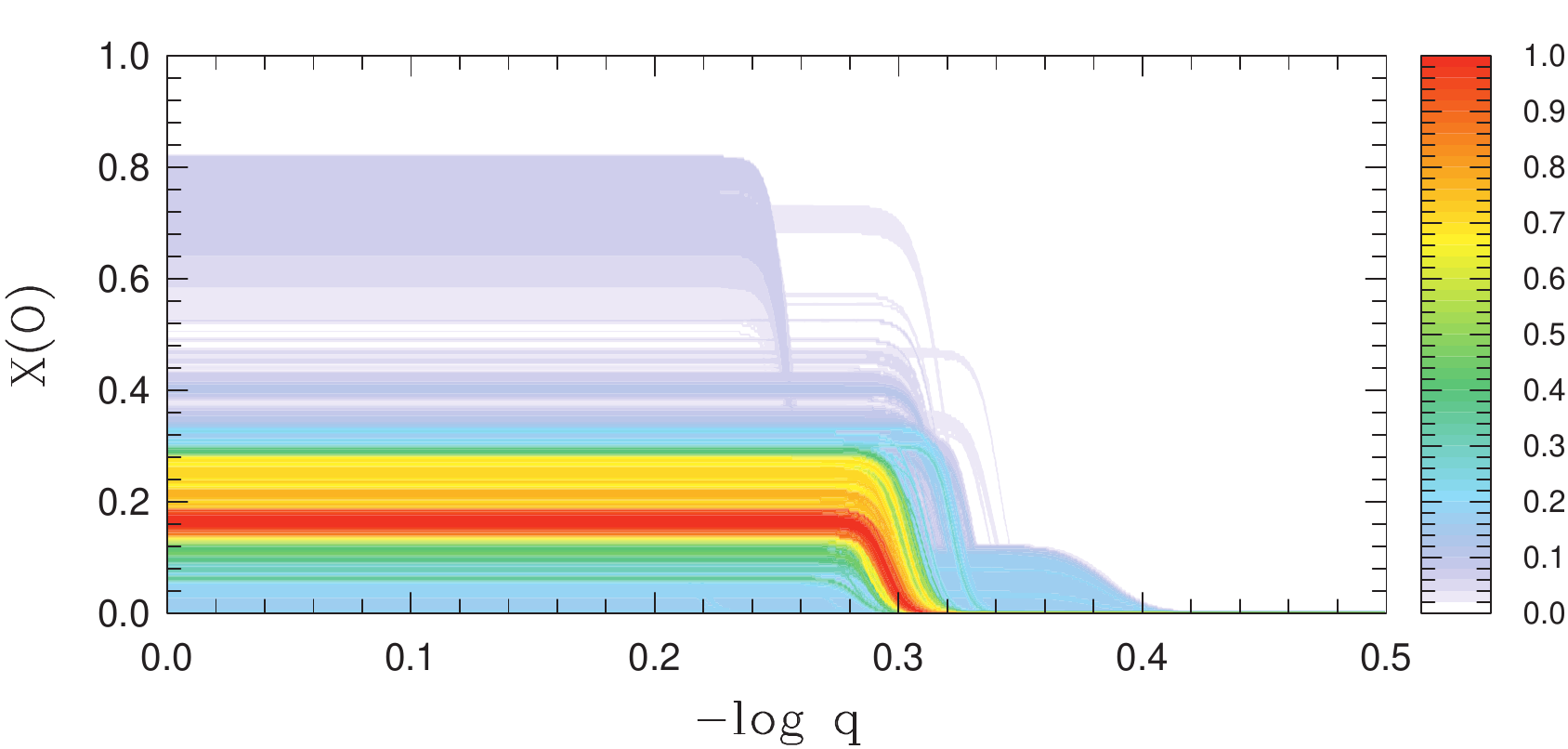}
\par\end{centering}
\caption{Probability distributions (color scale) normalized to one at maximum
as functions of the fractional mass depth, $\log q=\log[1-m(r)/M_{\star}]$,
obtained, from \emph{top} to \emph{bottom}, for the mass fraction
of hydrogen, helium, carbon, and oxygen inside TIC 278659026. The
red areas indicate the values, as functions of $\log q$, corresponding
to the best-matching seismic models. These distributions are derived
from the evaluation of 1,555,425 stellar models calculated during
exploration of parameter space. \label{fig:profiles}}
\end{figure}

As a last complementary check, we derive interstellar extinction from
the spectral energy distribution (SED) and colors by comparing all
available photometric measurements for TIC 278659026 to synthetic
measurements calculated from an appropriate model atmosphere (see Fig.~\ref{fig:sed}).
The photometry used in Fig.~\ref{fig:sed} (see Table \ref{tab:phot})
covers from the $u$-band to the infrared $J$, $H$, $K$ \citep[2MASS;][]{2006AJ....131.1163S}
and $W1$, $W2$ \citep[WISE;][]{2013yCat.2328....0C}. The Johnson
$V$ magnitude and colors for this specific test were taken from \citet{2013MNRAS.431..240O}.
The SkyMapper \citep{2018PASA...35...10W}, APASS $g$, $r$, $i$
\citep{2016yCat.2336....0H}, and \emph{Gaia} \citep{2018A&A...616A...1G}
magnitudes were also included. Details of the model spectra and the
fitting procedure are given by \citet{2018OAst...27...35H}. The angular
diameter obtained from this fit is $\theta=\left(3.42_{-0.06}^{+0.05}\right)\times10^{-11}$
rad, and interstellar reddening is found to be zero (i.e., $E$(B-V)
$\le0.009$; consistent with the value given in Table \ref{tab2}).
This approach also estimates a photometric effective temperature that
is well within $1\sigma$ agreement with other methods, giving $T_{{\rm eff}}=23600_{-400}^{+700}$
K. Using the high-precision parallax provided by \emph{Gaia} DR2,
$\varpi=4.910\pm0.051$, combined with the angular diameter and the
surface gravity, the stellar mass can be obtained through the relation
$M=g\,\Theta^{2}/(4G\,\varpi^{2}$). This results in a mass of $M=0.390\pm0.091$
$M_{\odot}$, based on the atmospheric parameters from \citet{2012MNRAS.427.2180N}
using a conservative estimate of 0.1 dex for the error on $\log g$,
instead of the statistical uncertainties quoted by the authors. This
value is remarkably consistent with the mass derived from the asteroseismic
solution. We close this discussion by noting that, interestingly,
the WISE W2 measurement rules out the presence of a companion earlier
than type M5 or M6 (assuming a $3\sigma$ excess at W2 above the expected
flux coming from the sdB photosphere).

\subsubsection{Potential implications in terms of evolution}

According to \citet{2002MNRAS.336..449H,2003MNRAS.341..669H}, such
a low-mass sdB star is most probably produced by the first stable
Roche lobe overflow (RLOF) channel, although there is presently no
indication of the presence of a stellar companion. The low mass inferred
suggests that TIC 278659026 must have started helium burning in nondegenerate
conditions, i.e., before reaching the critical mass that triggers
the helium flash. Stellar evolution calculations show that main sequence
stars with masses above $\sim2$ $M_{\odot}$ ignite helium before
electron degeneracy occurs, and consequently have less massive cores.
TIC 278659026 could, in this context, become the first evidence of
a rare sdB star that originates from a massive ($\gtrsim2$
$M_{\odot}$) red giant, an alternative formation channel investigated
by \citet{2008A&A...490..243H}. Finally, we note that the inferred
mass and the presence of pulsations at such high effective temperature
rule out the suggestion from \citet{2015MNRAS.450.3514K} that TIC
278659026 may be an ELM white dwarf progenitor.

\subsubsection{Internal structure}

Beyond the determination of the fundamental parameters characterizing
TIC 278659026, our asteroseismic analysis provides insight into the
internal structure of the star, in particular its chemical composition
and stratification. Interesting constraints on the double-layered
envelope structure are indeed obtained (Fig.~\ref{fig:hist-envl},
\ref{fig:hist-xh} and Table \ref{tab2}), along with measurements
of the helium-burning core properties (Fig.~\ref{fig:hist-core}
and Table \ref{tab2}). The various parameters defining these structures
fully determine the chemical composition distribution inside the best-fit
model for the four main constituents, H, He, C, and O, and Fig.~\ref{fig:profiles}
provides a more convenient way to visualize these profiles and their
uncertainties. This plot is constructed from the composition profiles
in the 1,555,425 models evaluated by the optimizer during its exploration
of the parameter space (see \citealp{2017A&A...598A.109G} who introduced
similar plots in a white dwarf context). Each model (and therefore
each composition profile) has a $S^{2}$ value attributed regarding
its ability to match the pulsation frequencies of TIC 278659026. Consequently,
probability distributions for the amount of H, He, O, and C (the complement
of He and O in the core) as functions of the fractional mass depth,
$\log q=\log(1-m(r)/M_{*})$, can be evaluated. We find from these
distributions that a well-defined region corresponding to best-fitting
models emerge for each element, thus materializing the chemical stratification
of TIC 278659026 as estimated from asteroseismology. Significantly
weaker secondary solutions can also be seen in these diagrams.

From the seismic solution obtained, we find that TIC 278659026 has
a thick envelope by sdB standards, with $\log q({\rm envl})=-2.11\pm0.11$,
where a double-layered structure for the He+H profile indicates that
gravitational settling has not yet completely segregated helium from
hydrogen\footnote{We stress that this result is not induced by construction because
the models also allow pure-hydrogen envelopes, even with the double-layer
parameterization. Single-layered envelopes are obtained when $\log q({\rm envl)}\le\log q({\rm H/diff})$.
Such configurations were within the search domain, but did not produce
optimal seismic solutions for TIC 278659026.}. Such a configuration is indeed expected for a cool sdB star such
as TIC 278659026, since envelope masses of EHB stars are strongly
correlated with effective temperature and element diffusion does not
have time in principle to affect the base of a thick envelope. The
mass fraction of hydrogen present in this mixed He+H region is not
really constrained, owing to a very weak sensitivy of the $g$-mode
pulsation frequencies to this parameter, although the optimal seismic
model gives $X({\rm H})_{{\rm envl}}\sim0.72$. We note that we imposed
this quantity to be close to the proportion expected in the solar
mixture (by exploring only a narrow range of values), assuming that
the sdB envelope is the remnant of the original main sequence stellar
envelope with unchanged composition. However, in light of Fig. \ref{fig:hist-xh},
we might argue that a global optimum may lie outside of the range
considered. Moreover, the evolution history of the progenitor of TIC
278659026 may not necessarily lead to a nearly solar H/He ratio in
the envelope. We have therefore, as a check, extended the analysis
to a wider range of $X({\rm H})_{{\rm envl}}$ values with additional
calculations described in Appendix A. The latter shows that  1) there
is no preferred value for $X({\rm H})_{{\rm envl}}$ in the $\sim0.58-0.86$
range that would lead to a significantly better fit; 2) models with
He-enriched envelopes above $Y\gtrsim0.42$ seem excluded; and 3)
our identified optimal solution remains entirely valid, although some
derived parameters may have slightly larger error estimates propagated
from the wider range considered for the envelope hydrogen content.
After this test, we therefore confidently keep the solution identified
from our more constrained calculations as the reference.

The inner core structure is also unveiled, pointing to a rather large
mixed core with a fractional mass $\log q({\rm core})=-0.295\pm0.013$,
corresponding to $M_{{\rm core}}=0.198\pm0.010$ $M_{\odot}$. This
size is reached while the estimated remaining helium mass fraction
in that region is $X({\rm He})_{{\rm core}}=0.575{}_{-0.027}^{+0.063}$.
TIC~278659026 is therefore close to the mid-stage of its helium-burning
phase, as it has transformed about 43\% of its helium into carbon
and oxygen. We point out that the value obtained for the core size
is comparable to some estimates already provided for this quantity
by \citet{2011A&A...530A...3C} for the star KIC 02697388 (solution
2) and by \citet{2010A&A...524A..63V} for the star KPD 0629-0016.
The latter in particular, although more massive as it has a canonical
mass of $0.471\pm0.002$ $M_{\odot}$, happens to be nearly at the
same stage of its core helium-burning phase, with $\sim59\%$ (in
mass) of helium left in the central region, while its estimated core
size is $\log q({\rm core})=-0.27\pm0.01$ ($M_{{\rm core}}=0.22\pm0.01$
$M_{\odot}$). Nonetheless, \citet{2010ApJ...718L..97V} also provided
a seismic estimate of the core mass for the sdB pulsator KPD 1943+4058,
but obtained a significantly larger measurement for this quantity. This
star may need to be reinvestigated with our most recent modeling
tools applied to the full data set now available; the \citealp{2010ApJ...718L..97V}
analysis was based on \emph{Kepler}'s Q0 light curve only. Remarkably,
and for the first time in a sdB star, our present seismic
modeling of TIC 278659026 also provides an estimate for the oxygen
mass fraction in the helium burning core. We find that $X({\rm O})_{{\rm core}}=0.16{}_{-0.05}^{+0.13}$
and consequently the estimated carbon mass fraction is $X({\rm C})_{{\rm core}}=0.27{}_{-0.14}^{+0.06}$.
These values, along with knowing the size of the core, are of high
interest as they are directly connected with the uncertain rate of
the $^{12}{\rm C}(\alpha,\gamma)^{16}{\rm O}$ nuclear reaction, which
is fundamental in many areas of astrophysics.

\subsubsection{Frequency match and mode identification}

\begin{figure}
\begin{centering}
\includegraphics[scale=0.45]{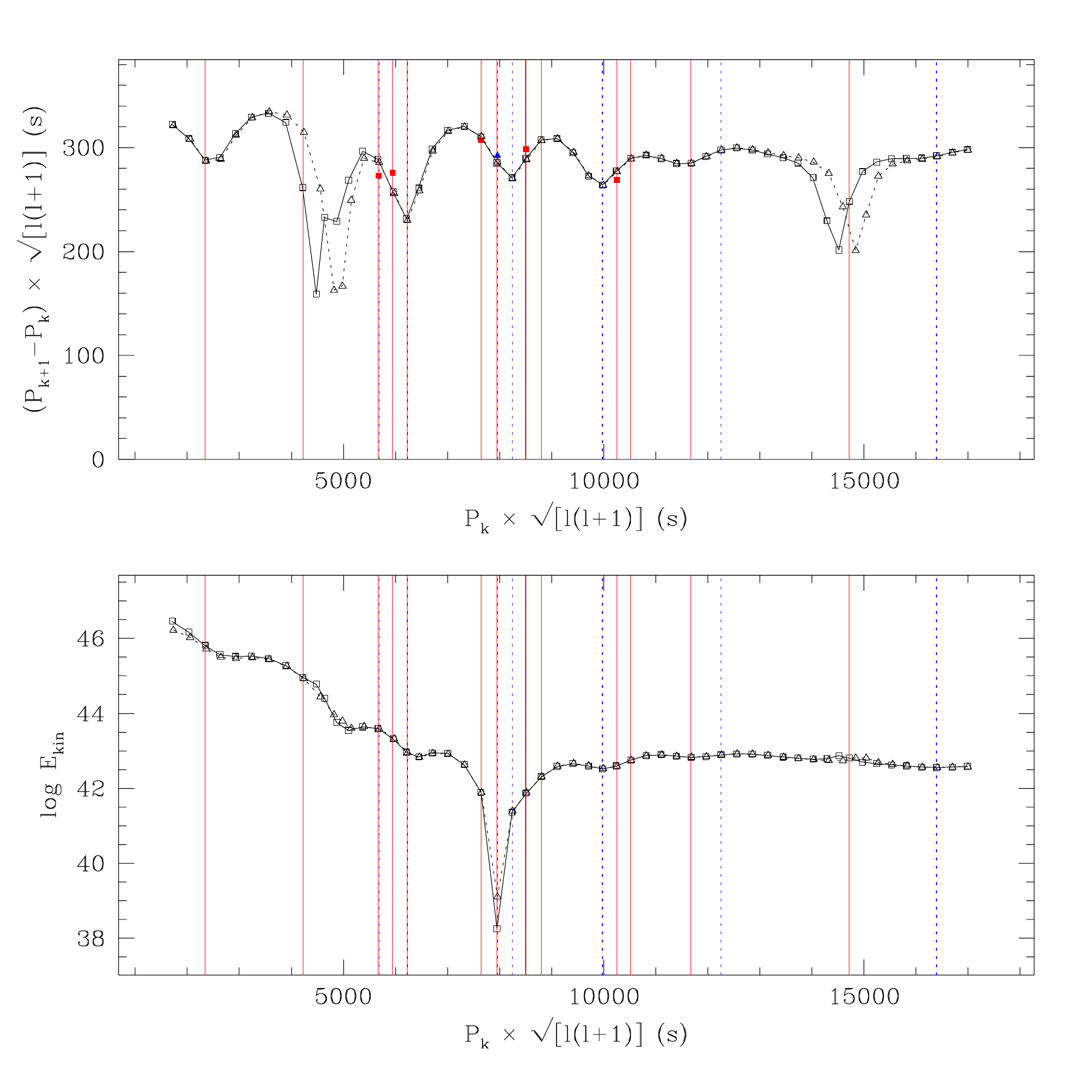}
\par\end{centering}
\caption{\emph{Top panel}: Period spectrum corresponding to the optimal seismic
model obtained for TIC 278659026 and represented in terms of the reduced
period spacing, $(P_{k+1}-P_{k}).\sqrt{\ell(\ell+1)}$ (in seconds)
as a function of reduced period, $P_{k}.\sqrt{\ell(\ell+1)}$ (in
seconds). Open squares connected by plain segments show the $\ell=1$
series of modes and open triangles connected by dotted segments are
$\ell=2$ modes. The range of radial orders covered spans $k=5$ to
$k=60$, as in Table 5. Plain and dotted vertical lines indicate the
reduced periods of the observed modes matched to $\ell=1$ and $\ell=2$
modes, respectively. When two observed modes of same degree have consecutive
radial orders, an observed reduced period spacing is also indicated
(filled squares and filled triangles for $\ell=1$ and $\ell=2$ modes,
respectively). \emph{Bottom panel}: Same as above, but for the logarithm
of the kinetic energy (or inertia), $\log E_{{\rm kin}}$, of the
modes as a function of reduced period. \label{fig:spectrum-fit} }
\end{figure}

The optimal seismic model uncovered for TIC 278659026 provides the
closest match to the observed frequencies obtained thus far for a
g-mode pulsating sdB star. With an average relative dispersion $|\overline{\Delta X/X}|=0.07\%$
($X=P$ or $\nu$), which corresponds on an absolute scale to $|\overline{\Delta P}|=3.16$
second and $|\overline{\Delta\nu}|=0.161$ $\mu$Hz for 20 frequencies
matched simultaneously, this fit outperforms previous comparable studies
presented by \citet{2010ApJ...718L..97V,2010A&A...524A..63V} and
\citet{2011A&A...530A...3C} by a factor $\sim3-5$ on the achieved
relative dispersion. This significant gain in matching precision is
clearly due to the improvements implemented in stellar models used
in the present analysis relative to these older studies, that is the
inclusion of a double-layered hydrogen-rich envelope (see Section
3.1). We point out that this ability to better reproduce the observed
oscillation spectrum is certainly the main factor that has allowed
us to extract more information about the internal structure of TIC
278659026. We also find that our best fit reaches an absolute precision
in frequency that is $\sim2.5$ times better than the formal resolution
of this \emph{TESS} observing run ($1/T=0.415$ $\mu$Hz). However,
since the actual uncertainty on each frequency measurement is on the
order of one-tenth of the formal resolution (Table \ref{tab1}), the
match is not perfect and significant room still remains to improve our seismic modeling further. A better description of the helium-burning
core boundary to produce profiles that comply more accurately to those
produced by, for example, semi-convection is among future improvements that
we plan to introduce.

All the details of the $g$-mode spectrum of the optimal model, frequency fit, and associated mode identification are given in
Table 5 and illustrated in Fig.~\ref{fig:spectrum-fit}. The table
provides, for each mode, the degree $\ell$ and radial order $k$,
the computed adiabatic frequency and period ($\nu_{{\rm th}}$ and
$P_{{\rm th}}$), their corresponding matched frequency and period
($\nu_{{\rm obs}}$ and $P_{{\rm obs}}$) when available, the logarithm
of the kinetic energy (or inertia) of the mode ($\log E_{{\rm kin}}$;
see, e.g., \citealp{2000ApJS..131..223C}), the first-order Ledoux
coefficient for slow rigid rotation ($C_{k\ell}$), and the relative
and absolute differences between the observed and computed values
$\Delta X/X$, $\Delta P$, and $\Delta\nu$, when available. We find
that the 20 independent frequencies of TIC 278659026 are well interpreted
as 13 $\ell=1$ and 7 $\ell=2$ $g$-modes of low-to-intermediate
radial orders ranging from $k=7-57$. The observed spectrum is clearly
incomplete, with many undetected eigenfrequencies that are present
in the model. Many of these unseen frequencies may be excited by the
driving engine, but to amplitude levels that are simply below the
detection threshold reached. They may also not be exited at all, but
linear nonadiabatic calculations indicate that all modes within the
instability range should in principle be driven \citep{2003ApJ...597..518F,2007MNRAS.378..379J,2014A&A...569A.123B}.
The mechanisms controlling mode amplitude saturation are not well
known. Moreover, amplitudes can change over long timescales in $g$-mode
sdB pulsators, probably because of nonlinear mode interactions, as documented
by \citet{2016A&A...594A..46Z,2018ApJ...853...98Z}. In this context,
bottom panel of Fig.~\ref{fig:spectrum-fit} suggests that modes
of lowest inertia (or kinetic energy), which are susceptible to be
excited to higher intrinsic amplitudes for a given amount of energy,
seem indeed to be those most seen in the spectrum of TIC 278659026.
Hence, the overall amplitude distribution (and detectability) of $g$-modes
in sdB pulsators probably has mode inertia as one of its controlling
factors, but other effects play an important role too.

Despite the relatively sparse distribution, several observed frequencies
are grouped into two or three modes of consecutive radial order. The top
panel of Fig.~\ref{fig:spectrum-fit} illustrates the pulsation spectrum
of the best-fit model and the matched observed modes in terms of their
reduced period spacing, $\Delta P_{k}=(P_{k+1}-P_{k}).\sqrt{\ell(\ell+1)}$,
plotted as a function of reduced period, $P_{k}.\sqrt{\ell(\ell+1)}$,
for the dipole and quadrupole series. This representation highlights
a series of oscillations and dips that occur around the theoretical
average reduced period spacing, $\Pi_{0}$, whose value is 278.4 s
for the optimal model\footnote{In the asymptotic limit, $\Pi_{0}=2\pi^{2}\left(\intop_{0}^{R}\frac{|N|}{r}dr\right)^{-1}$,
where $N$ is the Brunt-Väisälä frequency, $r$ is the distance from
the star's center, and $R$ is the radius of the star.}; these are typical of trapping effects generated by rapid changes
of the chemical composition, for example, at the core edge or at the mantle-envelope
transition \citep[see][and references therein]{2000ApJS..131..223C,2002ApJS..139..487C,2002ApJS..140..469C,2014ASPC..481..179C}.
We note that the period spacings derived from $\Pi_{0}$ are comparable
to those observed for the $g$-dominated mixed modes in core helium-burning
red giants \citep{2012ASPC..452..251O}, thus underlining that the
convectively mixed He-burning cores of EHB and red clump stars should
be similar. Moreover, we point out that trapping affecting the mixed
modes also occurs in red giant stars \citep{2011Natur.471..608B}.
As expected from asymptotic relations, all modes of the same radial
order but different degree $\ell$ almost always overlap in the reduced
period space and follow the same patterns. We note a slight distortion
to this rule around two dips in which the trapping occurs with a shift
of $\Delta k=1$ for $\ell=2$ modes compared to their $\ell=1$ counterpart.
This distortion appears related to the sharp transition at the core
boundary, which has the peculiarity to be located close to the inner
turning point of the gravity waves in their resonant cavity. Overall,
we find that the observed periods and their identification match the model spectrum properties well.
However, small discrepancies are
still noticeable, reflecting, as mentioned previously, the greatly
improved but still imperfect models used in this asteroseismic analysis.

\section{Summary and conclusions}

We have reported on the discovery of a new bright $g$-mode pulsating
sdB  star, TIC 278659026 (EC 21494-7078), by the NASA/\emph{TESS}
mission. This object is just one of the hundreds of evolved compact
stars (mostly hot subdwarfs and white dwarfs) monitored with the 120 s cadence
mode in the first \emph{TESS} sector. This effort is part of a larger
program driven by TASC Working Group 8 to survey thousands of bright
evolved compact stars for variability. The data consists of 27.88
days of photometry with a nearly continuous coverage, thus offering
a clear view of the pulsation spectrum of TIC 278659026. The analysis
of the time series revealed the presence of many frequencies in the
90\textendash 650 $\mu$Hz range, among which 20 are interpreted as
independent low-degree ($\ell=1$ and 2) gravity modes later used
to provide strong seismic constraints on the structure and fundamental
parameters of the star.

The asteroseismic investigation of TIC 278659026 was conducted with
our current modeling and optimization tools. These implement
a forward modeling approach developed over the past 15 years and
described in some detail in papers by \citet{2013A&A...553A..97V},
\citet{2016ApJS..223...10G}, and references therein. Compared to
former analyses of $g$-mode sdB pulsators, we used an improved version
of the Montréal 3G stellar models \citep{2010ApJ...718L..97V,2010A&A...524A..63V,2011A&A...530A...3C}
implementing a more complex envelope structure that allows for double-layered
hydrogen and helium profiles. Such profiles are characteristic of the
outcome of gravitational settling for the coolest hot subdwarfs, when
helium lacks time to completely sink below their relatively thick
envelopes during the lifetime of the star on the EHB. Former 3G models assumed pure hydrogen envelopes that were
not fully appropriate to analyze the coolest $g$-mode sdB pulsators.

The extensive search for a best-fit seismic model reproducing the
20 frequencies that characterize TIC 278659026 led to the identification
of a well-defined solution in close agreement with available constraints
on $T_{{\rm eff}}$ and $\log g$ derived from spectroscopy \citep{2012MNRAS.427.2180N}.
The solution also agrees remarkably well with the distance measured
independently from the \emph{Gaia} trigonometric parallax, through
a comparison with the seismic distance estimated from the model
properties (see Section 3.3.1 and Table \ref{tab2}), and with the
mass estimated from combining the fit of the SED
and the \emph{Gaia} parallax. All constitute important tests validating
the accuracy of the identified seismic model. Through this solution,
asteroseismology is giving us extensive information on fundamental
parameters and internal structure in TIC 278659026. In particular,
we find that this star has a mass of $0.391\pm0.009$ $M_{\odot}$,
which is significantly lower than the canonical mass of $0.47$ $M_{\odot}$
characterizing most sdBs, and could originate from a massive
($\gtrsim2$ $M_{\odot}$) red giant progenitor.

Other notable results include constraints on the chemical stratification
inside the star. We found, in particular, that TIC 278659026 has a
rather thick H-rich envelope, with $M(H)=0.0037\pm0.0010$ $M_{\odot}$
(computed from $\log q({\rm envl})=-2.11\pm0.11$; see Table \ref{tab2}),
whose structure includes a double-layered He and H distribution caused
by still ongoing gravitational settling of helium. The core, for its
part, is found to be more extended than typically predicted by standard
evolution models, reaching a mass of $0.198\pm0.010$ $M_{\odot}$
($\log q({\rm core})=-0.295\pm0.013$), while TIC 278659026 is at about
halfway in its evolution through the core helium-burning phase. The
mass fraction of helium remaining at the center of the star is estimated
to be $X({\rm He})_{{\rm core}}=0.575{}_{-0.027}^{+0.063}$. Therefore,
this star constitutes another case that suggests, along with three
other hot subdwarfs analyzed previously and recent evidence from white
dwarf asteroseismology, that helium-burning cores are larger than
predicted by current implementations of physical processes shaping
the core structure in stellar evolution calculations. The present
measurement thus provides a very useful quantitative constraint to
explore this issue. We also found that the oxygen mass fraction produced
in the core (and consequently the amount of carbon present) is constrained
for the first time in a sdB star (and by extension in a core helium
burning star) to an  interesting level of precision, $X({\rm O})_{{\rm core}}=0.16{}_{-0.05}^{+0.13}$
(and for carbon, $X({\rm C})_{{\rm core}}=0.27{}_{-0.14}^{+0.06}$).
These values, along with the determination of the core size mentioned
previously, are directly connected to the $^{12}{\rm C}(\alpha,\gamma)^{16}{\rm O}$
nuclear reaction, which is important in many areas of astrophysics
but whose rate is still highly uncertain. TIC 278659026, through the
present seismic analysis, may thus become an important test to improve
our knowledge on this issue as well.

We conclude this paper by emphasizing that this analysis of the hot
sdB star TIC 278659026 discovered to be pulsating by \emph{TESS},
and based on only one sector, is a demonstration that this instrument,
even if not technically optimized for asteroseismology of hot and
faint evolved compact stars, is providing outstanding data that can
significantly drive this field forward. The greatest advantage
of \emph{TESS} over previous similar projects (\emph{K2}, \emph{Kepler}, \emph{CoRoT},
and \emph{MOST}) is that it will survey many more hot subdwarf stars
during the two years of its main mission. Among these, many will be
pulsating sdB stars with a comparable seismic potential and this will
allow us to probe the interior and core structure of these stars over
wider ranges of masses and ages on the EHB. The
contribution of \emph{TESS}  to this research area should therefore prove
very significant.
\begin{acknowledgements}
We thank S. Jeffery for useful comments and suggestions that helped
improve this manuscript. Stéphane Charpinet acknowledges financial
support from the Centre National d'Études Spatiales (CNES, France)
and from the Agence Nationale de la Recherche (ANR, France) under
grant ANR-17-CE31-0018, funding the INSIDE project. This work was
granted access to the high-performance computing resources of the
CALMIP computing center under allocation numbers 2018-p0205 and 2019-p0205.
This paper includes data collected by the \emph{TESS} mission. Funding
for the \emph{TESS} mission is provided by the NASA Explorer Program.
Funding for the \emph{TESS} Asteroseismic Science Operations Centre
is provided by the Danish National Research Foundation (Grant agreement
no.: DNRF106), ESA PRODEX (PEA 4000119301) and Stellar Astrophysics
Centre (SAC) at Aarhus University. We thank the TESS team and staff
and TASC/TASOC for their support of the present work. G.F. acknowledges
the contribution of the Canada Research Chair Program. W.Z. acknowledges
the support from the National Natural Science Foundation of China
(NSFC) through the grant 11833002, the support from the China Postdoctoral
Science Foundation through the grant 2018M641244 and the LAMOST fellowship
as a Youth Researcher which is supported by the Special Funding for
Advanced Users, budgeted and administrated by the Center for Astronomical
Mega-Science, Chinese Academy of Sciences (CAMS). V.V.G. is an F.R.S.-FNRS
Research Associate. ZsB and \'AS acknowledge the financial support
of the GINOP-2.3.2-15-2016-00003, K-115709, K-113117, K-119517 and
PD-123910 grants of the Hungarian National Research, Development and
Innovation Office (NKFIH), and the Lend\"ulet Program of the Hungarian
Academy of Sciences, project No. LP2018-7/2018. DK acknowledges financial
support form the University of the Western Cape. I. P. acknowledges
funding by the Deutsche Forschungsgemeinschaft under grant GE2506/12-1.
AP and PK-S acknowledge support from the NCN grant no. 2016/21/B/ST9/01126.
ASB gratefully acknowledges financial support from the Polish National
Science Center under projects No. UMO-2017/26/E/ST9/00703 and UMO-2017/25/B
ST9/02218. SJM is a DECRA fellow supported by the Australian Research
Council (grant number DE180101104). KJB is supported by an NSF Astronomy
and Astrophysics Postdoctoral Fellowship under award AST-1903828.
We thank Andreas Irrgang and Simon Kreuzer for developing the SED
fitting tool. This work has made use of data from the European Space
Agency (ESA) mission \emph{Gaia} (\url{https://www.cosmos.esa.int/gaia}),
processed by the \emph{Gaia} Data Processing and Analysis Consortium
(DPAC, \url{https://www.cosmos.esa.int/web/gaia/dpac/consortium}).
Funding for the DPAC has been provided by national institutions, in
particular the institutions participating in the \emph{Gaia} Multilateral
Agreement. The national facility capability for SkyMapper has been
funded through ARC LIEF grant LE130100104 from the Australian Research
Council, awarded to the University of Sydney, the Australian National
University, Swinburne University of Technology, the University of
Queensland, the University of Western Australia, the University of
Melbourne, Curtin University of Technology, Monash University, and
the Australian Astronomical Observatory. SkyMapper is owned and operated
by The Australian National University's Research School of Astronomy
and Astrophysics. The survey data were processed and provided by the
SkyMapper Team at ANU. The SkyMapper node of the All-Sky Virtual Observatory
(ASVO) is hosted at the National Computational Infrastructure (NCI).
Development and support the SkyMapper node of the ASVO has been funded
in part by Astronomy Australia Limited (AAL) and the Australian Government
through the Commonwealth's Education Investment Fund (EIF) and National
Collaborative Research Infrastructure Strategy (NCRIS), particularly
the National eResearch Collaboration Tools and Resources (NeCTAR)
and the Australian National Data Service Projects (ANDS). This publication
makes use of data products from the Wide-field Infrared Survey Explorer,
which is a joint project of the University of California, Los Angeles,
and the Jet Propulsion Laboratory/California Institute of Technology,
funded by the National Aeronautics and Space Administration. This
publication makes use of data products from the Two Micron All Sky
Survey, which is a joint project of the University of Massachusetts
and the Infrared Processing and Analysis Center/California Institute
of Technology, funded by the National Aeronautics and Space Administration
and the National Science Foundation.
\end{acknowledgements}

\bibliographystyle{aa}
\bibliography{/home/stephane/Astro/Bibliotheque/Bibtex/charpinet-201901,/home/stephane/Astro/Bibliotheque/Bibtex/deeming,/home/stephane/Astro/Bibliotheque/Bibtex/hu-201312,/home/stephane/Astro/Bibliotheque/Bibtex/brassard-201312,/home/stephane/Astro/Bibliotheque/Bibtex/han-201309,/home/stephane/Astro/Bibliotheque/Bibtex/nemeth,/home/stephane/Astro/Bibliotheque/Bibtex/kilkenny-201309,/home/stephane/Astro/Bibliotheque/Bibtex/green-201309,/home/stephane/Astro/Bibliotheque/Bibtex/heber-reviews,/home/stephane/Astro/Bibliotheque/Bibtex/saffer,/home/stephane/Astro/Bibliotheque/Bibtex/dorman,/home/stephane/Astro/Bibliotheque/Bibtex/schuh,/home/stephane/Astro/Bibliotheque/Bibtex/baglin,/home/stephane/Astro/Bibliotheque/Bibtex/baran-201703,/home/stephane/Astro/Bibliotheque/Bibtex/ghasemi,/home/stephane/Astro/Bibliotheque/Bibtex/tic278659026,/home/stephane/Astro/Bibliotheque/Bibtex/gaia,/home/stephane/Astro/Bibliotheque/Bibtex/tess,/home/stephane/Astro/Bibliotheque/Bibtex/kepler,/home/stephane/Astro/Bibliotheque/Bibtex/most,/home/stephane/Astro/Bibliotheque/Bibtex/jeffery-201312,/home/stephane/Astro/Bibliotheque/Bibtex/nomenclature}

\begin{thebibliography}{90}
\expandafter\ifx\csname natexlab\endcsname\relax\def\natexlab#1{#1}\fi

\bibitem[{{Baglin} {et~al.}(2006){Baglin}, {Auvergne}, {Boisnard}, {Lam-Trong},
  {Barge}, {Catala}, {Deleuil}, {Michel}, \& {Weiss}}]{2006cosp...36.3749B}
{Baglin}, A., {Auvergne}, M., {Boisnard}, L., {et~al.} 2006, in COSPAR, Plenary
  Meeting, Vol.~36, 36th COSPAR Scientific Assembly, 3749--+

\bibitem[{{Baran} {et~al.}(2009){Baran}, {Oreiro}, {Pigulski}, {P{\'e}rez
  Hern{\'a}ndez}, {Ulla}, {Reed}, {Rodr{\'{\i}}guez-L{\'o}pez}, {Moskalik},
  {Kim}, {Chen}, {Crowe}, {Siwak}, {Armendarez}, {Binder}, {Choo}, {Dye},
  {Eggen}, {Garrido}, {Gonz{\'a}lez P{\'e}rez}, {Harms}, {Huang}, {Kozie{\l}},
  {Lee}, {MacDonald}, {Fox Machado}, {Monserrat}, {Stevick}, {Stewart},
  {Terry}, {Zhou}, \& {Zo{\l}a}}]{2009MNRAS.392.1092B}
{Baran}, A., {Oreiro}, R., {Pigulski}, A., {et~al.} 2009, \mnras, 392, 1092

\bibitem[{{Baran} {et~al.}(2017){Baran}, {Reed}, {{\O}stensen}, {Telting}, \&
  {Jeffery}}]{2017A&A...597A..95B}
{Baran}, A.~S., {Reed}, M.~D., {{\O}stensen}, R.~H., {Telting}, J.~H., \&
  {Jeffery}, C.~S. 2017, \aap, 597, A95

\bibitem[{{Bedding} {et~al.}(2011){Bedding}, {Mosser}, {Huber},
  {Montalb{\'a}n}, {Beck}, {Christensen-Dalsgaard}, {Elsworth},
  {Garc{\'{\i}}a}, {Miglio}, {Stello}, {White}, {De Ridder}, {Hekker}, {Aerts},
  {Barban}, {Belkacem}, {Broomhall}, {Brown}, {Buzasi}, {Carrier}, {Chaplin},
  {di Mauro}, {Dupret}, {Frandsen}, {Gilliland}, {Goupil}, {Jenkins},
  {Kallinger}, {Kawaler}, {Kjeldsen}, {Mathur}, {Noels}, {Silva Aguirre}, \&
  {Ventura}}]{2011Natur.471..608B}
{Bedding}, T.~R., {Mosser}, B., {Huber}, D., {et~al.} 2011, \nat, 471, 608

\bibitem[{{Bloemen} {et~al.}(2014){Bloemen}, {Hu}, {Aerts}, {Dupret},
  {{\O}stensen}, {Degroote}, {M{\"u}ller-Ringat}, \&
  {Rauch}}]{2014A&A...569A.123B}
{Bloemen}, S., {Hu}, H., {Aerts}, C., {et~al.} 2014, \aap, 569, A123

\bibitem[{{Borucki} {et~al.}(2010){Borucki}, {Koch}, {Basri}, {Batalha},
  {Brown}, {Caldwell}, {Caldwell}, {Christensen-Dalsgaard}, {Cochran},
  {DeVore}, {Dunham}, {Dupree}, {Gautier}, {Geary}, {Gilliland}, {Gould},
  {Howell}, {Jenkins}, {Kondo}, {Latham}, {Marcy}, {Meibom}, {Kjeldsen},
  {Lissauer}, {Monet}, {Morrison}, {Sasselov}, {Tarter}, {Boss}, {Brownlee},
  {Owen}, {Buzasi}, {Charbonneau}, {Doyle}, {Fortney}, {Ford}, {Holman},
  {Seager}, {Steffen}, {Welsh}, {Rowe}, {Anderson}, {Buchhave}, {Ciardi},
  {Walkowicz}, {Sherry}, {Horch}, {Isaacson}, {Everett}, {Fischer}, {Torres},
  {Johnson}, {Endl}, {MacQueen}, {Bryson}, {Dotson}, {Haas}, {Kolodziejczak},
  {Van Cleve}, {Chandrasekaran}, {Twicken}, {Quintana}, {Clarke}, {Allen},
  {Li}, {Wu}, {Tenenbaum}, {Verner}, {Bruhweiler}, {Barnes}, \&
  {Prsa}}]{2010Sci...327..977B}
{Borucki}, W.~J., {Koch}, D., {Basri}, G., {et~al.} 2010, Science, 327, 977

\bibitem[{{Brassard} \& {Charpinet}(2008)}]{2008Ap&SS.316..107B}
{Brassard}, P. \& {Charpinet}, S. 2008, \apss, 316, 107

\bibitem[{{Brassard} \& {Fontaine}(2008)}]{2008ASPC..392..261B}
{Brassard}, P. \& {Fontaine}, G. 2008, in ASPCS, Vol. 392, Hot Subdwarf Stars
  and Related Objects, ed. U.~{Heber}, C.~S. {Jeffery}, \& R.~{Napiwotzki}, 261

\bibitem[{{Brassard} \& {Fontaine}(2009)}]{2009JPhCS.172a2016B}
{Brassard}, P. \& {Fontaine}, G. 2009, Journal of Physics Conference Series,
  172, 012016

\bibitem[{{Brassard} {et~al.}(2001){Brassard}, {Fontaine}, {Bill{\`e}res},
  {Charpinet}, {Liebert}, \& {Saffer}}]{2001ApJ...563.1013B}
{Brassard}, P., {Fontaine}, G., {Bill{\`e}res}, M., {et~al.} 2001, \apj, 563,
  1013

\bibitem[{{Brassard} {et~al.}(1992){Brassard}, {Pelletier}, {Fontaine}, \&
  {Wesemael}}]{1992ApJS...80..725B}
{Brassard}, P., {Pelletier}, C., {Fontaine}, G., \& {Wesemael}, F. 1992, \apjs,
  80, 725

\bibitem[{{Charpinet} {et~al.}(2009){Charpinet}, {Brassard}, {Fontaine},
  {Green}, {Van Grootel}, {Randall}, \& {Chayer}}]{2009AIPC.1170..585C}
{Charpinet}, S., {Brassard}, P., {Fontaine}, G., {et~al.} 2009, in American
  Institute of Physics Conference Series, Vol. 1170, American Institute of
  Physics Conference Series, ed. J.~A. {Guzik} \& P.~A. {Bradley}, 585--596

\bibitem[{{Charpinet} {et~al.}(2014{\natexlab{a}}){Charpinet}, {Brassard}, {Van
  Grootel}, \& {Fontaine}}]{2014ASPC..481..179C}
{Charpinet}, S., {Brassard}, P., {Van Grootel}, V., \& {Fontaine}, G.
  2014{\natexlab{a}}, in Astronomical Society of the Pacific Conference Series,
  Vol. 481, 6th Meeting on Hot Subdwarf Stars and Related Objects, ed. V.~{van
  Grootel}, E.~{Green}, G.~{Fontaine}, \& S.~{Charpinet}, 179

\bibitem[{{Charpinet} {et~al.}(2001){Charpinet}, {Fontaine}, \&
  {Brassard}}]{2001PASP..113..775C}
{Charpinet}, S., {Fontaine}, G., \& {Brassard}, P. 2001, \pasp, 113, 775

\bibitem[{{Charpinet} {et~al.}(1997){Charpinet}, {Fontaine}, {Brassard},
  {Chayer}, {Rogers}, {Iglesias}, \& {Dorman}}]{1997ApJ...483L.123C}
{Charpinet}, S., {Fontaine}, G., {Brassard}, P., {et~al.} 1997, \apjl, 483,
  L123

\bibitem[{{Charpinet} {et~al.}(1996){Charpinet}, {Fontaine}, {Brassard}, \&
  {Dorman}}]{1996ApJ...471L.103C}
{Charpinet}, S., {Fontaine}, G., {Brassard}, P., \& {Dorman}, B. 1996, \apjl,
  471, L103

\bibitem[{{Charpinet} {et~al.}(2000){Charpinet}, {Fontaine}, {Brassard}, \&
  {Dorman}}]{2000ApJS..131..223C}
{Charpinet}, S., {Fontaine}, G., {Brassard}, P., \& {Dorman}, B. 2000, \apjs,
  131, 223

\bibitem[{{Charpinet} {et~al.}(2002{\natexlab{a}}){Charpinet}, {Fontaine},
  {Brassard}, \& {Dorman}}]{2002ApJS..139..487C}
{Charpinet}, S., {Fontaine}, G., {Brassard}, P., \& {Dorman}, B.
  2002{\natexlab{a}}, \apjs, 139, 487

\bibitem[{{Charpinet} {et~al.}(2002{\natexlab{b}}){Charpinet}, {Fontaine},
  {Brassard}, \& {Dorman}}]{2002ApJS..140..469C}
{Charpinet}, S., {Fontaine}, G., {Brassard}, P., \& {Dorman}, B.
  2002{\natexlab{b}}, \apjs, 140, 469

\bibitem[{{Charpinet} {et~al.}(2005){Charpinet}, {Fontaine}, {Brassard},
  {Green}, \& {Chayer}}]{2005A&A...437..575C}
{Charpinet}, S., {Fontaine}, G., {Brassard}, P., {Green}, E.~M., \& {Chayer},
  P. 2005, \aap, 437, 575

\bibitem[{{Charpinet} {et~al.}(2015){Charpinet}, {Giammichele}, {Brassard},
  {Van Grootel}, \& {Fontaine}}]{2015ASPC..493..151C}
{Charpinet}, S., {Giammichele}, N., {Brassard}, P., {Van Grootel}, V., \&
  {Fontaine}, G. 2015, in Astronomical Society of the Pacific Conference
  Series, Vol. 493, 19th European Workshop on White Dwarfs, ed. P.~{Dufour},
  P.~{Bergeron}, \& G.~{Fontaine}, 151

\bibitem[{{Charpinet} {et~al.}(2018){Charpinet}, {Giammichele}, {Zong}, {Van
  Grootel}, {Brassard}, \& {Fontaine}}]{2018OAst...27..112C}
{Charpinet}, S., {Giammichele}, N., {Zong}, W., {et~al.} 2018, Open Astronomy,
  27, 112

\bibitem[{{Charpinet} {et~al.}(2010){Charpinet}, {Green}, {Baglin}, {Van
  Grootel}, {Fontaine}, {Vauclair}, {Chaintreuil}, {Weiss}, {Michel},
  {Auvergne}, {Catala}, {Samadi}, \& {Baudin}}]{2010A&A...516L...6C}
{Charpinet}, S., {Green}, E.~M., {Baglin}, A., {et~al.} 2010, \aap, 516, L6

\bibitem[{{Charpinet} {et~al.}(2014{\natexlab{b}}){Charpinet}, {Van Grootel},
  {Brassard}, \& {Fontaine}}]{2014IAUS..301..397C}
{Charpinet}, S., {Van Grootel}, V., {Brassard}, P., \& {Fontaine}, G.
  2014{\natexlab{b}}, in IAU Symposium, Vol. 301, Precision Asteroseismology,
  ed. J.~A. {Guzik}, W.~J. {Chaplin}, G.~{Handler}, \& A.~{Pigulski}, 397--398

\bibitem[{{Charpinet} {et~al.}(2014{\natexlab{c}}){Charpinet}, {Van Grootel},
  {Brassard}, {Fontaine}, {Green}, \& {Randall}}]{2014ASPC..481..105C}
{Charpinet}, S., {Van Grootel}, V., {Brassard}, P., {et~al.}
  2014{\natexlab{c}}, in Astronomical Society of the Pacific Conference Series,
  Vol. 481, 6th Meeting on Hot Subdwarf Stars and Related Objects, ed. V.~{van
  Grootel}, E.~{Green}, G.~{Fontaine}, \& S.~{Charpinet}, 105

\bibitem[{{Charpinet} {et~al.}(2011){Charpinet}, {Van Grootel}, {Fontaine},
  {Green}, {Brassard}, {Randall}, {Silvotti}, {{\O}stensen}, {Kjeldsen},
  {Christensen-Dalsgaard}, {Kawaler}, {Clarke}, {Li}, \&
  {Wohler}}]{2011A&A...530A...3C}
{Charpinet}, S., {Van Grootel}, V., {Fontaine}, G., {et~al.} 2011, \aap, 530,
  A3

\bibitem[{{Charpinet} {et~al.}(2008){Charpinet}, {Van Grootel}, {Reese},
  {Fontaine}, {Green}, {Brassard}, \& {Chayer}}]{2008A&A...489..377C}
{Charpinet}, S., {Van Grootel}, V., {Reese}, D., {et~al.} 2008, \aap, 489, 377

\bibitem[{{Copperwheat} {et~al.}(2011){Copperwheat}, {Morales-Rueda}, {Marsh},
  {Maxted}, \& {Heber}}]{2011MNRAS.415.1381C}
{Copperwheat}, C.~M., {Morales-Rueda}, L., {Marsh}, T.~R., {Maxted}, P.~F.~L.,
  \& {Heber}, U. 2011, \mnras, 415, 1381

\bibitem[{{Cutri} \& {et al.}(2013)}]{2013yCat.2328....0C}
{Cutri}, R.~M. \& {et al.} 2013, VizieR Online Data Catalog, 2328

\bibitem[{{Deeming}(1976)}]{1976Ap&SS..42..257D}
{Deeming}, T.~J. 1976, \apss, 42, 257

\bibitem[{{Dorman} {et~al.}(1993){Dorman}, {Rood}, \&
  {O'Connell}}]{1993ApJ...419..596D}
{Dorman}, B., {Rood}, R.~T., \& {O'Connell}, R.~W. 1993, \apj, 419, 596

\bibitem[{{Fontaine} {et~al.}(2019){Fontaine}, {Bergeron}, {Brassard},
  {Charpinet}, {Randall}, {Van Grootel}, {Latour}, \&
  {Green}}]{2019ApJ...880...79F}
{Fontaine}, G., {Bergeron}, P., {Brassard}, P., {et~al.} 2019, \apj, 880, 79

\bibitem[{{Fontaine} {et~al.}(2003){Fontaine}, {Brassard}, {Charpinet},
  {Green}, {Chayer}, {Bill{\`e}res}, \& {Randall}}]{2003ApJ...597..518F}
{Fontaine}, G., {Brassard}, P., {Charpinet}, S., {et~al.} 2003, \apj, 597, 518

\bibitem[{{Fontaine} {et~al.}(2012){Fontaine}, {Brassard}, {Charpinet},
  {Green}, {Randall}, \& {Van Grootel}}]{2012A&A...539A..12F}
{Fontaine}, G., {Brassard}, P., {Charpinet}, S., {et~al.} 2012, \aap, 539, A12

\bibitem[{{Gaia Collaboration} {et~al.}(2018){Gaia Collaboration}, {Brown},
  {Vallenari}, {Prusti}, {de Bruijne}, {Babusiaux}, {Bailer-Jones}, {Biermann},
  {Evans}, {Eyer}, {Jansen}, {Jordi}, {Klioner}, {Lammers}, {Lindegren},
  {Luri}, {Mignard}, {Panem}, {Pourbaix}, {Randich}, {Sartoretti}, {Siddiqui},
  {Soubiran}, {van Leeuwen}, {Walton}, {Arenou}, {Bastian}, {Cropper},
  {Drimmel}, {Katz}, {Lattanzi}, {Bakker}, {Cacciari}, {Casta{\~n}eda},
  {Chaoul}, {Cheek}, {De Angeli}, {Fabricius}, {Guerra}, {Holl}, {Masana},
  {Messineo}, {Mowlavi}, {Nienartowicz}, {Panuzzo}, {Portell}, {Riello},
  {Seabroke}, {Tanga}, {Th{\'e}venin}, {Gracia-Abril}, {Comoretto},
  {Garcia-Reinaldos}, {Teyssier}, {Altmann}, {Andrae}, {Audard},
  {Bellas-Velidis}, {Benson}, {Berthier}, {Blomme}, {Burgess}, {Busso},
  {Carry}, {Cellino}, {Clementini}, {Clotet}, {Creevey}, {Davidson}, {De
  Ridder}, {Delchambre}, {Dell'Oro}, {Ducourant},
  {Fern{\'a}ndez-Hern{\'a}ndez}, {Fouesneau}, {Fr{\'e}mat}, {Galluccio},
  {Garc{\'\i}a-Torres}, {Gonz{\'a}lez-N{\'u}{\~n}ez}, {Gonz{\'a}lez-Vidal},
  {Gosset}, {Guy}, {Halbwachs}, {Hambly}, {Harrison}, {Hern{\'a}ndez},
  {Hestroffer}, {Hodgkin}, {Hutton}, {Jasniewicz}, {Jean-Antoine-Piccolo},
  {Jordan}, {Korn}, {Krone-Martins}, {Lanzafame}, {Lebzelter}, {L{\"o}ffler},
  {Manteiga}, {Marrese}, {Mart{\'\i}n-Fleitas}, {Moitinho}, {Mora}, {Muinonen},
  {Osinde}, {Pancino}, {Pauwels}, {Petit}, {Recio-Blanco}, {Richards},
  {Rimoldini}, {Robin}, {Sarro}, {Siopis}, {Smith}, {Sozzetti}, {S{\"u}veges},
  {Torra}, {van Reeven}, {Abbas}, {Abreu Aramburu}, {Accart}, {Aerts},
  {Altavilla}, {{\'A}lvarez}, {Alvarez}, {Alves}, {Anderson}, {Andrei},
  {Anglada Varela}, {Antiche}, {Antoja}, {Arcay}, {Astraatmadja}, {Bach},
  {Baker}, {Balaguer-N{\'u}{\~n}ez}, {Balm}, {Barache}, {Barata}, {Barbato},
  {Barblan}, {Barklem}, {Barrado}, {Barros}, {Barstow}, {Bartholom{\'e}
  Mu{\~n}oz}, {Bassilana}, {Becciani}, {Bellazzini}, {Berihuete}, {Bertone},
  {Bianchi}, {Bienaym{\'e}}, {Blanco-Cuaresma}, {Boch}, {Boeche}, {Bombrun},
  {Borrachero}, {Bossini}, {Bouquillon}, {Bourda}, {Bragaglia}, {Bramante},
  {Breddels}, {Bressan}, {Brouillet}, {Br{\"u}semeister}, {Brugaletta},
  {Bucciarelli}, {Burlacu}, {Busonero}, {Butkevich}, {Buzzi}, {Caffau},
  {Cancelliere}, {Cannizzaro}, {Cantat-Gaudin}, {Carballo}, {Carlucci},
  {Carrasco}, {Casamiquela}, {Castellani}, {Castro-Ginard}, {Charlot},
  {Chemin}, {Chiavassa}, {Cocozza}, {Costigan}, {Cowell}, {Crifo}, {Crosta},
  {Crowley}, {Cuypers}, {Dafonte}, {Damerdji}, {Dapergolas}, {David}, {David},
  {de Laverny}, {De Luise}, {De March}, {de Martino}, {de Souza}, {de Torres},
  {Debosscher}, {del Pozo}, {Delbo}, {Delgado}, {Delgado}, {Di Matteo},
  {Diakite}, {Diener}, {Distefano}, {Dolding}, {Drazinos}, {Dur{\'a}n},
  {Edvardsson}, {Enke}, {Eriksson}, {Esquej}, {Eynard Bontemps}, {Fabre},
  {Fabrizio}, {Faigler}, {Falc{\~a}o}, {Farr{\`a}s Casas}, {Federici},
  {Fedorets}, {Fernique}, {Figueras}, {Filippi}, {Findeisen}, {Fonti},
  {Fraile}, {Fraser}, {Fr{\'e}zouls}, {Gai}, {Galleti}, {Garabato},
  {Garc{\'\i}a-Sedano}, {Garofalo}, {Garralda}, {Gavel}, {Gavras}, {Gerssen},
  {Geyer}, {Giacobbe}, {Gilmore}, {Girona}, {Giuffrida}, {Glass}, {Gomes},
  {Granvik}, {Gueguen}, {Guerrier}, {Guiraud}, {Guti{\'e}rrez-S{\'a}nchez},
  {Haigron}, {Hatzidimitriou}, {Hauser}, {Haywood}, {Heiter}, {Helmi}, {Heu},
  {Hilger}, {Hobbs}, {Hofmann}, {Holland}, {Huckle}, {Hypki}, {Icardi},
  {Jan{\ss}en}, {Jevardat de Fombelle}, {Jonker}, {Juh{\'a}sz}, {Julbe},
  {Karampelas}, {Kewley}, {Klar}, {Kochoska}, {Kohley}, {Kolenberg},
  {Kontizas}, {Kontizas}, {Koposov}, {Kordopatis}, {Kostrzewa-Rutkowska},
  {Koubsky}, {Lambert}, {Lanza}, {Lasne}, {Lavigne}, {Le Fustec}, {Le
  Poncin-Lafitte}, {Lebreton}, {Leccia}, {Leclerc}, {Lecoeur-Taibi},
  {Lenhardt}, {Leroux}, {Liao}, {Licata}, {Lindstr{\o}m}, {Lister}, {Livanou},
  {Lobel}, {L{\'o}pez}, {Managau}, {Mann}, {Mantelet}, {Marchal}, {Marchant},
  {Marconi}, {Marinoni}, {Marschalk{\'o}}, {Marshall}, {Martino}, {Marton},
  {Mary}, {Massari}, {Matijevi{\v{c}}}, {Mazeh}, {McMillan}, {Messina},
  {Michalik}, {Millar}, {Molina}, {Molinaro}, {Moln{\'a}r}, {Montegriffo},
  {Mor}, {Morbidelli}, {Morel}, {Morris}, {Mulone}, {Muraveva}, {Musella},
  {Nelemans}, {Nicastro}, {Noval}, {O'Mullane}, {Ord{\'e}novic},
  {Ord{\'o}{\~n}ez-Blanco}, {Osborne}, {Pagani}, {Pagano}, {Pailler},
  {Palacin}, {Palaversa}, {Panahi}, {Pawlak}, {Piersimoni}, {Pineau}, {Plachy},
  {Plum}, {Poggio}, {Poujoulet}, {Pr{\v{s}}a}, {Pulone}, {Racero}, {Ragaini},
  {Rambaux}, {Ramos-Lerate}, {Regibo}, {Reyl{\'e}}, {Riclet}, {Ripepi}, {Riva},
  {Rivard}, {Rixon}, {Roegiers}, {Roelens}, {Romero-G{\'o}mez}, {Rowell},
  {Royer}, {Ruiz-Dern}, {Sadowski}, {Sagrist{\`a} Sell{\'e}s}, {Sahlmann},
  {Salgado}, {Salguero}, {Sanna}, {Santana-Ros}, {Sarasso}, {Savietto},
  {Schultheis}, {Sciacca}, {Segol}, {Segovia}, {S{\'e}gransan}, {Shih},
  {Siltala}, {Silva}, {Smart}, {Smith}, {Solano}, {Solitro}, {Sordo}, {Soria
  Nieto}, {Souchay}, {Spagna}, {Spoto}, {Stampa}, {Steele},
  {Steidelm{\"u}ller}, {Stephenson}, {Stoev}, {Suess}, {Surdej}, {Szabados},
  {Szegedi-Elek}, {Tapiador}, {Taris}, {Tauran}, {Taylor}, {Teixeira},
  {Terrett}, {Teyssand ier}, {Thuillot}, {Titarenko}, {Torra Clotet}, {Turon},
  {Ulla}, {Utrilla}, {Uzzi}, {Vaillant}, {Valentini}, {Valette}, {van Elteren},
  {Van Hemelryck}, {van Leeuwen}, {Vaschetto}, {Vecchiato}, {Veljanoski},
  {Viala}, {Vicente}, {Vogt}, {von Essen}, {Voss}, {Votruba}, {Voutsinas},
  {Walmsley}, {Weiler}, {Wertz}, {Wevers}, {Wyrzykowski}, {Yoldas},
  {{\v{Z}}erjal}, {Ziaeepour}, {Zorec}, {Zschocke}, {Zucker}, {Zurbach}, \&
  {Zwitter}}]{2018A&A...616A...1G}
{Gaia Collaboration}, {Brown}, A.~G.~A., {Vallenari}, A., {et~al.} 2018, \aap,
  616, A1

\bibitem[{{Gaia Collaboration} {et~al.}(2016){Gaia Collaboration}, {Prusti},
  {de Bruijne}, {Brown}, {Vallenari}, {Babusiaux}, {Bailer-Jones}, {Bastian},
  {Biermann}, {Evans}, \& et~al.}]{2016A&A...595A...1G}
{Gaia Collaboration}, {Prusti}, T., {de Bruijne}, J.~H.~J., {et~al.} 2016,
  \aap, 595, A1

\bibitem[{{Geier} \& {Heber}(2012)}]{2012A&A...543A.149G}
{Geier}, S. \& {Heber}, U. 2012, \aap, 543, A149

\bibitem[{{Geier} {et~al.}(2010){Geier}, {Heber}, {Podsiadlowski}, {Edelmann},
  {Napiwotzki}, {Kupfer}, \& {M{\"u}ller}}]{2010A&A...519A..25G}
{Geier}, S., {Heber}, U., {Podsiadlowski}, P., {et~al.} 2010, \aap, 519, A25

\bibitem[{{Ghasemi} {et~al.}(2017){Ghasemi}, {Moravveji}, {Aerts}, {Safari}, \&
  {Vu{\v c}kovi{\'c}}}]{2017MNRAS.465.1518G}
{Ghasemi}, H., {Moravveji}, E., {Aerts}, C., {Safari}, H., \& {Vu{\v
  c}kovi{\'c}}, M. 2017, \mnras, 465, 1518

\bibitem[{{Giammichele} {et~al.}(2017){Giammichele}, {Charpinet}, {Brassard},
  \& {Fontaine}}]{2017A&A...598A.109G}
{Giammichele}, N., {Charpinet}, S., {Brassard}, P., \& {Fontaine}, G. 2017,
  \aap, 598, A109

\bibitem[{{Giammichele} {et~al.}(2018){Giammichele}, {Charpinet}, {Fontaine},
  {Brassard}, {Green}, {Van Grootel}, {Bergeron}, {Zong}, \&
  {Dupret}}]{2018Natur.554...73G}
{Giammichele}, N., {Charpinet}, S., {Fontaine}, G., {et~al.} 2018, \nat, 554,
  73

\bibitem[{{Giammichele} {et~al.}(2016){Giammichele}, {Fontaine}, {Brassard}, \&
  {Charpinet}}]{2016ApJS..223...10G}
{Giammichele}, N., {Fontaine}, G., {Brassard}, P., \& {Charpinet}, S. 2016,
  \apjs, 223, 10

\bibitem[{{Gilliland} {et~al.}(2010){Gilliland}, {Brown},
  {Christensen-Dalsgaard}, {Kjeldsen}, {Aerts}, {Appourchaux}, {Basu},
  {Bedding}, {Chaplin}, {Cunha}, {De Cat}, {De Ridder}, {Guzik}, {Handler},
  {Kawaler}, {Kiss}, {Kolenberg}, {Kurtz}, {Metcalfe}, {Monteiro}, {Szab{\'o}},
  {Arentoft}, {Balona}, {Debosscher}, {Elsworth}, {Quirion}, {Stello},
  {Su{\'a}rez}, {Borucki}, {Jenkins}, {Koch}, {Kondo}, {Latham}, {Rowe}, \&
  {Steffen}}]{2010PASP..122..131G}
{Gilliland}, R.~L., {Brown}, T.~M., {Christensen-Dalsgaard}, J., {et~al.} 2010,
  \pasp, 122, 131

\bibitem[{{Green} {et~al.}(2003){Green}, {Fontaine}, {Reed}, {Callerame},
  {Seitenzahl}, {White}, {Hyde}, {{\O}stensen}, {Cordes}, {Brassard}, {Falter},
  {Jeffery}, {Dreizler}, {Schuh}, {Giovanni}, {Edelmann}, {Rigby}, \&
  {Bronowska}}]{2003ApJ...583L..31G}
{Green}, E.~M., {Fontaine}, G., {Reed}, M.~D., {et~al.} 2003, \apjl, 583, L31

\bibitem[{{Han} {et~al.}(2003){Han}, {Podsiadlowski}, {Maxted}, \&
  {Marsh}}]{2003MNRAS.341..669H}
{Han}, Z., {Podsiadlowski}, P., {Maxted}, P.~F.~L., \& {Marsh}, T.~R. 2003,
  \mnras, 341, 669

\bibitem[{{Han} {et~al.}(2002){Han}, {Podsiadlowski}, {Maxted}, {Marsh}, \&
  {Ivanova}}]{2002MNRAS.336..449H}
{Han}, Z., {Podsiadlowski}, P., {Maxted}, P.~F.~L., {Marsh}, T.~R., \&
  {Ivanova}, N. 2002, \mnras, 336, 449

\bibitem[{{Heber}(2016)}]{2016PASP..128h2001H}
{Heber}, U. 2016, \pasp, 128, 082001

\bibitem[{{Heber} {et~al.}(2018){Heber}, {Irrgang}, \&
  {Schaffenroth}}]{2018OAst...27...35H}
{Heber}, U., {Irrgang}, A., \& {Schaffenroth}, J. 2018, Open Astronomy, 27, 35

\bibitem[{{Henden} {et~al.}(2016){Henden}, {Templeton}, {Terrell}, {Smith},
  {Levine}, \& {Welch}}]{2016yCat.2336....0H}
{Henden}, A.~A., {Templeton}, M., {Terrell}, D., {et~al.} 2016, VizieR Online
  Data Catalog, 2336

\bibitem[{{H{\o}g} {et~al.}(2000){H{\o}g}, {Fabricius}, {Makarov}, {Urban},
  {Corbin}, {Wycoff}, {Bastian}, {Schwekendiek}, \&
  {Wicenec}}]{2000A&A...355L..27H}
{H{\o}g}, E., {Fabricius}, C., {Makarov}, V.~V., {et~al.} 2000, \aap, 355, L27

\bibitem[{{Howell} {et~al.}(2014){Howell}, {Sobeck}, {Haas}, {Still},
  {Barclay}, {Mullally}, {Troeltzsch}, {Aigrain}, {Bryson}, {Caldwell},
  {Chaplin}, {Cochran}, {Huber}, {Marcy}, {Miglio}, {Najita}, {Smith},
  {Twicken}, \& {Fortney}}]{2014PASP..126..398H}
{Howell}, S.~B., {Sobeck}, C., {Haas}, M., {et~al.} 2014, \pasp, 126, 398

\bibitem[{{Hu} {et~al.}(2008){Hu}, {Dupret}, {Aerts}, {Nelemans}, {Kawaler},
  {Miglio}, {Montalban}, \& {Scuflaire}}]{2008A&A...490..243H}
{Hu}, H., {Dupret}, M.-A., {Aerts}, C., {et~al.} 2008, \aap, 490, 243

\bibitem[{{Hu} {et~al.}(2010){Hu}, {Glebbeek}, {Thoul}, {Dupret}, {Stancliffe},
  {Nelemans}, \& {Aerts}}]{2010A&A...511A..87H}
{Hu}, H., {Glebbeek}, E., {Thoul}, A.~A., {et~al.} 2010, \aap, 511, A87

\bibitem[{{Hu} {et~al.}(2009){Hu}, {Nelemans}, {Aerts}, \&
  {Dupret}}]{2009A&A...508..869H}
{Hu}, H., {Nelemans}, G., {Aerts}, C., \& {Dupret}, M.-A. 2009, \aap, 508, 869

\bibitem[{{Jeffery} \& {Saio}(2006)}]{2006MNRAS.371..659J}
{Jeffery}, C.~S. \& {Saio}, H. 2006, \mnras, 371, 659

\bibitem[{{Jeffery} \& {Saio}(2007)}]{2007MNRAS.378..379J}
{Jeffery}, C.~S. \& {Saio}, H. 2007, \mnras, 378, 379

\bibitem[{{Jim{\'e}nez-Esteban} {et~al.}(2011){Jim{\'e}nez-Esteban},
  {Caballero}, \& {Solano}}]{2011A&A...525A..29J}
{Jim{\'e}nez-Esteban}, F.~M., {Caballero}, J.~A., \& {Solano}, E. 2011, \aap,
  525, A29

\bibitem[{{Kawka} {et~al.}(2015){Kawka}, {Vennes}, {O'Toole}, {N{\'e}meth},
  {Burton}, {Kotze}, \& {Buckley}}]{2015MNRAS.450.3514K}
{Kawka}, A., {Vennes}, S., {O'Toole}, S., {et~al.} 2015, \mnras, 450, 3514

\bibitem[{{Kern} {et~al.}(2017){Kern}, {Reed}, {Baran}, {{\O}stensen}, \&
  {Telting}}]{2017MNRAS.465.1057K}
{Kern}, J.~W., {Reed}, M.~D., {Baran}, A.~S., {{\O}stensen}, R.~H., \&
  {Telting}, J.~H. 2017, \mnras, 465, 1057

\bibitem[{{Ketzer} {et~al.}(2017){Ketzer}, {Reed}, {Baran}, {N{\'e}meth},
  {Telting}, {{\O}stensen}, \& {Jeffery}}]{2017MNRAS.467..461K}
{Ketzer}, L., {Reed}, M.~D., {Baran}, A.~S., {et~al.} 2017, \mnras, 467, 461

\bibitem[{{Kilkenny} {et~al.}(2010){Kilkenny}, {Fontaine}, {Green}, \&
  {Schuh}}]{2010IBVS.5927....1K}
{Kilkenny}, D., {Fontaine}, G., {Green}, E.~M., \& {Schuh}, S. 2010,
  Information Bulletin on Variable Stars, 5927, 1

\bibitem[{{Kilkenny} {et~al.}(1997){Kilkenny}, {Koen}, {O'Donoghue}, \&
  {Stobie}}]{1997MNRAS.285..640K}
{Kilkenny}, D., {Koen}, C., {O'Donoghue}, D., \& {Stobie}, R.~S. 1997, \mnras,
  285, 640

\bibitem[{{N{\'e}meth} {et~al.}(2012){N{\'e}meth}, {Kawka}, \&
  {Vennes}}]{2012MNRAS.427.2180N}
{N{\'e}meth}, P., {Kawka}, A., \& {Vennes}, S. 2012, \mnras, 427, 2180

\bibitem[{{O'Donoghue} {et~al.}(2013){O'Donoghue}, {Kilkenny}, {Koen},
  {Hambly}, {MacGillivray}, \& {Stobie}}]{2013MNRAS.431..240O}
{O'Donoghue}, D., {Kilkenny}, D., {Koen}, C., {et~al.} 2013, \mnras, 431, 240

\bibitem[{{{\O}stensen} {et~al.}(2011){{\O}stensen}, {Silvotti}, {Charpinet},
  {Oreiro}, {Bloemen}, {Baran}, {Reed}, {Kawaler}, {Telting}, {Green},
  {O'Toole}, {Aerts}, {G{\"a}nsicke}, {Marsh}, {Breedt}, {Heber}, {Koester},
  {Quint}, {Kurtz}, {Rodr{\'{\i}}guez-L{\'o}pez}, {Vu{\v c}kovi{\'c}},
  {Ottosen}, {Frimann}, {Somero}, {Wilson}, {Thygesen}, {Lindberg}, {Kjeldsen},
  {Christensen-Dalsgaard}, {Allen}, {McCauliff}, \&
  {Middour}}]{2011MNRAS.414.2860O}
{{\O}stensen}, R.~H., {Silvotti}, R., {Charpinet}, S., {et~al.} 2011, \mnras,
  414, 2860

\bibitem[{{{\O}stensen} {et~al.}(2010){{\O}stensen}, {Silvotti}, {Charpinet},
  {Oreiro}, {Handler}, {Green}, {Bloemen}, {Heber}, {G{\"a}nsicke}, {Marsh},
  {Kurtz}, {Telting}, {Reed}, {Kawaler}, {Aerts}, {Rodr{\'{\i}}guez-L{\'o}pez},
  {Vu{\v c}kovi{\'c}}, {Ottosen}, {Liimets}, {Quint}, {Van Grootel}, {Randall},
  {Gilliland}, {Kjeldsen}, {Christensen-Dalsgaard}, {Borucki}, {Koch}, \&
  {Quintana}}]{2010MNRAS.409.1470O}
{{\O}stensen}, R.~H., {Silvotti}, R., {Charpinet}, S., {et~al.} 2010, \mnras,
  409, 1470

\bibitem[{{{\O}stensen} {et~al.}(2014){{\O}stensen}, {Telting}, {Reed},
  {Baran}, {Nemeth}, \& {Kiaeerad}}]{2014A&A...569A..15O}
{{\O}stensen}, R.~H., {Telting}, J.~H., {Reed}, M.~D., {et~al.} 2014, \aap,
  569, A15

\bibitem[{{O'Toole}(2012)}]{2012ASPC..452..251O}
{O'Toole}, S.~J. 2012, in Astronomical Society of the Pacific Conference
  Series, Vol. 452, Fifth Meeting on Hot Subdwarf Stars and Related Objects,
  ed. D.~{Kilkenny}, C.~S. {Jeffery}, \& C.~{Koen}, 251

\bibitem[{{Pablo} {et~al.}(2012){Pablo}, {Kawaler}, {Reed}, {Bloemen},
  {Charpinet}, {Hu}, {Telting}, {{\O}stensen}, {Baran}, {Green}, {Hermes},
  {Barclay}, {O'Toole}, {Mullally}, {Kurtz}, {Christensen-Dalsgaard},
  {Caldwell}, {Christiansen}, \& {Kinemuchi}}]{2012MNRAS.422.1343P}
{Pablo}, H., {Kawaler}, S.~D., {Reed}, M.~D., {et~al.} 2012, \mnras, 422, 1343

\bibitem[{{Randall} {et~al.}(2006{\natexlab{a}}){Randall}, {Fontaine}, {Green},
  {Brassard}, {Kilkenny}, {Crause}, {Terndrup}, {Daane}, {Kiss}, {Jacob},
  {Bedding}, {For}, {Quirion}, \& {Chayer}}]{2006ApJ...643.1198R}
{Randall}, S.~K., {Fontaine}, G., {Green}, E.~M., {et~al.} 2006{\natexlab{a}},
  \apj, 643, 1198

\bibitem[{{Randall} {et~al.}(2006{\natexlab{b}}){Randall}, {Green}, {Fontaine},
  {Brassard}, {Terndrup}, {Brown}, {Fontaine}, {Zacharias}, \&
  {Chayer}}]{2006ApJ...645.1464R}
{Randall}, S.~K., {Green}, E.~M., {Fontaine}, G., {et~al.} 2006{\natexlab{b}},
  \apj, 645, 1464

\bibitem[{{Randall} {et~al.}(2009){Randall}, {Van Grootel}, {Fontaine},
  {Charpinet}, \& {Brassard}}]{2009A&A...507..911R}
{Randall}, S.~K., {Van Grootel}, V., {Fontaine}, G., {Charpinet}, S., \&
  {Brassard}, P. 2009, \aap, 507, 911

\bibitem[{{Reed} {et~al.}(2011){Reed}, {Baran}, {Quint}, {Kawaler}, {O'Toole},
  {Telting}, {Charpinet}, {Rodr{\'{\i}}guez-L{\'o}pez}, {{\O}stensen},
  {Provencal}, {Johnson}, {Thompson}, {Allen}, {Middour}, {Kjeldsen}, \&
  {Christensen-Dalsgaard}}]{2011MNRAS.414.2885R}
{Reed}, M.~D., {Baran}, A., {Quint}, A.~C., {et~al.} 2011, \mnras, 414, 2885

\bibitem[{{Reed} {et~al.}(2014){Reed}, {Foster}, {Telting}, {{\O}stensen},
  {Farris}, {Oreiro}, \& {Baran}}]{2014MNRAS.440.3809R}
{Reed}, M.~D., {Foster}, H., {Telting}, J.~H., {et~al.} 2014, \mnras, 440, 3809

\bibitem[{{Reed} {et~al.}(2019){Reed}, {Telting}, {Ketzer}, {Crooke}, {Baran},
  {Vos}, {N{\'e}meth}, {{\O}stensen}, \& {Jeffery}}]{2019MNRAS.483.2282R}
{Reed}, M.~D., {Telting}, J.~H., {Ketzer}, L., {et~al.} 2019, \mnras, 483, 2282

\bibitem[{{Ricker} {et~al.}(2014){Ricker}, {Winn}, {Vanderspek}, {Latham},
  {Bakos}, {Bean}, {Berta-Thompson}, {Brown}, {Buchhave}, {Butler}, {Butler},
  {Chaplin}, {Charbonneau}, {Christensen-Dalsgaard}, {Clampin}, {Deming},
  {Doty}, {De Lee}, {Dressing}, {Dunham}, {Endl}, {Fressin}, {Ge}, {Henning},
  {Holman}, {Howard}, {Ida}, {Jenkins}, {Jernigan}, {Johnson}, {Kaltenegger},
  {Kawai}, {Kjeldsen}, {Laughlin}, {Levine}, {Lin}, {Lissauer}, {MacQueen},
  {Marcy}, {McCullough}, {Morton}, {Narita}, {Paegert}, {Palle}, {Pepe},
  {Pepper}, {Quirrenbach}, {Rinehart}, {Sasselov}, {Sato}, {Seager},
  {Sozzetti}, {Stassun}, {Sullivan}, {Szentgyorgyi}, {Torres}, {Udry}, \&
  {Villasenor}}]{2014SPIE.9143E..20R}
{Ricker}, G.~R., {Winn}, J.~N., {Vanderspek}, R., {et~al.} 2014, in \procspie,
  Vol. 9143, Space Telescopes and Instrumentation 2014: Optical, Infrared, and
  Millimeter Wave, 914320

\bibitem[{{Saffer} {et~al.}(1994){Saffer}, {Bergeron}, {Koester}, \&
  {Liebert}}]{1994ApJ...432..351S}
{Saffer}, R.~A., {Bergeron}, P., {Koester}, D., \& {Liebert}, J. 1994, \apj,
  432, 351

\bibitem[{{Schuh} {et~al.}(2006){Schuh}, {Huber}, {Dreizler}, {Heber},
  {O'Toole}, {Green}, \& {Fontaine}}]{2006A&A...445L..31S}
{Schuh}, S., {Huber}, J., {Dreizler}, S., {et~al.} 2006, \aap, 445, L31

\bibitem[{{Skrutskie} {et~al.}(2006){Skrutskie}, {Cutri}, {Stiening},
  {Weinberg}, {Schneider}, {Carpenter}, {Beichman}, {Capps}, {Chester},
  {Elias}, {Huchra}, {Liebert}, {Lonsdale}, {Monet}, {Price}, {Seitzer},
  {Jarrett}, {Kirkpatrick}, {Gizis}, {Howard}, {Evans}, {Fowler}, {Fullmer},
  {Hurt}, {Light}, {Kopan}, {Marsh}, {McCallon}, {Tam}, {Van Dyk}, \&
  {Wheelock}}]{2006AJ....131.1163S}
{Skrutskie}, M.~F., {Cutri}, R.~M., {Stiening}, R., {et~al.} 2006, \aj, 131,
  1163

\bibitem[{{Telting} {et~al.}(2014){Telting}, {Baran}, {Nemeth}, {{\O}stensen},
  {Kupfer}, {Macfarlane}, {Heber}, {Aerts}, \& {Geier}}]{2014A&A...570A.129T}
{Telting}, J.~H., {Baran}, A.~S., {Nemeth}, P., {et~al.} 2014, \aap, 570, A129

\bibitem[{{Van Grootel} {et~al.}(2013){Van Grootel}, {Charpinet}, {Brassard},
  {Fontaine}, \& {Green}}]{2013A&A...553A..97V}
{Van Grootel}, V., {Charpinet}, S., {Brassard}, P., {Fontaine}, G., \& {Green},
  E.~M. 2013, \aap, 553, A97

\bibitem[{{Van Grootel} {et~al.}(2008{\natexlab{a}}){Van Grootel}, {Charpinet},
  {Fontaine}, \& {Brassard}}]{2008A&A...483..875V}
{Van Grootel}, V., {Charpinet}, S., {Fontaine}, G., \& {Brassard}, P.
  2008{\natexlab{a}}, \aap, 483, 875

\bibitem[{{Van Grootel} {et~al.}(2008{\natexlab{b}}){Van Grootel}, {Charpinet},
  {Fontaine}, {Brassard}, {Green}, {Chayer}, \&
  {Randall}}]{2008A&A...488..685V}
{Van Grootel}, V., {Charpinet}, S., {Fontaine}, G., {et~al.}
  2008{\natexlab{b}}, \aap, 488, 685

\bibitem[{{Van Grootel} {et~al.}(2010{\natexlab{a}}){Van Grootel}, {Charpinet},
  {Fontaine}, {Brassard}, {Green}, {Randall}, {Silvotti}, {{\O}stensen},
  {Kjeldsen}, {Christensen-Dalsgaard}, {Borucki}, \&
  {Koch}}]{2010ApJ...718L..97V}
{Van Grootel}, V., {Charpinet}, S., {Fontaine}, G., {et~al.}
  2010{\natexlab{a}}, \apjl, 718, L97

\bibitem[{{Van Grootel} {et~al.}(2010{\natexlab{b}}){Van Grootel}, {Charpinet},
  {Fontaine}, {Green}, \& {Brassard}}]{2010A&A...524A..63V}
{Van Grootel}, V., {Charpinet}, S., {Fontaine}, G., {Green}, E.~M., \&
  {Brassard}, P. 2010{\natexlab{b}}, \aap, 524, A63

\bibitem[{{Walker} {et~al.}(2003){Walker}, {Matthews}, {Kuschnig}, {Johnson},
  {Rucinski}, {Pazder}, {Burley}, {Walker}, {Skaret}, {Zee}, {Grocott},
  {Carroll}, {Sinclair}, {Sturgeon}, \& {Harron}}]{2003PASP..115.1023W}
{Walker}, G., {Matthews}, J., {Kuschnig}, R., {et~al.} 2003, \pasp, 115, 1023

\bibitem[{{Wolf} {et~al.}(2018){Wolf}, {Onken}, {Luvaul}, {Schmidt}, {Bessell},
  {Chang}, {Da Costa}, {Mackey}, {Martin-Jones}, {Murphy}, {Preston}, {Scalzo},
  {Shao}, {Smillie}, {Tisserand}, {White}, \& {Yuan}}]{2018PASA...35...10W}
{Wolf}, C., {Onken}, C.~A., {Luvaul}, L.~C., {et~al.} 2018, \pasa, 35, e010

\bibitem[{{Zong} {et~al.}(2018){Zong}, {Charpinet}, {Fu}, {Vauclair}, {Niu}, \&
  {Su}}]{2018ApJ...853...98Z}
{Zong}, W., {Charpinet}, S., {Fu}, J.-N., {et~al.} 2018, \apj, 853, 98

\bibitem[{{Zong} {et~al.}(2016{\natexlab{a}}){Zong}, {Charpinet}, \&
  {Vauclair}}]{2016A&A...594A..46Z}
{Zong}, W., {Charpinet}, S., \& {Vauclair}, G. 2016{\natexlab{a}}, \aap, 594,
  A46

\bibitem[{{Zong} {et~al.}(2016{\natexlab{b}}){Zong}, {Charpinet}, {Vauclair},
  {Giammichele}, \& {Van Grootel}}]{2016A&A...585A..22Z}
{Zong}, W., {Charpinet}, S., {Vauclair}, G., {Giammichele}, N., \& {Van
  Grootel}, V. 2016{\natexlab{b}}, \aap, 585, A22

\end{thebibliography}

\newpage

\footnotesize
\onecolumn
\begin{center}
\begin{longtable}{ccccccccccccl}
\caption{Mode identification and details of the optimal frequency match obtained for TIC 278659026.
The mean relative dispersion of the fit is $\overline{\Delta X/X} = 0.07 \%$ ($X=P$ or $\nu$),
corresponding to $\overline{\Delta P} = 3.16 $ s, $\overline{\Delta \nu} = 0.161 $ $\mu$Hz, and $S^2=0.139$.}
\label{tab3} \\
\hline\hline
 & & $\nu_{\rm obs}$ & $\nu_{\rm th}$ & $P_{\rm obs}$ & $P_{\rm th}$ & $\log E$ & $C_{kl} $ & $\Delta X/X$ & $\Delta P$  & $\Delta \nu$ & Amplitude & Id.\tabularnewline
 $l$ & $k$ & ($\mu$Hz) & ($\mu$Hz) & (s) & (s) & (erg) & & ($\%$) & (s) & ($\mu$Hz) & (\%) & \tabularnewline
\hline

 & \tabularnewline
\endfirsthead

\caption{continued.}\\
\hline\hline
 & & $\nu_{\rm obs}$ & $\nu_{\rm th}$ & $P_{\rm obs}$ & $P_{\rm th}$ & $\log E$ & $C_{kl} $ & $\Delta X/X$ & $\Delta P$  & $\Delta \nu$ & Amplitude & Id.\tabularnewline
 $l$ & $k$ & ($\mu$Hz) & ($\mu$Hz) & (s) & (s) & (erg) & & ($\%$) & (s) & ($\mu$Hz) & (\%) & \tabularnewline
\hline

 & \tabularnewline
\endhead

 & \tabularnewline
\hline

\endfoot

 & \tabularnewline
\hline

\endlastfoot

1 & $5$ & ... & 827.455 & ... & 1208.52 & 46.456 & 0.4849 & ... & ... & ... & ... &\tabularnewline
1 & $6$ & ... & 696.183 & ... & 1436.40 & 46.165 & 0.4846 & ... & ... & ... & ... &\tabularnewline
1 & $7$ & 604.580 & 604.358 & 1654.04 & 1654.65 & 45.812 & 0.4849 & $-0.04$ & $-0.61$ & $+0.221$ & $0.0212$ & $f_{19}$\tabularnewline
1 & $8$ & ... & 538.122 & ... & 1858.32 & 45.570 & 0.4859 & ... & ... & ... & ... &\tabularnewline
1 & $9$ & ... & 484.536 & ... & 2063.83 & 45.519 & 0.4881 & ... & ... & ... & ... &\tabularnewline
1 & $10$ & ... & 437.573 & ... & 2285.33 & 45.530 & 0.4899 & ... & ... & ... & ... &\tabularnewline
1 & $11$ & ... & 397.132 & ... & 2518.06 & 45.463 & 0.4910 & ... & ... & ... & ... &\tabularnewline
1 & $12$ & ... & 363.148 & ... & 2753.70 & 45.273 & 0.4913 & ... & ... & ... & ... &\tabularnewline
1 & $13$ & 335.033 & 335.244 & 2984.78 & 2982.91 & 44.964 & 0.4891 & $+0.06$ & $+1.88$ & $-0.211$ & $0.0735$ & $f_{4}$\tabularnewline
1 & $14$ & ... & 315.659 & ... & 3167.98 & 44.776 & 0.4707 & ... & ... & ... & ... &\tabularnewline
1 & $15$ & ... & 304.833 & ... & 3280.48 & 44.407 & 0.4733 & ... & ... & ... & ... &\tabularnewline
1 & $16$ & ... & 290.261 & ... & 3445.18 & 43.753 & 0.4846 & ... & ... & ... & ... &\tabularnewline
1 & $17$ & ... & 277.235 & ... & 3607.05 & 43.545 & 0.4880 & ... & ... & ... & ... &\tabularnewline
1 & $18$ & ... & 263.368 & ... & 3796.96 & 43.643 & 0.4934 & ... & ... & ... & ... &\tabularnewline
1 & $19$ & 249.269 & 249.584 & 4011.74 & 4006.66 & 43.594 & 0.4946 & $+0.13$ & $+5.07$ & $-0.315$ & $0.3184$ & $f_{1}$\tabularnewline
1 & $20$ & 237.829 & 237.490 & 4204.70 & 4210.70 & 43.322 & 0.4944 & $-0.14$ & $-6.00$ & $+0.339$ & $0.0200$ & $f_{21}$\tabularnewline
1 & $21$ & 227.285 & 227.646 & 4399.77 & 4392.79 & 42.960 & 0.4930 & $+0.16$ & $+6.98$ & $-0.361$ & $0.0810$ & $f_{3}$\tabularnewline
1 & $22$ & ... & 219.469 & ... & 4556.46 & 42.841 & 0.4935 & ... & ... & ... & ... &\tabularnewline
1 & $23$ & ... & 210.919 & ... & 4741.16 & 42.947 & 0.4957 & ... & ... & ... & ... &\tabularnewline
1 & $24$ & ... & 201.920 & ... & 4952.46 & 42.933 & 0.4968 & ... & ... & ... & ... &\tabularnewline
1 & $25$ & ... & 193.185 & ... & 5176.39 & 42.635 & 0.4971 & ... & ... & ... & ... &\tabularnewline
1 & $26$ & 185.014 & 185.086 & 5404.99 & 5402.89 & 41.899 & 0.4970 & $+0.04$ & $+2.10$ & $-0.072$ & $0.0452$ & $f_{7}$\tabularnewline
1 & $27$ & 177.869 & 177.864 & 5622.13 & 5622.27 & 38.251 & 0.4964 & $-0.00$ & $-0.14$ & $+0.005$ & $0.0223$ & $f_{16}$\tabularnewline
1 & $28$ & ... & 171.701 & ... & 5824.09 & 41.361 & 0.4955 & ... & ... & ... & ... &\tabularnewline
1 & $29$ & 166.256 & 166.232 & 6014.82 & 6015.70 & 41.870 & 0.4959 & $-0.01$ & $-0.87$ & $+0.024$ & $0.0266$ & $f_{12}$\tabularnewline
1 & $30$ & 160.622 & 160.757 & 6225.81 & 6220.57 & 42.306 & 0.4969 & $+0.08$ & $+5.24$ & $-0.135$ & $0.0655$ & $f_{6}$\tabularnewline
1 & $31$ & ... & 155.327 & ... & 6438.01 & 42.584 & 0.4975 & ... & ... & ... & ... &\tabularnewline
1 & $32$ & ... & 150.236 & ... & 6656.19 & 42.661 & 0.4975 & ... & ... & ... & ... &\tabularnewline
1 & $33$ & ... & 145.671 & ... & 6864.76 & 42.592 & 0.4971 & ... & ... & ... & ... &\tabularnewline
1 & $34$ & ... & 141.693 & ... & 7057.50 & 42.521 & 0.4966 & ... & ... & ... & ... &\tabularnewline
1 & $35$ & 137.912 & 138.039 & 7251.01 & 7244.32 & 42.595 & 0.4970 & $+0.09$ & $+6.69$ & $-0.127$ & $0.0192$ & $f_{23}$\tabularnewline
1 & $36$ & 134.387 & 134.398 & 7441.22 & 7440.61 & 42.756 & 0.4976 & $+0.01$ & $+0.61$ & $-0.011$ & $0.0375$ & $f_{10}$\tabularnewline
1 & $37$ & ... & 130.797 & ... & 7645.43 & 42.869 & 0.4980 & ... & ... & ... & ... &\tabularnewline
1 & $38$ & ... & 127.350 & ... & 7852.39 & 42.893 & 0.4980 & ... & ... & ... & ... &\tabularnewline
1 & $39$ & ... & 124.116 & ... & 8057.00 & 42.858 & 0.4980 & ... & ... & ... & ... &\tabularnewline
1 & $40$ & 121.144 & 121.092 & 8254.62 & 8258.21 & 42.828 & 0.4979 & $-0.04$ & $-3.59$ & $+0.053$ & $0.0194$ & $f_{22}$\tabularnewline
1 & $41$ & ... & 118.207 & ... & 8459.72 & 42.846 & 0.4980 & ... & ... & ... & ... &\tabularnewline
1 & $42$ & ... & 115.396 & ... & 8665.81 & 42.891 & 0.4982 & ... & ... & ... & ... &\tabularnewline
1 & $43$ & ... & 112.660 & ... & 8876.27 & 42.921 & 0.4983 & ... & ... & ... & ... &\tabularnewline
1 & $44$ & ... & 110.034 & ... & 9088.07 & 42.914 & 0.4984 & ... & ... & ... & ... &\tabularnewline
1 & $45$ & ... & 107.544 & ... & 9298.48 & 42.880 & 0.4984 & ... & ... & ... & ... &\tabularnewline
1 & $46$ & ... & 105.193 & ... & 9506.34 & 42.838 & 0.4983 & ... & ... & ... & ... &\tabularnewline
1 & $47$ & ... & 102.970 & ... & 9711.59 & 42.804 & 0.4982 & ... & ... & ... & ... &\tabularnewline
1 & $48$ & ... & 100.876 & ... & 9913.21 & 42.782 & 0.4979 & ... & ... & ... & ... &\tabularnewline
1 & $49$ & ... & 98.962 & ... & 10104.90 & 42.791 & 0.4969 & ... & ... & ... & ... &\tabularnewline
1 & $50$ & ... & 97.399 & ... & 10267.04 & 42.872 & 0.4945 & ... & ... & ... & ... &\tabularnewline
1 & $51$ & 96.073 & 96.065 & 10408.72 & 10409.58 & 42.806 & 0.4955 & $-0.01$ & $-0.86$ & $+0.008$ & $0.0324$ & $f_{11}$\tabularnewline
1 & $52$ & ... & 94.473 & ... & 10585.06 & 42.701 & 0.4976 & ... & ... & ... & ... &\tabularnewline
1 & $53$ & ... & 92.757 & ... & 10780.91 & 42.658 & 0.4983 & ... & ... & ... & ... &\tabularnewline
1 & $54$ & ... & 91.047 & ... & 10983.32 & 42.628 & 0.4986 & ... & ... & ... & ... &\tabularnewline
1 & $55$ & ... & 89.383 & ... & 11187.83 & 42.596 & 0.4987 & ... & ... & ... & ... &\tabularnewline
1 & $56$ & ... & 87.776 & ... & 11392.59 & 42.570 & 0.4987 & ... & ... & ... & ... &\tabularnewline
1 & $57$ & ... & 86.225 & ... & 11597.63 & 42.560 & 0.4988 & ... & ... & ... & ... &\tabularnewline
1 & $58$ & ... & 84.716 & ... & 11804.14 & 42.568 & 0.4989 & ... & ... & ... & ... &\tabularnewline
1 & $59$ & ... & 83.243 & ... & 12013.04 & 42.582 & 0.4990 & ... & ... & ... & ... &\tabularnewline
1 & $60$ & ... & 81.808 & ... & 12223.77 & 42.584 & 0.4991 & ... & ... & ... & ... &\tabularnewline

 & & & & & & & & & & & \tabularnewline
2 & $5$  & ... & 1413.333 & ... & 707.55 & 46.223 & 0.1524 & ... & ... & ... & ... &\tabularnewline
2 & $6$  & ... & 1192.361 & ... & 838.67 & 46.024 & 0.1517 & ... & ... & ... & ... &\tabularnewline
2 & $7$  & ... & 1036.894 & ... & 964.42 & 45.718 & 0.1517 & ... & ... & ... & ... &\tabularnewline
2 & $8$  & ... & 924.614 & ... & 1081.53 & 45.496 & 0.1526 & ... & ... & ... & ... &\tabularnewline
2 & $9$  & ... & 833.753 & ... & 1199.40 & 45.463 & 0.1548 & ... & ... & ... & ... &\tabularnewline
2 & $10$ & ... & 753.739 & ... & 1326.72 & 45.492 & 0.1568 & ... & ... & ... & ... &\tabularnewline
2 & $11$ & ... & 684.500 & ... & 1460.92 & 45.439 & 0.1581 & ... & ... & ... & ... &\tabularnewline
2 & $12$ & ... & 625.989 & ... & 1597.47 & 45.255 & 0.1587 & ... & ... & ... & ... &\tabularnewline
2 & $13$ & ... & 577.092 & ... & 1732.83 & 44.925 & 0.1587 & ... & ... & ... & ... &\tabularnewline
2 & $14$ & ... & 537.254 & ... & 1861.32 & 44.446 & 0.1568 & ... & ... & ... & ... &\tabularnewline
2 & $15$ & ... & 508.264 & ... & 1967.48 & 43.968 & 0.1497 & ... & ... & ... & ... &\tabularnewline
2 & $16$ & ... & 491.646 & ... & 2033.98 & 43.796 & 0.1426 & ... & ... & ... & ... &\tabularnewline
2 & $17$ & ... & 475.730 & ... & 2102.03 & 43.597 & 0.1508 & ... & ... & ... & ... &\tabularnewline
2 & $18$ & ... & 453.771 & ... & 2203.75 & 43.645 & 0.1583 & ... & ... & ... & ... &\tabularnewline
2 & $19$ & 430.875 & 430.663 & 2320.86 & 2322.00 & 43.589 & 0.1607 & $-0.05$ & $-1.14$ & $+0.212$ & $0.0686$ & $f_{5}$\tabularnewline
2 & $20$ & ... & 410.071 & ... & 2438.60 & 43.320 & 0.1609 & ... & ... & ... & ... &\tabularnewline
2 & $21$ & 393.032 & 393.215 & 2544.32 & 2543.14 & 42.959 & 0.1597 & $+0.05$ & $+1.18$ & $-0.183$ & $0.0251$ & $f_{13}$\tabularnewline
2 & $22$ & ... & 379.225 & ... & 2636.96 & 42.831 & 0.1601 & ... & ... & ... & ... &\tabularnewline
2 & $23$ & ... & 364.632 & ... & 2742.49 & 42.933 & 0.1623 & ... & ... & ... & ... &\tabularnewline
2 & $24$ & ... & 349.200 & ... & 2863.69 & 42.921 & 0.1634 & ... & ... & ... & ... &\tabularnewline
2 & $25$ & ... & 334.162 & ... & 2992.56 & 42.625 & 0.1638 & ... & ... & ... & ... &\tabularnewline
2 & $26$ & ... & 320.185 & ... & 3123.20 & 41.880 & 0.1637 & ... & ... & ... & ... &\tabularnewline
2 & $27$ & 307.720 & 307.692 & 3249.70 & 3250.00 & 39.115 & 0.1631 & $-0.01$ & $-0.30$ & $+0.028$ & $0.0377$ & $f_{9}$\tabularnewline
2 & $28$ & 296.822 & 297.018 & 3369.02 & 3366.80 & 41.398 & 0.1622 & $+0.07$ & $+2.22$ & $-0.196$ & $0.0224$ & $f_{15}$\tabularnewline
2 & $29$ & ... & 287.589 & ... & 3477.18 & 41.882 & 0.1625 & ... & ... & ... & ... &\tabularnewline
2 & $30$ & ... & 278.168 & ... & 3594.95 & 42.310 & 0.1636 & ... & ... & ... & ... &\tabularnewline
2 & $31$ & ... & 268.801 & ... & 3720.22 & 42.589 & 0.1642 & ... & ... & ... & ... &\tabularnewline
2 & $32$ & ... & 260.000 & ... & 3846.16 & 42.669 & 0.1642 & ... & ... & ... & ... &\tabularnewline
2 & $33$ & ... & 252.096 & ... & 3966.74 & 42.601 & 0.1638 & ... & ... & ... & ... &\tabularnewline
2 & $34$ & 245.609 & 245.207 & 4071.52 & 4078.19 & 42.526 & 0.1633 & $-0.16$ & $-6.67$ & $+0.402$ & $0.0240$ & $f_{14}$\tabularnewline
2 & $35$ & ... & 238.896 & ... & 4185.93 & 42.594 & 0.1636 & ... & ... & ... & ... &\tabularnewline
2 & $36$ & ... & 232.611 & ... & 4299.02 & 42.755 & 0.1643 & ... & ... & ... & ... &\tabularnewline
2 & $37$ & ... & 226.389 & ... & 4417.17 & 42.870 & 0.1647 & ... & ... & ... & ... &\tabularnewline
2 & $38$ & ... & 220.427 & ... & 4536.64 & 42.895 & 0.1647 & ... & ... & ... & ... &\tabularnewline
2 & $39$ & ... & 214.831 & ... & 4654.82 & 42.861 & 0.1647 & ... & ... & ... & ... &\tabularnewline
2 & $40$ & ... & 209.599 & ... & 4771.03 & 42.830 & 0.1646 & ... & ... & ... & ... &\tabularnewline
2 & $41$ & ... & 204.611 & ... & 4887.33 & 42.846 & 0.1647 & ... & ... & ... & ... &\tabularnewline
2 & $42$ & 199.913 & 199.750 & 5002.18 & 5006.25 & 42.890 & 0.1649 & $-0.08$ & $-4.07$ & $+0.163$ & $0.1706$ & $f_{2}$\tabularnewline
2 & $43$ & ... & 195.017 & ... & 5127.75 & 42.920 & 0.1650 & ... & ... & ... & ... &\tabularnewline
2 & $44$ & ... & 190.472 & ... & 5250.12 & 42.914 & 0.1651 & ... & ... & ... & ... &\tabularnewline
2 & $45$ & ... & 186.157 & ... & 5371.82 & 42.879 & 0.1651 & ... & ... & ... & ... &\tabularnewline
2 & $46$ & ... & 182.075 & ... & 5492.25 & 42.836 & 0.1651 & ... & ... & ... & ... &\tabularnewline
2 & $47$ & ... & 178.202 & ... & 5611.60 & 42.799 & 0.1651 & ... & ... & ... & ... &\tabularnewline
2 & $48$ & ... & 174.519 & ... & 5730.05 & 42.769 & 0.1651 & ... & ... & ... & ... &\tabularnewline
2 & $49$ & ... & 171.034 & ... & 5846.80 & 42.745 & 0.1649 & ... & ... & ... & ... &\tabularnewline
2 & $50$ & ... & 167.810 & ... & 5959.11 & 42.742 & 0.1645 & ... & ... & ... & ... &\tabularnewline
2 & $51$ & ... & 165.063 & ... & 6058.31 & 42.808 & 0.1632 & ... & ... & ... & ... &\tabularnewline
2 & $52$ & ... & 162.855 & ... & 6140.43 & 42.809 & 0.1630 & ... & ... & ... & ... &\tabularnewline
2 & $53$ & ... & 160.348 & ... & 6236.44 & 42.693 & 0.1645 & ... & ... & ... & ... &\tabularnewline
2 & $54$ & ... & 157.537 & ... & 6347.71 & 42.638 & 0.1651 & ... & ... & ... & ... &\tabularnewline
2 & $55$ & ... & 154.709 & ... & 6463.76 & 42.600 & 0.1653 & ... & ... & ... & ... &\tabularnewline
2 & $56$ & ... & 151.951 & ... & 6581.06 & 42.571 & 0.1654 & ... & ... & ... & ... &\tabularnewline
2 & $57$ & 149.432 & 149.277 & 6692.01 & 6698.94 & 42.560 & 0.1655 & $-0.10$ & $-6.93$ & $+0.155$ & $0.0416$ & $f_{8}$\tabularnewline
2 & $58$ & ... & 146.674 & ... & 6817.86 & 42.568 & 0.1656 & ... & ... & ... & ... &\tabularnewline
2 & $59$ & ... & 144.128 & ... & 6938.26 & 42.581 & 0.1657 & ... & ... & ... & ... &\tabularnewline
2 & $60$ & ... & 141.647 & ... & 7059.78 & 42.583 & 0.1657 & ... & ... & ... & ... &\tabularnewline

\end{longtable}
\end{center}
\twocolumn
\normalsize

\begin{appendix}

\section{Exploring a wider range of envelope H/He mass ratios}

\begin{figure*}[tp]
\begin{centering}
\includegraphics[scale=0.33]{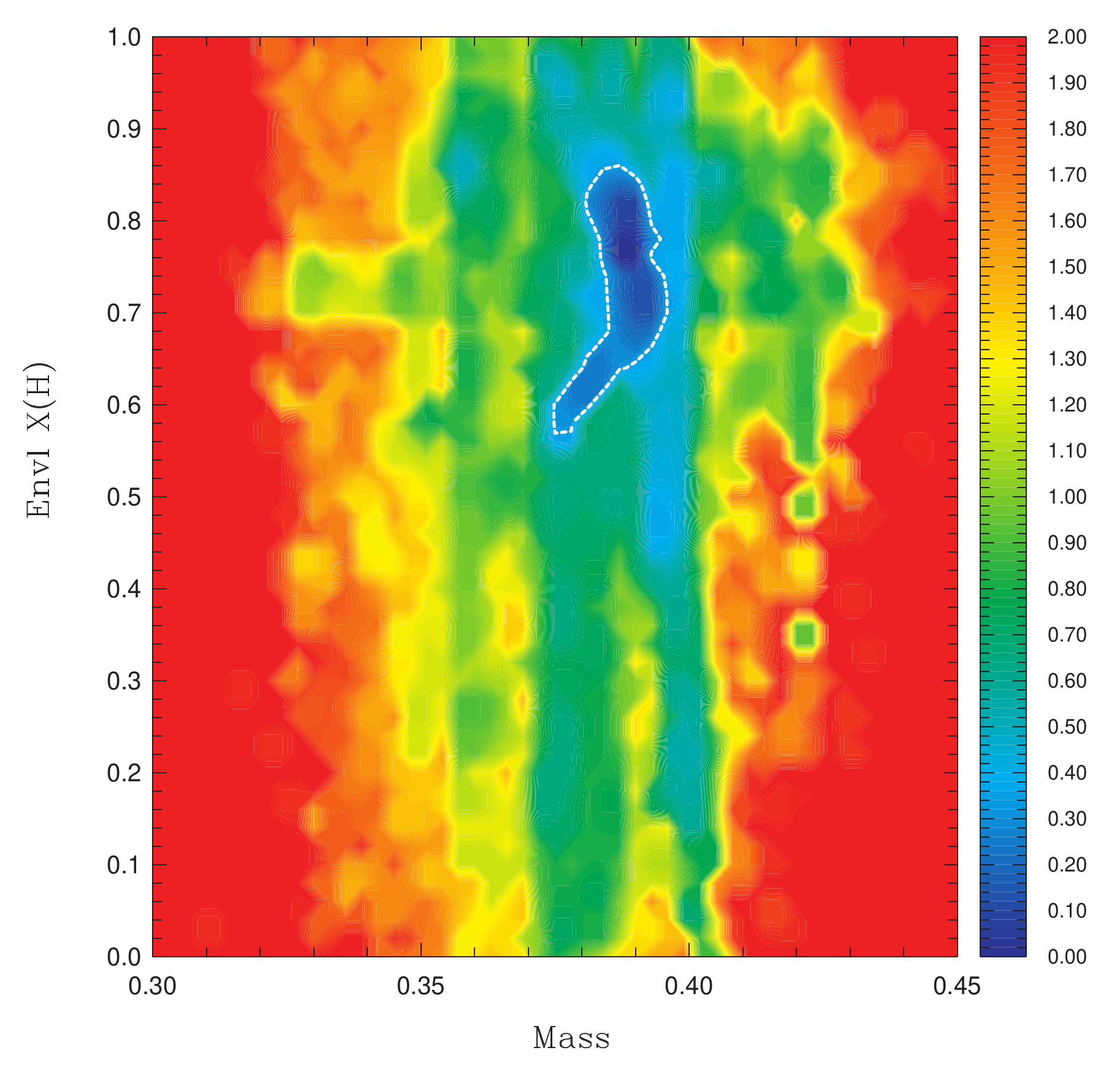}\includegraphics[scale=0.33]{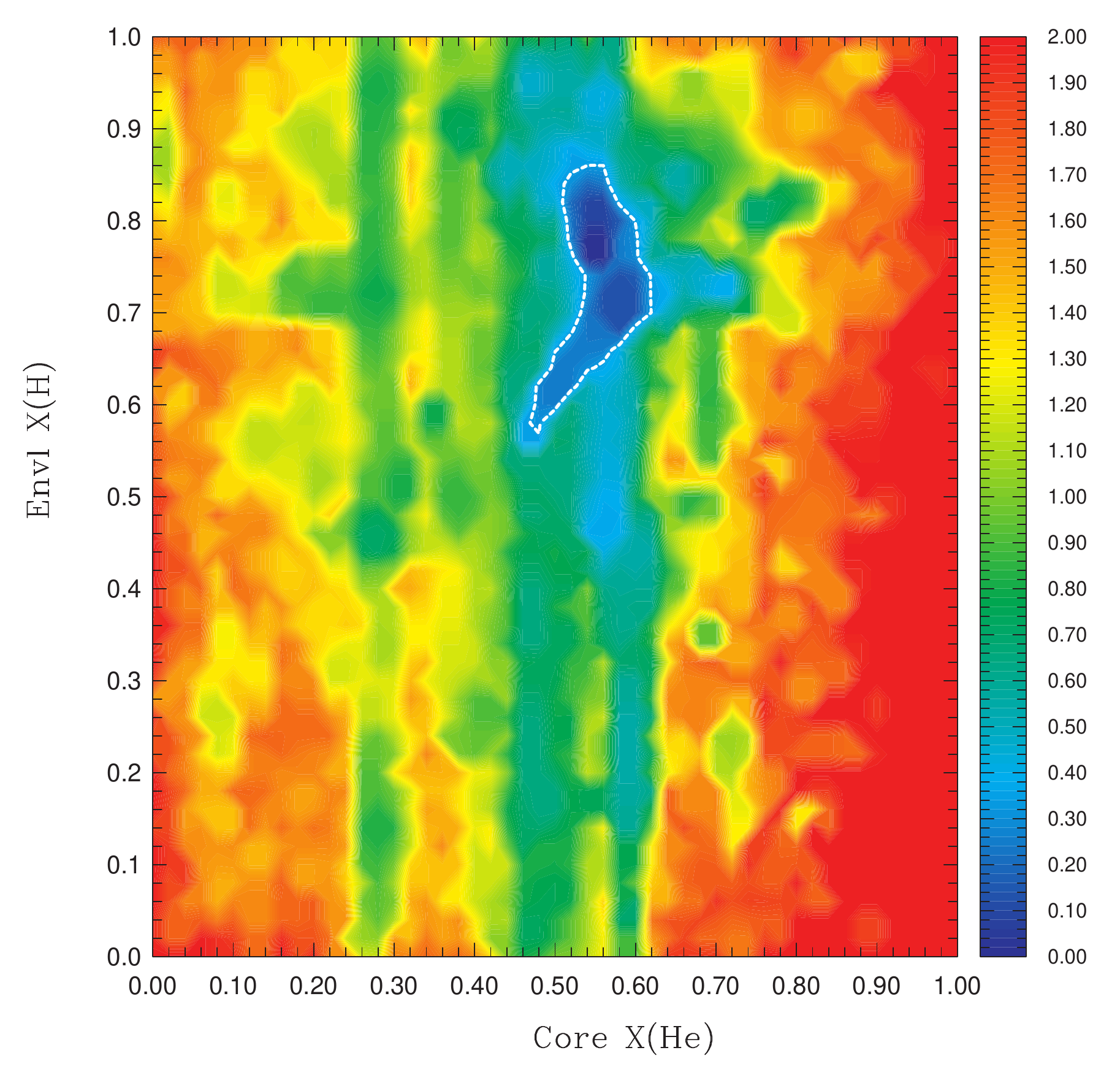}\includegraphics[scale=0.33]{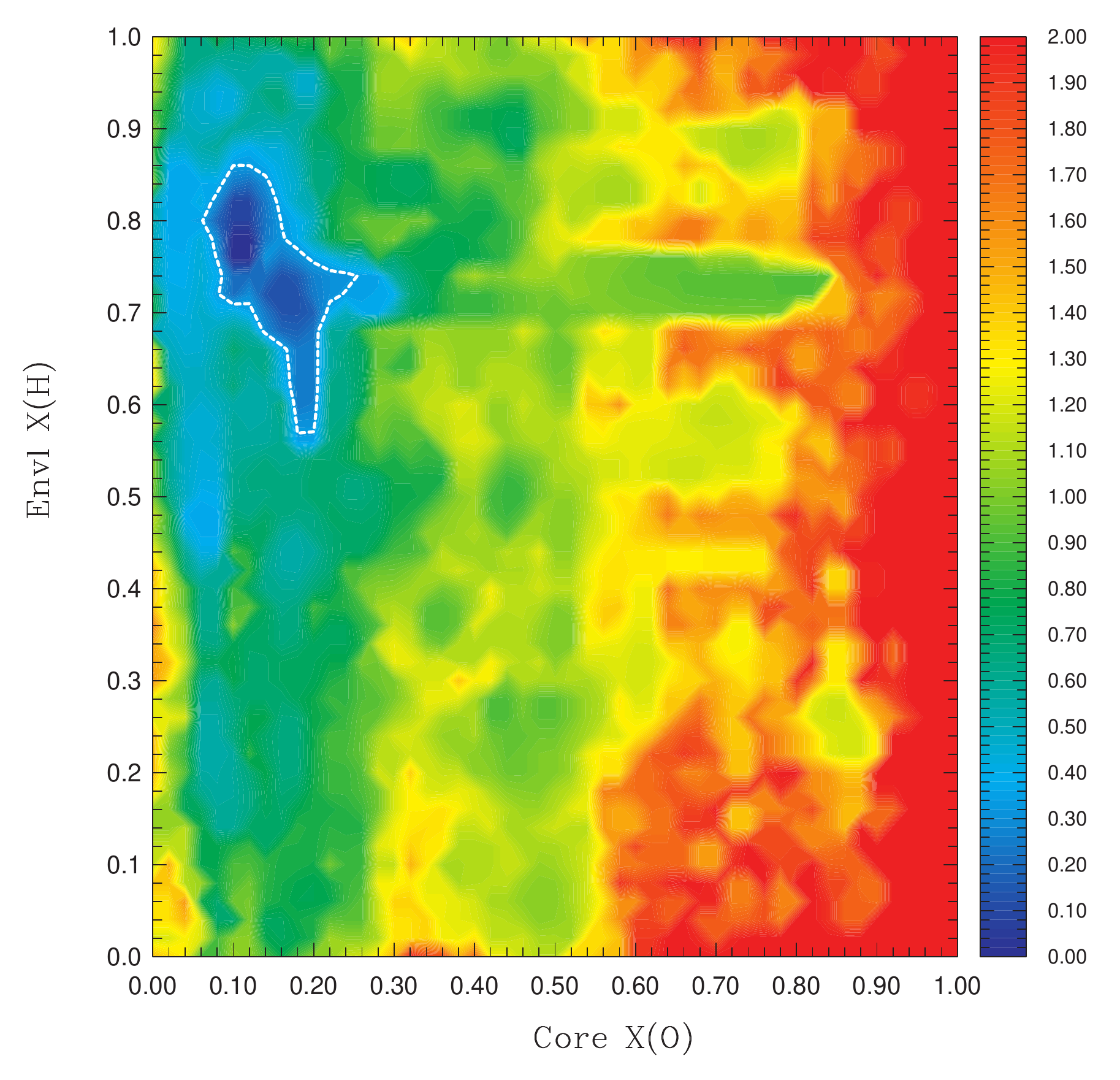}\\
\includegraphics[scale=0.33]{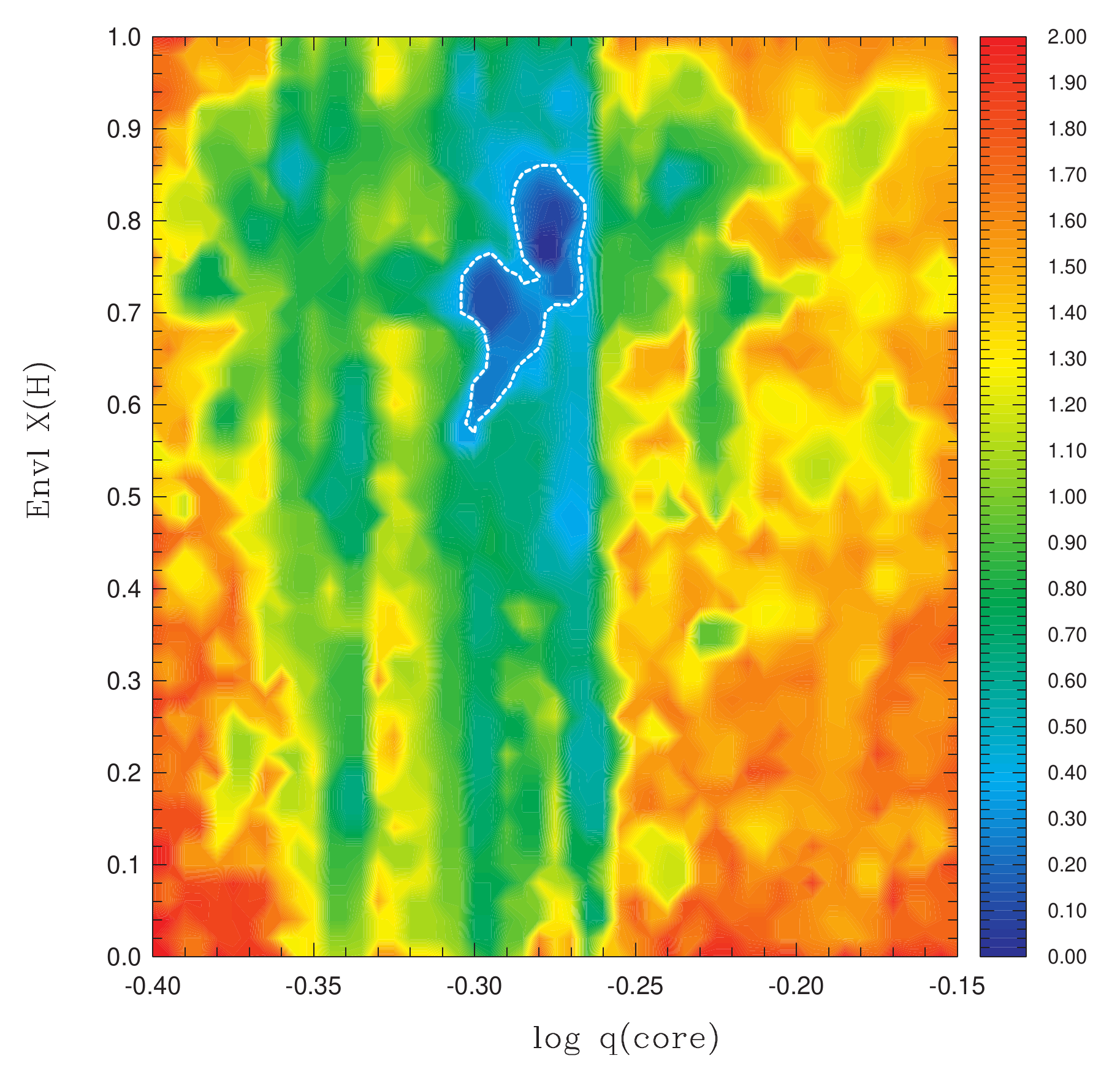}\includegraphics[scale=0.33]{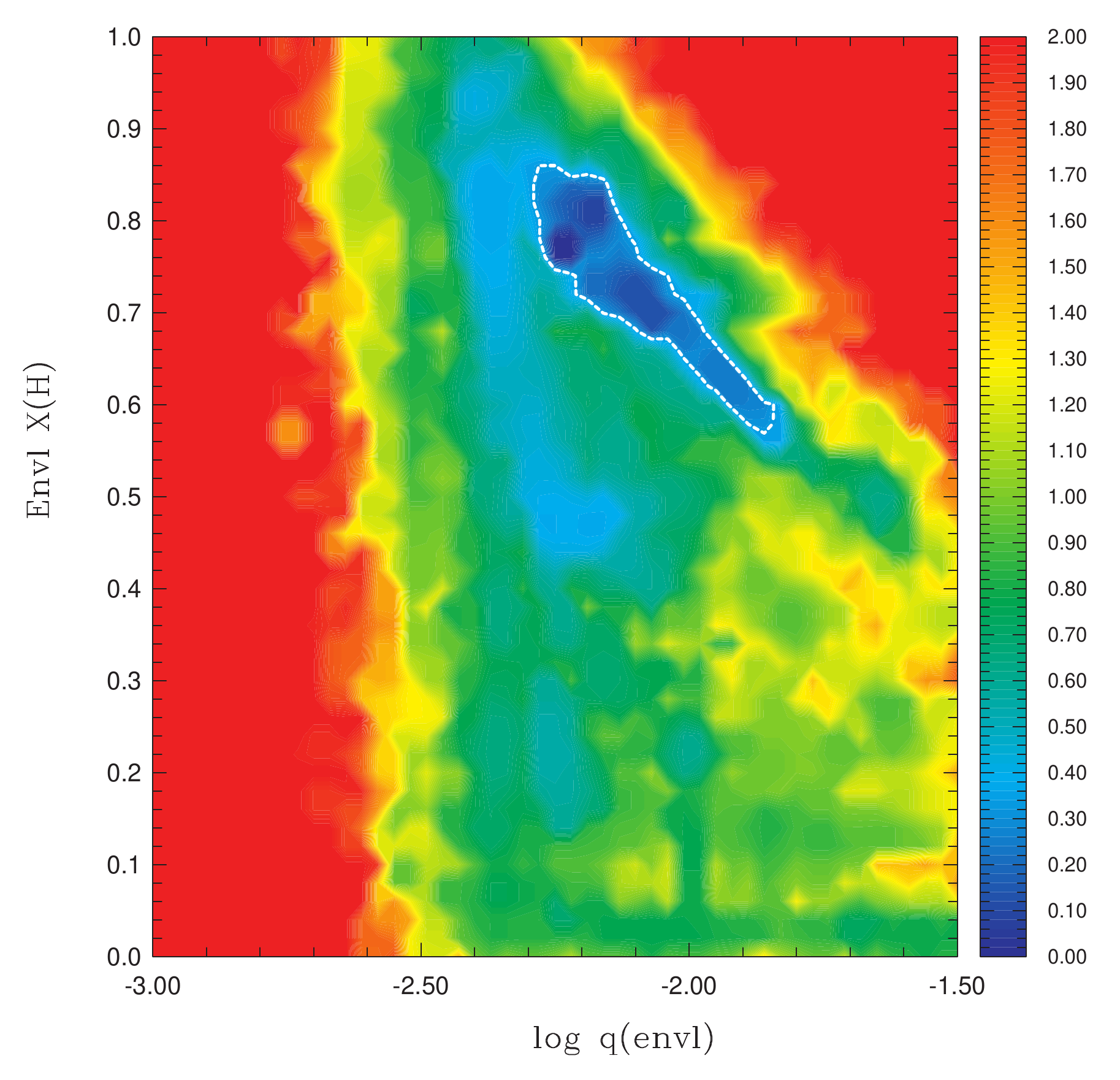}\includegraphics[scale=0.33]{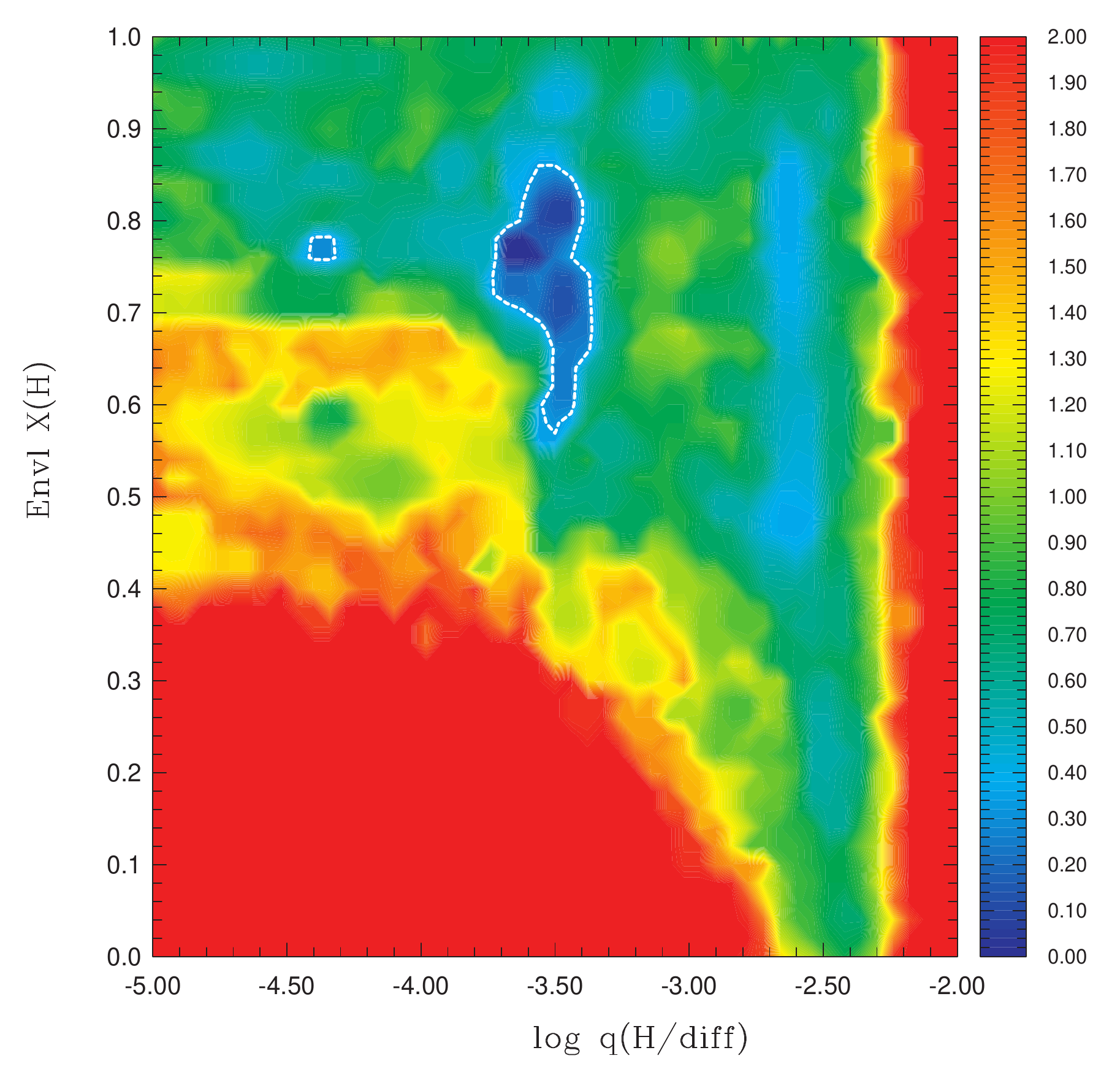}\\
\includegraphics[scale=0.33]{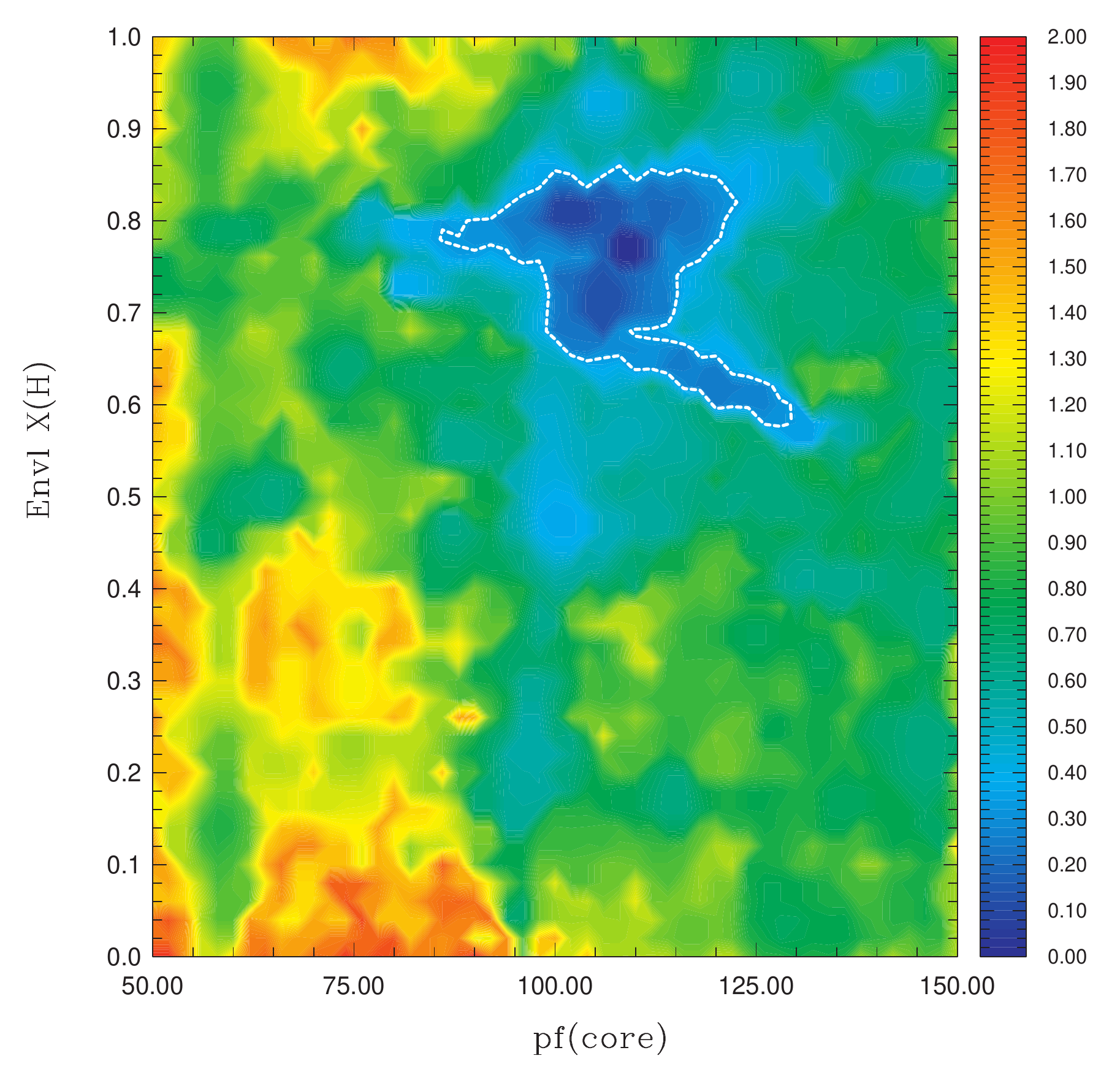}\includegraphics[scale=0.33]{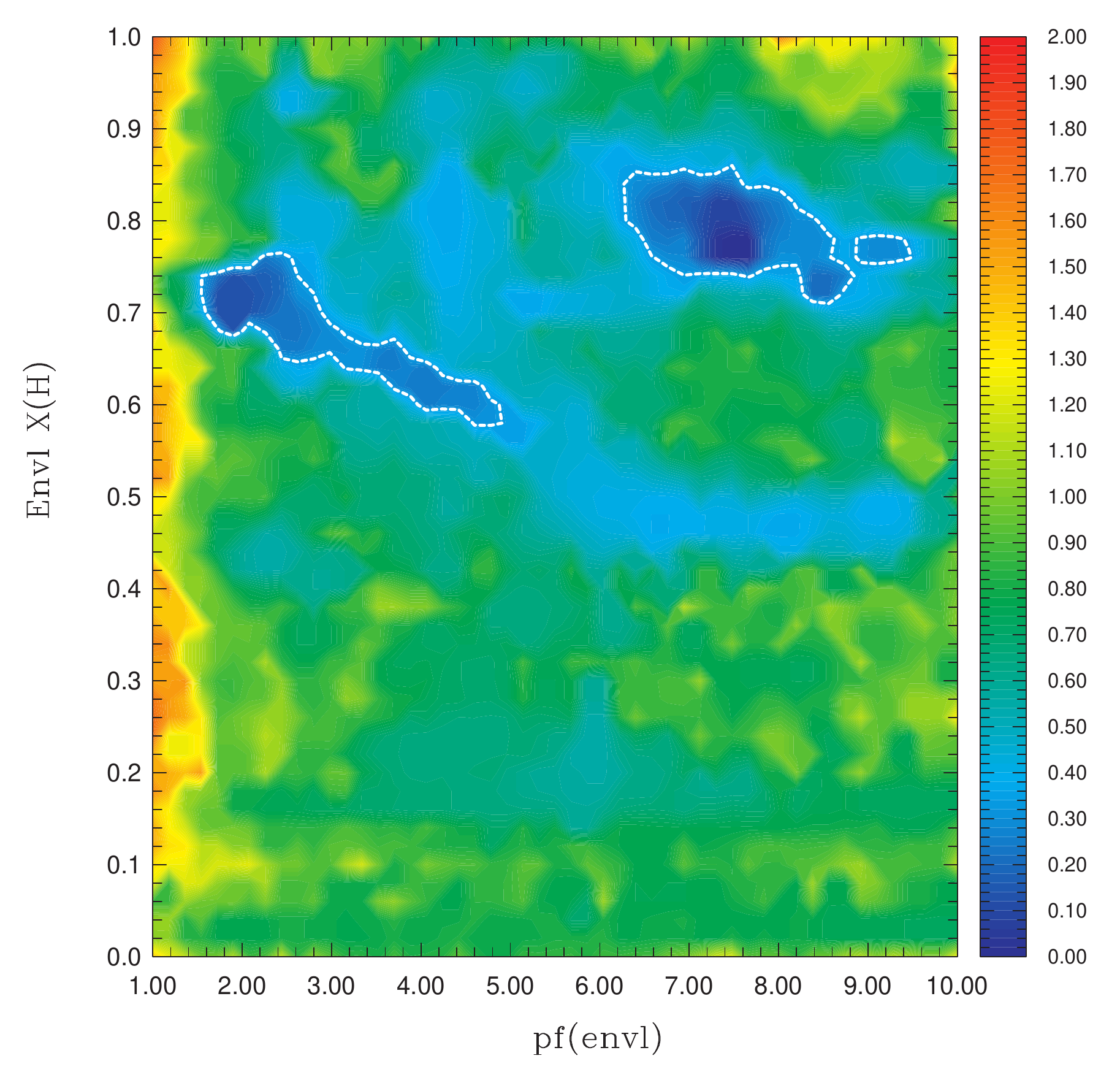}\includegraphics[scale=0.33]{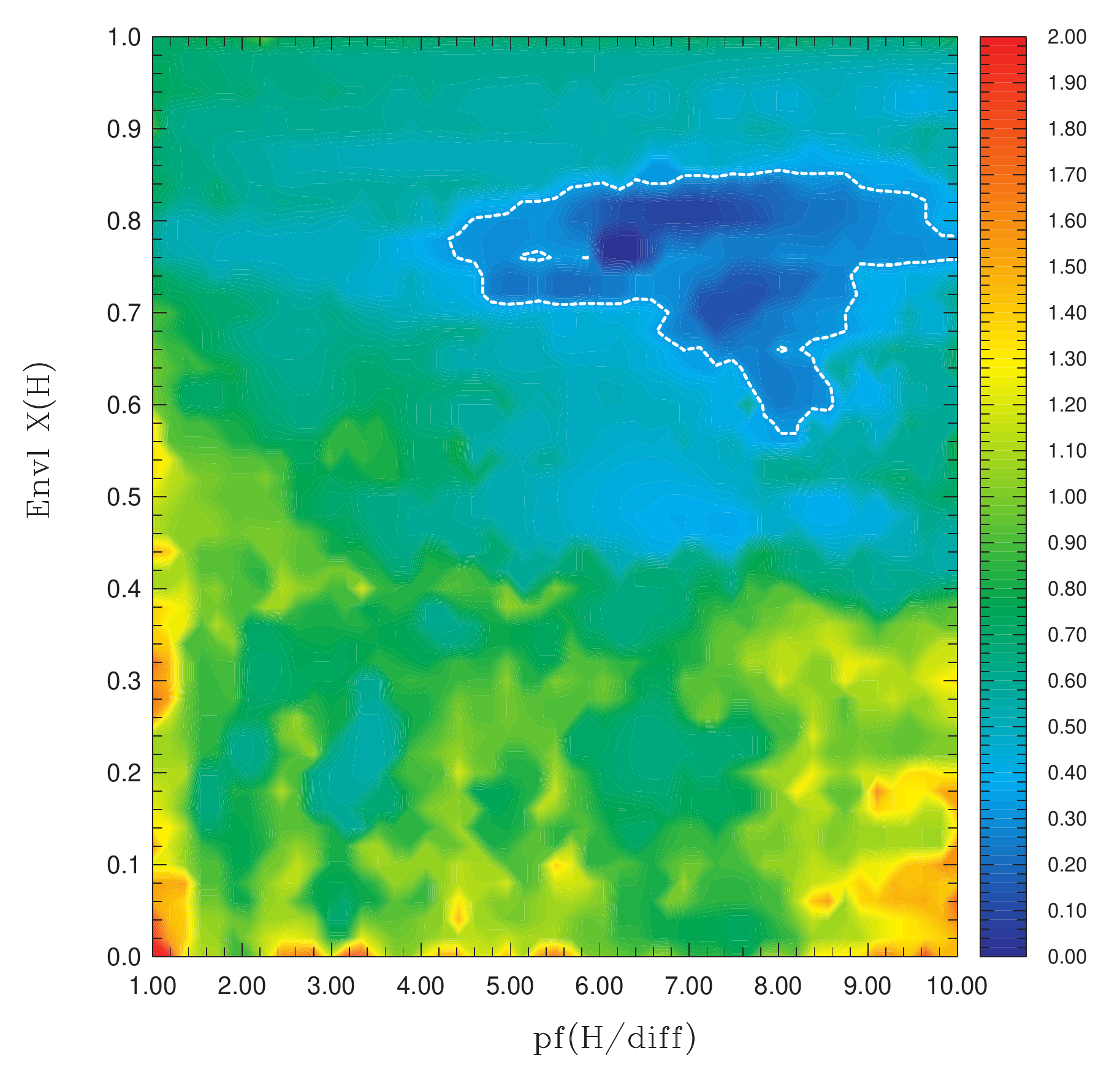}
\par\end{centering}
\caption{$\log S^{2}$-projection maps for pairs of model parameters showing
location and shape of best-fit regions in parameter space (see Sect.
3.2 and text for details). All pairs include the $X({\rm H})_{{\rm envl}}$
parameter, thus showing correlations that may exist between this quantity
and other model parameters. The quantity $S^{2}$ is normalized to one
at minimum and the color scale is logarithmic. Dark blue indicates
the best-fit regions and the dotted contour line is an estimate of
the $1\sigma$ confidence level, obtained in a similar way to that
described in \citet{2001ApJ...563.1013B}. \label{figA1}}
\end{figure*}

The model parameter space searched for a best-fit seismic solution
that matches the frequencies of TIC 278659026 was originally defined
to cover the widest possible range assumed relevant for typical
sdB stars. In this context, the quantity $X({\rm H})_{{\rm envl}}$,
which sets the hydrogen mass fraction (and consequently the H/He mass
ratio; see Fig.\ref{fig4}) at the bottom of the envelope, was kept
confined within a narrow range between 0.70 and 0.75, assuming that
the main sequence progenitor probably had an H-rich envelope of roughly
solar composition that remained unspoiled before reaching the hot
sdB stage. However, the results from the seismic analysis suggests
that TIC 278659026 has a relatively low mass and may not be a typical
sdB star. This star may have undergone a nondegenerate ignition in
the helium core, thus possibly originating from a massive progenitor.
If produced, for instance, through a RLOF phase, the
original composition of the envelope may have been altered to significantly
differ from standard evolution expectations, where $X({\rm H})_{{\rm envl}}$
is possibly well outside the narrow range originally defined. Moreover,
the lack of constraint from the seismic solution for this parameter
in the 0.70 \textendash{} 0.75 range (see Fig. \ref{fig5} and \ref{fig:hist-xh})
leaves open the possibility that a better seismic solution could reside
outside of the parameter space originally searched.

To address this issue, we present additional computations extending
the search to a much wider range of values for the $X({\rm H})_{{\rm envl}}$
parameter. A new optimization was launched within the ranges given
in the third column of Table \ref{tab:ranges}, except that $X({\rm H})_{{\rm envl}}$
was now allowed to vary between 0 and 1 (instead of 0.70 and 0.75
in the original search), thus covering the entire range of possible
values for this parameter. Owing to the larger parameter space open
for exploration, this calculation was performed on the high-performance
cluster OLYMPE at the CALMIP computing center using 960 CPU cores
in parallel,  which ran for approximately two days and required the computation
of 1,863,344 seismic models; thus this added to the 1,555,425 models already
calculated in this study. The results are illustrated in Fig. \ref{figA1}.

First, we find no significant improvement in terms of quality of fit
for the best solutions within this extended parameter space. The model
formerly identified still holds up as among the best, i.e., within
the $1\sigma$ contours shown in Fig. \ref{figA1}. Hence, no clear
optimal value exists for $X({\rm H})_{{\rm envl}}$ that could have
been missed because of the narrow range originally imposed. Instead, Fig.
\ref{figA1} reveals that best-fit models are found within a flat
and elongated region that covers the range $X({\rm H})_{{\rm envl}}\sim0.58-0.86$.
Models with envelope hydrogen mass fractions above and below these
limits appear to produce lower quality solutions. Therefore, a significantly
helium-enriched envelope (with $Y\gtrsim0.42$) for TIC 278659026
seems excluded.

The impact for the other model parameters of allowing a wider range
of values for $X({\rm H})_{{\rm envl}}$ is limited (we refer again
to Fig. \ref{figA1}). The best-fit stellar masses remain in the $\sim0.375-0.40$
$M_{\odot}$ range, which is almost entirely contained in the errors
quoted in Table \ref{tab2} for this quantity. Interestingly, only
models with $X({\rm H})_{{\rm envl}}\lesssim0.62$ result in slightly
lower mass solutions ($\lesssim0.38$$M_{\odot}$), which arguably
may be less consistent with other independent measurements for this
quantity. Similarly agreeing with values and errors given in Table
\ref{tab2}, the inferred helium and oxygen mass fractions in the
core, $X({\rm He})_{{\rm core}}$ and $X({\rm O})_{{\rm core}}$,
remain within the $0.48-0.62$ and $0.08-0.29$ ranges, respectively,
while the core boundary fractional mass depth, $\log q({\rm core})$,
is in the interval $[-0.31,-0.27]$. Slightly wider ranges for the
main parameters defining the envelope structure are obtained. The
best-fit fractional mass depth for the core-envelope boundary, $\log q({\rm envl})$,
ranges from $-2.30$ to $-1.80$; there is a clear correlation with $X({\rm H})_{{\rm envl}}$
, i.e., less hydrogen is associated with a slightly deeper transition.
The inferred location for the transition with pure-hydrogen layers,
$\log q({\rm H/diff})$, is in the interval $[-3.70,-3.30]$.

We conclude from this extended search that the parameters derived
from the original study restraining $X({\rm H})_{{\rm envl}}$ to
the narrow interval $[0.70,0.75]$ are not fundamentally changed when
widening this range. Therefore, the asteroseismic solution was not
artificially constrained by this prior. The main impact, if we allow
the H/He mass ratio to differ significantly from typical solar composition,
is slightly larger uncertainties for some derived parameters.

\end{appendix}
\end{document}